\documentclass{aa}  

\usepackage{graphicx}
\usepackage{txfonts}
\usepackage[colorlinks=true,allcolors=blue]{hyperref}
\usepackage{xcolor}
\usepackage{textgreek}
\usepackage{orcidlink}
\usepackage{bbding}
\graphicspath{Immagini/}

\begin{document} 

\newcommand{\solarM}{\,\mathrm{M}_\odot}
\newcommand{\solarL}{\,\mathrm{L}_\odot}
\newcommand{\E}[1]{\times10^{#1}}
\newcommand{\nH}{\,\mathrm{cm}^{-2}}
\newcommand{\ergcms}{\,\mathrm{erg}\,\mathrm{cm}^{-2}\,\mathrm{s}^{-1}}
\newcommand{\asec}{\,\mathrm{arcsec}}
\newcommand{\amin}{\,\mathrm{arcmin}}
\newcommand{\magnitude}{\,\mathrm{mag}}
\newcommand{\ergs}{\,\mathrm{erg}\,\mathrm{s}^{-1}}
\newcommand{\pdotunit}{\,\mathrm{s}\,\mathrm{s}^{-1}}
\newcommand{\LX}{L_\mathrm{X}}
\newcommand{\LEdd}{L_\mathrm{Edd}}
\newcommand{\pspin}{P_\mathrm{spin}}
\newcommand{\porb}{P_\mathrm{orb}}
\newcommand{\asini}{a_\mathrm{x}\sin i}
\newcommand{\FX}{F_\mathrm{X}}
\newcommand{\apm}[2]{_{-#1}^{+#2}}
\newcommand{\xmm}{\emph{XMM-Newton}}
\newcommand{\swift}{\emph{Swift}}
\newcommand{\chandra}{\emph{Chandra}}
\newcommand{\erosita}{\emph{eROSITA}}
\newcommand{\nustar}{\emph{NuSTAR}}
\newcommand{\src}{J1257}
\newcommand{\srclong}{2MASX\,J12571076+2724177}
\newcommand{\salt}{\emph{SALT}}
\newcommand{\rxte}{\emph{RXTE}}
\newcommand{\matt}[1]{\textcolor{red}{#1 - MI}}
\newcommand{\as}[1]{\textcolor{blue}{#1 - AS}}
\newcommand{\giallo}[1]{\textcolor{orange}{#1 - GLI}}
\newcommand{\paolo}[1]{\textcolor{green}{#1 - PE}}
\newcommand{\rob}[1]{\textcolor{cyan}{#1 - RA}}
\newcommand{\gio}[1]{\textcolor{purple}{#1 - GM}}
\newcommand{\rev}[1]{\textbf{#1}}

\defcitealias{Liu2021}{L21}

   \title{A supermassive black hole under the radar: repeating X-ray variability in a Seyfert galaxy}
   \titlerunning{Repeating X-ray variability in \src}
   \authorrunning{M. Imbrogno et al.}

   \author{Matteo Imbrogno
          \inst{1,2,3}\thanks{These authors contributed equally.}
          \orcidlink{0000-0001-8688-9784}
          \and
          Andrea Sacchi
          \inst{4}$^\star$
          \orcidlink{0000-0002-7295-5661}
          \and
          Giovanni Miniutti
          \inst{5}
          \orcidlink{0000-0003-0707-4531}
          \and
          Francesco Tombesi
          \inst{1,2,6}
          \orcidlink{0000-0002-6562-8654}
          \and
          Gian Luca Israel
          \inst{2}
          \orcidlink{0000-0001-5480-6438}
          \and
          Enrico Piconcelli
          \inst{2}
          \orcidlink{0000-0001-9095-2782}
          \and
          Roberta Amato
          \inst{2}
          \orcidlink{0000-0003-0593-4681}
          }

   \institute{Dipartimento di Fisica, Università degli Studi di Roma “Tor Vergata”, via della Ricerca Scientifica 1, I-00133 Rome, Italy\\ 
              \email{matteo.imbrogno@inaf.it}
         \and 
             INAF -- Osservatorio Astronomico di Roma, via Frascati 33, I-00078 Monte Porzio Catone (RM), Italy
         \and 
            Dipartimento di Fisica, Università degli Studi di Roma “La Sapienza”, piazzale Aldo Moro 5, I-00185 Rome, Italy
         \and 
             Center for Astrophysics $\vert$ Harvard \& Smithsonian, 60 Garden Street, Cambridge, MA 02138, USA
         \and 
             Centro de Astrobiología (CAB), CSIC-INTA, Camino Bajo del Castillo s/n, 28692 Villanueva de la Cañada, Madrid, Spain
         \and 
             INFN -- Roma Tor Vergata, via della Ricerca Scientifica 1, I-00133 Rome, Italy
             }

   \date{Received MONTH XX, YYYY; accepted MONTH XX, YYYY}

   \abstract{
   In the last few years, a few supermassive black holes (SMBHs) have shown short-term (of the order of hours) X-ray variability. Given the limited size of the sample, every new addition to this class of SMBHs can bring invaluable information. Within the context of an automated search for X-ray sources showing flux variability in the \chandra\ archive, we identified peculiar variability patterns in \srclong\ (\src), a SMBH in the Coma cluster, during observations performed in 2020. We investigated the long-term evolution of the flux, together with the evolution of the spectral parameters throughout the \chandra\ and \xmm\ observations, which cover a time span of approximately 20 years. We found that \src\ has repeatedly shown peculiar variability over the last 20 years, on typical timescales of $\simeq20-30$\,ks. From our spectral analysis, we found hints of a softer-when-brighter behaviour and of two well-separated flux states. We suggest that \src\ might represent a new addition to the ever-growing size of relatively low mass SMBHs ($M\simeq10^6-10^7\solarM$) showing extreme, possibly quasi-periodic X-ray variability on short time scales. The available dataset does not allow for a definitive classification of the nature of the variability. However, given the observed properties, it could either represent a quasi-periodic oscillation at a particularly low frequency or be associated with quasi-periodic eruptions in an AGN with peculiar spectral properties.
   }

   \maketitle
%

\section{Introduction}\label{sec:introduction}

The study of supermassive black holes (SMBHs) has recently been revolutionised by the detection of different types of (extreme) X-ray variability phenomena. A few SMBHs have shown quasi-periodic oscillations \citep[QPOs; for the original definition and a more recent review, see][]{vanderKlis1989,Ingram2019} at frequencies as low as a few $10^{-4}$\,Hz in their X-ray flux \citep[see e.g.][]{Smith2021,Zhang2023}. First detected in (Galactic) X-ray binaries \citep{Motch1983}, different models have been proposed to describe how QPOs are produced. The general consensus is that they arise from instabilities in the accretion process and/or geometrical changes of the accretion disk along the line of sight \citep[see e.g.][]{Stella1998,Stella1999,Tagger1999,Ingram2009}. In the case of accreting BHs in Galactic X-ray binaries showing QPOs at frequencies $\gtrsim1$\,Hz, it has been proposed that the QPO frequency can be used as an independent tool to estimate the mass of the accreting BH, with $\nu_{\mathrm{QPO}}\propto1/M_\mathrm{BH}$ \citep[see e.g.][]{Abramowicz1988,Remillard2006,Motta2018}. Later, it has been suggested that this relation can be extended to the case of candidate intermediate-mass BHs in ultraluminous X-ray sources \citep{Casella2008} and SMBHs \citep[see e.g.][]{Zhou2015,Smith2018,Smith2021} showing (sub-)mHz QPOs\footnote{A note of warning is needed. Tempting as it is, one should use these relations with care and always verify the results with independent estimates, since (sub-)mHz QPOs are also detected in X-ray binaries hosting both stellar-mass ($M_\mathrm{BH}\lesssim10-20\solarM$) BHs \citep[see e.g.][]{Vikhlinin1994,Morgan1997} and neutron stars \citep[see e.g.][]{Lyu2015,Sidoli2016,Imbrogno2024}.}.

\cite{GonzalezMartin2012} analysed the time variability of a sample of 104 AGN and confirmed only the QPO detected in the source RE\,J1034+396, suggesting that most previous detections of QPOs in SMBHs were unreliable. Later, \cite{Zhang2023} reached a similar conclusion by analysing active galactic nuclei (AGN) data with (tentative) QPOs using Gaussian Processes, again confirming the presence of QPOs only in the source RE\,J1034+396 \citep{Gierlinski2008}. Thanks to its high significance and the high number of observations showing QPOs, RE\,J1034+396 has been extensively studied in the last few years \citep[see e.g.][]{Xia2024,Xia2025}. It should be noted that the (re)analysis of \cite{Zhang2023} does not include sources such as 2XMM\,J123103.2+110648 \citep{Lin2013}, whose QPO has a highly significant ($\sim5\sigma$) detection. At the same time, after their work, other (highly) significant QPOs have been reported, such as Mrk\,142 \citep{Zhong2024} and 1ES\,1927+654 \citep{Masterson2025}. The latter is an AGN that exhibited extreme variability in the recent past \citep{Ricci2021,Masterson2022}.

While QPOs are detected in X-ray binaries and SMBHs alike, in recent years a dozen SMBHs have shown other unique X-ray phenomena, such as quasi-periodic eruptions (QPEs). These high-amplitude, soft X-ray flares typically last 1-10~h and recur quasi-periodically every few-100~h. QPEs shine at X-ray luminosities of $\simeq10^{42}-10^{43}\ergs$ and show a thermal spectrum with typical temperatures $kT \simeq 100$-$200$~eV \citep{Miniutti2019,Giustini2020,Arcodia2021,Chakraborty2021,Quintin2023,Arcodia2024,Nicholl2024,Chakraborty2025,Arcodia2025,Hernandez-Garcia2025}. 

The origin of QPEs is still debated. Another phenomenon unique to SMBHs is the occurrence of tidal disruption events \citep[TDEs; see][]{Rees1988,Komossa2015,Gezari2021}, during which a star in the proximity of the SMBH gets ripped by the tidal forces of the latter. A link between QPEs and TDEs is emerging from several lines of evidence, ranging from host properties \citep{Wevers2024} and X-ray long-term variability \citep[e.g.][]{Shu2018,Miniutti2023,Chakraborty2021} to the detection of QPEs in optically selected, spectroscopically confirmed TDEs \citep[see e.g.][]{Quintin2023,Nicholl2024,Chakraborty2025}. At the same time, models invoking the repeated interaction between the SMBH accretion disk and a stellar-mass compact object orbiting around the SMBH itself are gaining significant traction \citep[e.g.][]{Linial2023,Franchini2023}. In these systems, known as extreme-mass-ratio inspirals \citep[EMRIs; see e.g.][]{AmaroSeoane2015,AmaroSeoane2018}, the compact object is inspiraling towards the SMBH due to the loss of energy through gravitational waves. QPEs might therefore represent the electromagnetic counterparts or precursors of some of the most promising sources of gravitational waves in the 0.1--100\,mHz range, to be observed with future space-based interferometric detectors such as LISA \citep[e.g.][]{AmaroSeoane2017,Colpi2024,Kejriwal2024,Lui2025arXiv}.

In this paper, we report on the discovery of repeating X-ray variability in the Seyfert galaxy \srclong\ (\src\ hereafter). We identified this source within the framework of the CATS@BAR project \citep[][see also \citealp{Esposito2013a,Esposito2013,Esposito2015,Sidoli2016,Sidoli2017,Bartlett2017}]{Israel2016a}, aimed at the search for new variable X-ray sources in the \chandra\ archive. \src\ is located in the Coma cluster, at a redshift $z=0.02068$ \citep{Ahn2012}. This source is also present in the sample of local AGN showing broad-H$\alpha$ variability studied by \cite{Liu2021}. Following Equation 6 of \cite{Xiao2011}, which combines the measured luminosity and the broad-H$\alpha$ line width of an AGN to estimate its virial mass, \cite{Liu2021} derived a mass $\log M_\mathrm{BH}/M_\odot\sim6.3$, in the low-mass tail of the SMBH population.

The article is structured as follows: in Sect.~\ref{sec:ObsDataReduction}, we describe the observation analysed in this paper and the data reduction process. In Sect.~\ref{sec:results}, we present the results of our timing and spectral analysis. We discuss possible interpretations for our results in Sect.~\ref{sec:discussion} and draw our conclusions in Sect.~\ref{sec:Conclusions}.    

\section{Observations and data reduction}\label{sec:ObsDataReduction}

The field of \src\ has been observed 23 times with relatively deep, pointed X-ray observations from \xmm\ \citep{Jansen2001} and \chandra\ \citep{Weisskopf2000}. We discarded those observations in which \src\ falls in a chip gap 
or (in the case of \xmm) the background level is too high. We list the remaining 17 observations (8 from \chandra, 9 from \xmm) that we analysed in this work in Table~\ref{tab:XrayObs}, where we also report the net exposure time after data reduction and flare filtering. \swift\ \citep{Burrows2005} also observed \src\ field 4 times, with 2 observations in 2017 and 2 observations in 2022. While in 2017 \swift's total observing time amounts to $\sim120$\,s, in 2022 \swift\ observed \src\ for a total of $\sim1500$\,s. Given the short exposure times, we did not consider these observations for this work. We also checked the \erosita's archive, but no data are available for \src, as its location is not within the footprint of the eROSITA-DE Data Release 1. To extract the source events we centered the \chandra\ and \xmm\ source regions at the \textit{GAIA} coordinates ($\mathrm{R.A.}=12^\mathrm{h}57^\mathrm{m}10\fs76$, $\mathrm{Dec}=27^\circ24'17\farcs7$, J2000; \citealt{GAIA2020}). 

\begin{table}
    \centering
    \caption{Log of \xmm\ and \chandra\ observations analysed in this work.
    }
    \resizebox{\columnwidth}{!}{
    \begin{tabular}{ccccc}
    \hline
    \hline
     Satellite & ObsID & Start Date & Exposure time\tablefootmark{a} & Net Counts\tablefootmark{a} 
     \\ 
      &  &  & (ks) &\\ 
    \hline
    \xmm\ & 0124710101 & 2000 Jun 21 & 23.7 / 34.6 / 34.6 & 1182 / 411 / 376
    \\
    \xmm\ & 0403150201 & 2006 Jun 11 & -- / 44.9 / 44.8 & -- / 216 / 232  
    \\
    \xmm\ & 0403150101 & 2006 Jun 14 & -- / 41.2 / 41.7 & -- / 212 / 189 
    \\
    \xmm\ & 0652310401 & 2010 Jun 24 & 8.6 / -- / -- & 996 / -- / --
    \\
    \chandra\ & 12887 & 2010 Nov 11 & 43.4 & 318
    \\
    \xmm\ & 0652310801 & 2010 Dec 03 & 4.4 / 9.0 / 9.0 & 461 / 249 / 202 
    \\
    \xmm\ & 0652310901 & 2010 Dec 05 & -- / 11.1 / 11.1 & -- / 607 / 557  
    \\
    \xmm\ & 0652311001 & 2010 Dec 11 & -- / 5.9 / -- & -- / 234 / -- 
    \\
    \xmm\ & 0691610201 & 2012 Jun 02 & 24.1 / -- / 37.2 & 2035 / -- / 723 
    \\
    \xmm\ & 0691610301 & 2012 Jun 04 & 17.7 / -- / 32.3 & 1039 / -- / 429 
    \\
    \chandra\tablefootmark{b} & 22648 & 2020 Mar 03 & 32.6 & 451
    \\
    \chandra\tablefootmark{b} & 22649 & 2020 Mar 04 & 34.1 & 483 
    \\
    \chandra\tablefootmark{b} & 23182 & 2020 Mar 04 & 30.7 & 523
    \\
    \chandra\ & 22930 & 2020 Aug 11 & 36.8 & 759  
    \\
    \chandra\ & 23361 & 2020 Nov 02 & 18.8 & 72 
    \\
    \chandra\ & 24853 & 2020 Nov 03 & 26.7 & 323 
    \\
    \chandra\ & 24854 & 2020 Nov 08 & 7.9 & 175 
    \\
    \hline
    \hline
    \end{tabular}
    }
    \tablefoot{
    \tablefoottext{a}{In the case of \xmm, the times (net counts) are reported in the order EPIC pn/EPIC MOS1/EPIC MOS2, and after cleaning from background flaring. A dash means that the data from the corresponding camera has not been analysed, either because data are not available or because of the high background.}
    \tablefoottext{b}{In the text, we refer to these three observations as dataset C1.}
    }
    \label{tab:XrayObs}
\end{table}

\subsection{Chandra}

We downloaded the \chandra\ \src\ observations from the \chandra\ Search \& Retrieval\footnote{\url{https://cda.cfa.harvard.edu/chaser/}} (ChaSeR). For the data reduction and the extraction of source events, light curves, and energy spectra, we used the \chandra\ Interactive Analysis of Observations (\textsc{CIAO}) software v4.16.0 \citep{Fruscione2006} and v4.11.3 of the calibration database. We reprocessed the observations with the task \texttt{chandra\_repro}. All observations were performed in \texttt{VFAINT} mode. Therefore, we set the \texttt{checkvfpha} flag to \texttt{yes}. For all the observations except ObsID 12887, we used a circular source extraction region of radius 4\arcsec, while to estimate the background contribution, we used a nearby circular region with a radius of 60\arcsec, free of other X-ray sources. During observation 12887, the source is off-axis and appears elliptical. Therefore, for this observation, we considered a 7\arcsec$\times$5\arcsec\ elliptical source extraction region. For both timing and spectral analysis, we only considered events in the 0.5--7\,keV band. We generated source and background energy spectra, together with spectral redistribution matrices and ancillary response files, with the task \texttt{specextract} and verified that the source contribution to the total spectrum was at least 95\%. We discarded observation 23361 for spectral analysis due to its poor statistics (72 detected photons), but retained it for timing analysis to study the source variability. We then extract source events and background-corrected light curves for timing analysis using the tasks \texttt{dmcopy} and \texttt{dmextract}, respectively. Applying the barycentric correction to the \chandra\ data would lead to a change in the photon time of arrivals of $\lesssim500$\,s \citep[see e.g.][]{Backer1986}, that is $\lesssim2-2.5\%$ of the timescale of the possible modulation of \src, comparable with our uncertainty on the estimated period. For this reason, and since we are not interested in high-precision timing analysis, we did not barycenter the data.

\subsection{XMM-Newton}

For each \xmm\ observation, we considered data coming from both EPIC pn \citep{Struder2001} and EPIC MOS \citep{Turner2001} cameras. Due to the source falling in a chip gap or the presence of high background in one or two cameras, simultaneous data from all three cameras are available only for two observations (ObsIDs 0124710101 and 0652310801). In Table~\ref{tab:XrayObs} we report the details of the analysed cameras for each observation. To prepare the raw \xmm\ data for both timing and spectral analysis, we used \textsc{SAS} \citep{Gabriel2004} v21.0.0 with the latest \xmm\ calibrations and applied standard data reduction procedures. EPIC pn and EPIC MOS data were reduced using the \texttt{epproc} and \texttt{emproc} tasks, respectively. We selected only the events with $\texttt{PATTERN}\leq4$ from the EPIC pn data and events with $\texttt{PATTERN}\leq12$ from the EPIC MOS data. We extracted the high-energy ($E > 10$\,keV) light curves of the entire field of view to verify the presence of high-background particle flares. We filtered out time intervals in which the PN (MOS) background count rate was higher than 0.4 (0.35) cts/s using the task \texttt{tabgtigen}. For observations 0652310701, 0652310401, 0691610201, and 0691610301, given the higher background level, for the EPIC pn data we chose a threshold of 0.5, 0.5, 0.45, and 0.45 cts/s, respectively.

We considered events in the 0.3--10\,keV band for both our timing and spectral analysis. We extracted source events from a circular region with a radius of 15\arcsec\ centred at the \textit{GAIA} coordinates. The source always falls near a chip border or in a corner of the chip; therefore, for background extraction, we considered a nearby circular region with a radius of 40\arcsec\ in the same CCD and free of other X-ray sources. To correct for the background and for the vignetting given by the off-axis position of the source, we produced background-subtracted light curves using the task \texttt{epiclccorr}. Again, given the timescales of the variability and the maximum amplitude of the correction to the photon time of arrivals, the events were not barycentered. We used the tasks \texttt{rmfgen} and \texttt{arfgen} to create response matrices and ancillary files, respectively. Spectra have been rebinned to have at least 1 count per energy bin. 


\section{Data analysis and results}\label{sec:results}

In the following, to compute the source luminosity, we converted \src\ redshift in distances assuming a standard flat $\Lambda$CDM cosmology ($\Omega_\mathrm{M}=0.3$, $\Omega_\Lambda=0.7$, $H_0=70$\,km\,s$^{-1}$\,Mpc$^{-1}$). Using the NASA Extragalactic Database (NED) Cosmology Calculator\footnote{\url{https://ned.ipac.caltech.edu/help/cosmology_calc.html}}, we found that \src\ is located at a distance $d\simeq90$\,Mpc, which is the value we use throughout the paper. Unless otherwise specified, the reported errors correspond to 1$\sigma$ (68.3\%) confidence ranges.

\subsection{X-ray timing analysis}\label{sec:timing}

We analysed the light curves of the observations reported in Table~\ref{tab:XrayObs} using the \textsc{XRONOS} task \texttt{lcurve}\footnote{\url{https://heasarc.gsfc.nasa.gov/docs/software/xronos/help/lcurve.html}}. We considered only events in the 0.3--10\,keV and the 0.5--7\,keV bands for \xmm\ and \chandra's data, respectively. Most of the timing analysis relies upon \chandra's 2020 observations, which provide the best coverage of the possible recurrent X-ray variability. We report in this Section a selection of light curves as exemplary of \src\ phenomenology. The light curves of the observations not discussed in detail in this Section are shown in Appendix~\ref{appendix:timing}, where one can see that a certain degree of variability is almost always present.

\subsubsection{Recurrent modulating behaviour}\label{sec:modulatingbehaviour}

\begin{figure}
    \centering    \includegraphics[width=0.95\columnwidth]{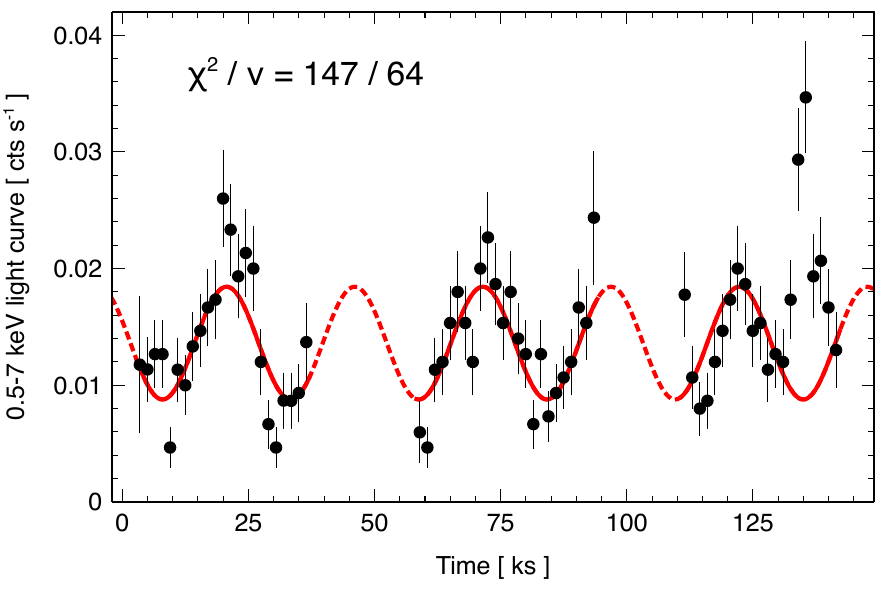}
    \caption{\chandra\ light curve (black points) in the 0.5--7\,keV band of ObsIDs 22648, 22649, and 23182. The bin time is 1500\,s. The best-fit sinusoidal is shown in red. The dotted line shows the model's extrapolation during time intervals with no data.}
    \label{fig:3BestChandraLC}
\end{figure}

In Fig.~\ref{fig:3BestChandraLC} we show the light curve in the 0.5--7\,keV band of the \chandra\ ObsIDs 22648, 22649, and 23182, performed within 2 days in 2020. In the following, we will refer to these observations as dataset C1. A modulation on the timescale of a few tens of kiloseconds is clearly visible. To estimate the period of the modulation, we fitted the C1 light curve with a sinusoid function, plus a constant to model the mean flux level. We obtained a best-fit period of $P=25.4\pm0.4$\,ks, with $\chi^2/\mathrm{dof}=147/64$. The overall bad quality of the fit is probably driven by the presence of a peak in the last 10\,ks of the light curve. The peak appears to happen where one would expect a minimum of the modulation. As one can see from Fig.~\ref{fig:069201XMMhr} and Fig.~\ref{fig:C12887}, \src\ showed at least twice a flare-like event when the flux is low (see Section~\ref{sec:flare}). A similar event could have happened during the last 10\,ks of the C1 light curve. We discuss in more detail the implications of this in Sect.~\ref{sec:qpe_g}, where we consider the possibility that this source represents a QPE candidate. Here, we note that if we add a Gaussian component to model the flare, the fit significantly improves, with $\chi^2/\mathrm{dof}=75/61$.

\begin{figure}
    \centering
    \includegraphics[width=\columnwidth]{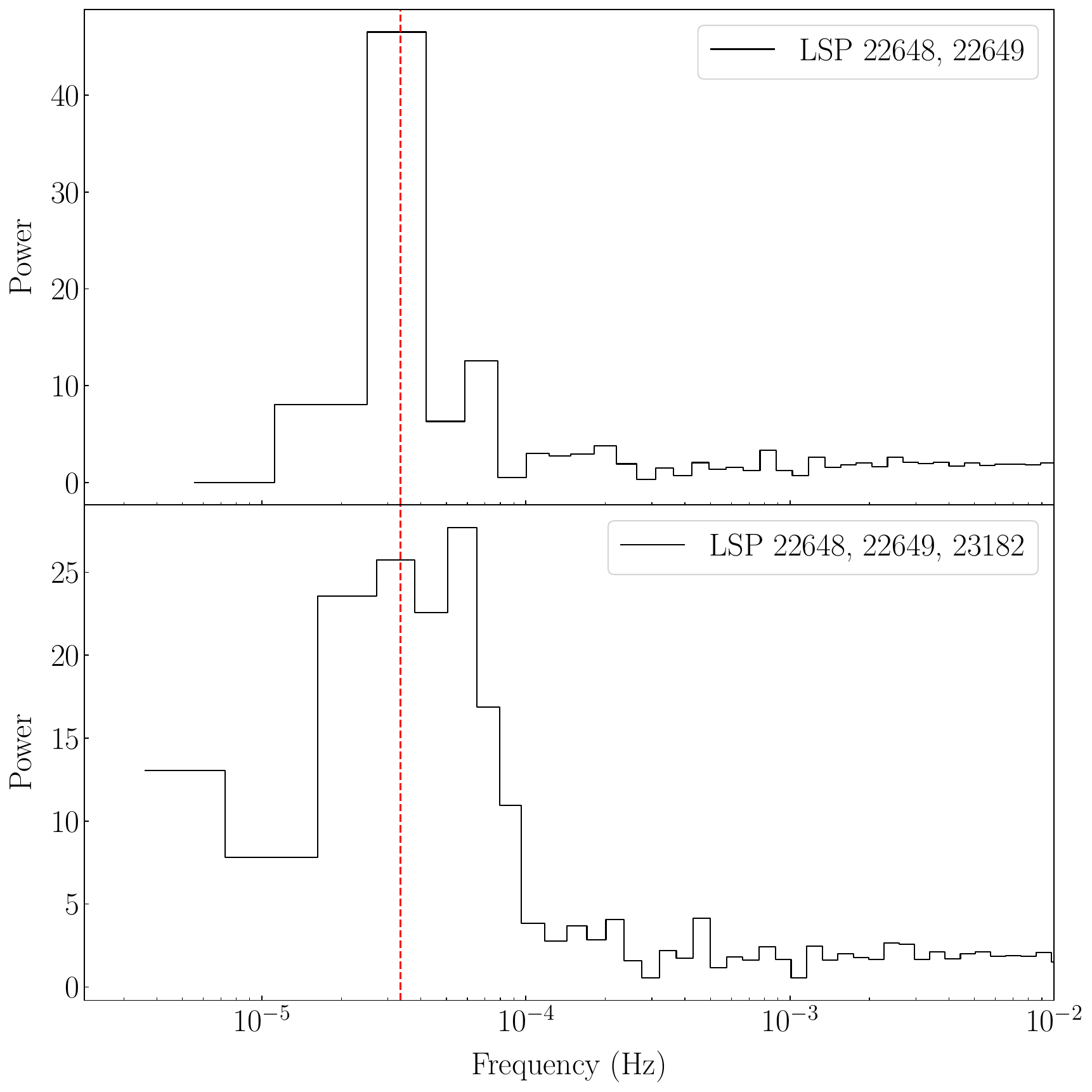}
    \caption{LSPs of the C1 dataset. Top panel: LSP of observations 22648 and 22649. Bottom panel: LSP of the whole C1 dataset (bottom). The LSPs have been rebinned with a logarithmic factor of 1.14. The red dashed lines show the frequency of the peak found in the top LSP. At frequencies higher than $\nu\simeq0.01$\,Hz, white noise dominates. Therefore, for visual purposes, we show the LSPs only up to 0.01\,Hz.}
    \label{fig:discoveryPDSs}
\end{figure}

In Fig.~\ref{fig:discoveryPDSs}, we show the Lomb-Scargle periodograms \citep[LSP;][]{Lomb1976,Scargle1982} of the C1 dataset. The two LSPs, computed with the version implemented in the \texttt{Stingray} and \texttt{hendrics} Python packages \citep{Huppenkothen2019,Huppenkothen2019a,Bachetti2018,Bachetti2024} and considering the data from ObsIDs 22648 and 22649 (top panel) and from the whole C1 dataset (bottom panel), both show a peak at a frequency of $\simeq3.3\E{-5}$\,Hz ($\simeq$30\,ks). The LSP of the whole C1 dataset shows a broad peak, implying that the modulation is quasi-coherent in nature. 
Estimating the significance of this peak is tricky. At such low frequencies, one cannot use the standard white noise assumption and compute a frequency-independent threshold \citep[see][]{vanderKlis1989a,Israel1996}. At the same time, it is difficult to verify the presence of red noise in the LSPs in Fig.~\ref{fig:discoveryPDSs}. When trying to add a power-law component to model it, the parameters are not constrained, likely because the red noise dominates only in the very first bins. Therefore, we lack the necessary statistics to test and quantify the significance of such a component. Finally, we note that we cannot exclude that red noise is the actual origin of the observed possible modulation. However, the presence of a peak in both LSPs would be hard to explain in this case.

\begin{figure}
    \centering
    \includegraphics[width=0.95\columnwidth]{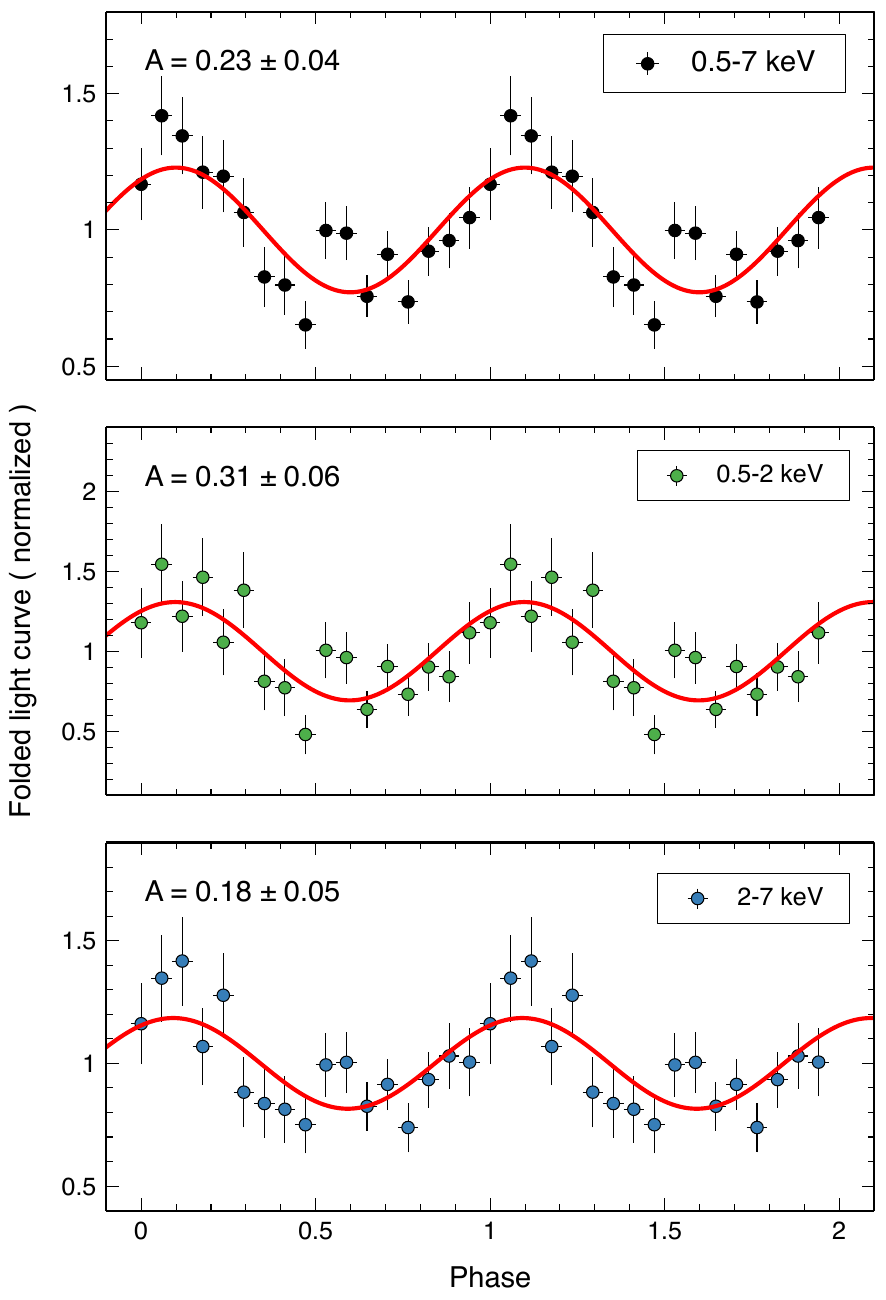}
    \caption{Phase-folded profile of the \chandra's C1 light curve. Top panel: folded light curve in the whole 0.5--7\,keV band. Middle panel: folded light curve in the soft 0.5--2\,keV band. Bottom panel: folded light curve in the hard 2--7\,keV band. The input period in the three panels is the same as that of the best-fit sinusoid shown in Fig.~\ref{fig:3BestChandraLC} ($P=25.4\pm0.4$\,ks). $A$ is the semi-amplitude of the modulation. Two cycles are shown in each panel for clarity.}
    \label{fig:2020ChandraEfold}
\end{figure}

Fig.~\ref{fig:2020ChandraEfold} shows the 0.5--7\,keV, 0.5--2\,keV, and 2--7\,keV folded light curve of all the \chandra's C1 observations, phase-folded at the 25.4\,ks period derived from the sinusoid fit of C1. In each panel, we also show the value of the semi-amplitude $A$ of the modulation, derived from fitting the data with the function $C+A\sin[2\pi(x-x_0)/P]$. $C$ is the normalised mean flux level and $P$ the period of the modulation, both fixed to 1, while $x_0$ is the phase of the modulation. The amplitude $A_\mathrm{soft}=(31\pm6)\%$ is larger in the soft band than in the hard band $A_\mathrm{hard}=(18\pm5)\%$, although we must note that the amplitudes in the two bands are consistent within 3$\sigma$. The folded profile, especially in the whole 0.5--7\,keV band, appears not strictly sinusoidal. Given the short exposure times compared to the length of the modulation, it is difficult to assess whether the flux goes to background level for a prolonged time, as expected for QPEs, or if the flux keeps oscillating, like in a QPO.

From a visual inspection of the light curves reported in Appendix~\ref{appendix:timing}, it is interesting to note that \src\ has shown a similar variability behaviour on other occasions. Throughout 2020, \src\ flux is characterised by a modulating behaviour. For example, during ObsID 22930 (Figure~\ref{fig:C22930}) one can see two peaks separated by $\simeq20$\,ks. During ObsID 0124710101, two flares on top of a steady rise in the flux are also present (Figure~\ref{fig:X0124710101}). However, the short duration of the exposures prevents us from performing a detailed analysis of the observed variability. 

Finally, we must note that adding the other \chandra\ observations performed in 2020 (i.e., ObsIDs 22930, 23361, 24853, and 24854) to construct the folded profile progressively diminishes the amplitude of the modulation. The trend continues if we add the older \chandra\ and \xmm\ observations. However, this is not surprising and should not be directly considered as an indication of the absence of a real signal. Given the present uncertainty on the period of the possible modulation, we cannot perform a phase connection between observations that were performed months (or even years) before or after the C1 observations. Additionally, if the modulation is quasi-periodic in nature, the period could change significantly among the different epochs, contributing to the loss of amplitude in the folded profile.

\subsubsection{Flare-like behaviour}\label{sec:flare}

\begin{figure}
    \centering
    \includegraphics[width=\columnwidth]{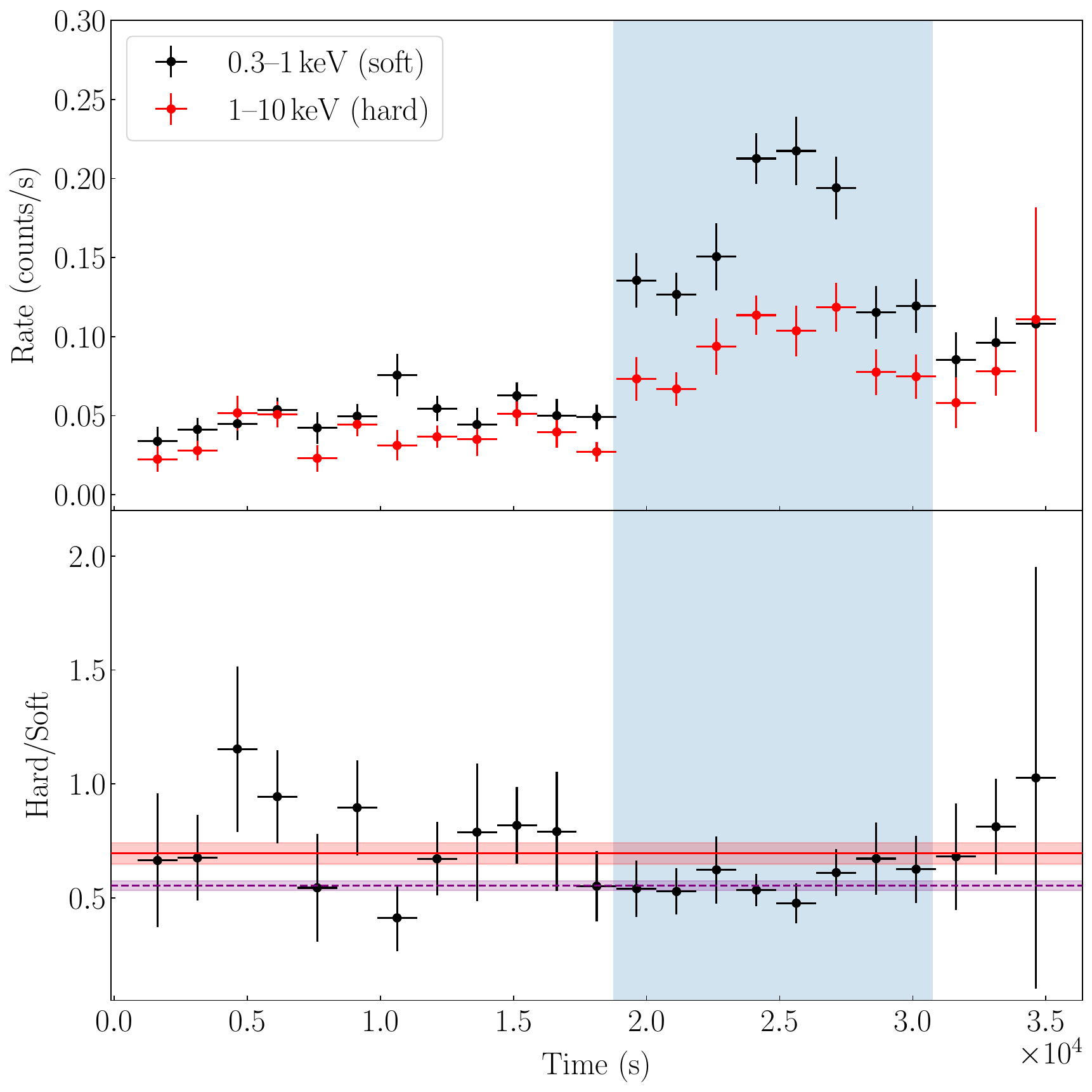}
    \caption{Light curve of \src\ during the \xmm\ observation 0691610201. Top panel: light curve in the soft (0.3--1\,keV) and hard (1--10\,keV) band. Bottom panel: hardness ratio of the hard count rate over the soft count rate. The red (dashed purple) line shows the result of the fit of the hardness ratio with a constant outside (within) the high-flux phase (highlighted in blue). The corresponding shaded region shows the 1$\sigma$ confidence interval. The bin time in both panels is 1500\,s.}
    \label{fig:069201XMMhr}
\end{figure}

\src\ showed (at least twice) a flaring-like behaviour, in which the count rate suddenly increased by a factor of 5, that is ObsID 12887 and ObsID 0691610201. The light curve of ObsID 12887 is shown in Fig.~\ref{fig:C12887}. Here, we focus on ObsID 0691610201, which has a higher statistic and allows for a better study of the energy-dependent emission. In the upper panel Fig.~\ref{fig:069201XMMhr}, we present the \xmm\ light curve from ObsID 0691610201 in the soft (0.3--1\,keV, black points) and hard (1--10\,keV, red points) band, together with the hardness ratio (hard/soft) in the lower panel. We tried different energy cuts. Based on the shape of the energy spectrum (see e.g. Figure~\ref{fig:xspec}), we selected a soft band from 0.3\,keV to 0.6\,keV and a hard band from 1.5\,keV to 10\,keV or from 2\,keV to 10\,keV. These choices did not change our results. Therefore, we opted for the cut at 1\,keV between the two bands to have a similar number of photons and similar statistics. 

We identified a low-flux state and a high-flux state (highlighted in blue in Figure \ref{fig:069201XMMhr}). After the first $\simeq18$\,ks, during which the count rate has a mean value of $\simeq0.05$\,counts/s, it rapidly increases of a factor $\sim5$ in about 5\,ks. After a decrease in the following $5-6$\,ks, this high-flux state ends, but right after it, the count rate seems to rise again. Unfortunately, this rise coincides with the end of the observation. Therefore, we cannot verify whether \src\ is showing the same modulation we detected in the previously shown \chandra\ observations or a different kind of variability. It is interesting to note, however, that the flare during ObsID 0691610201 has a similar duration ($\sim10$\,ks) and peak/out-of-peak flux ratio in the 0.5--7\,keV band ($\sim5$) of the flare detected at the end of ObsID 23182, shown in Fig.~\ref{fig:3BestChandraLC}. These similarities could suggest that the two flares are produced by the same mechanism. 

To model the evolution of the hardness ratio, we considered the values outside (within) the high-flux state and fitted the hardness ratio with a constant. The result of our fit is shown by the red (purple) solid (dashed) line in Fig.~\ref{fig:069201XMMhr}. The flux seems to get softer during the high-flux state, but the values of the hardness ratio outside ($HR_\mathrm{low}=0.69\pm0.05$) and during ($HR_\mathrm{high}=0.55\pm0.02$) this phase are consistent within 3$\sigma$. Therefore, we cannot claim any significant change in the hardness ratio during the observation. We studied the hardness ratio for all the \xmm\ and \chandra\ observations reported in Table~\ref{tab:XrayObs}. For the \xmm\ observations, we used the same 0.3--1/1--10\,keV bands, while for \chandra\ observations we used a soft 0.5--2\,keV band and a hard 2--7\,keV to met our requirement on an equal number of photons in both bands. As for ObsID 0691610201, we always found that a constant is sufficient to model the evolution of the hardness ratio. 


\subsection{X-ray spectral analysis}
The X-ray spectra were analysed with the spectral fitting package \texttt{XSPEC} \citep{Arnaud1996} version 12.12.1. The spectra were rebinned to have at least 1 count per energy bin, and the modified Cash statistic \citep[W-stat;][]{Cash1979} was adopted. All X-ray spectra were corrected for foreground interstellar absorption adopting the model \textsc{TBabs}, with $N_\textup{H}$ fixed to the Galactic value $8.4\times10^{19}$ cm$^{-2}$ \citep{HI4PICollaboration2016}, using abundances from \citet{Wilms2000}, with the photoelectric absorption cross-sections from \citet{Verner1996}. All reported uncertainties correspond to $1\sigma$, all reported fluxes and luminosities are corrected for absorption.

J1257 shows significant variability in its spectral behaviour (see Table~\ref{tab:spec}): while a simple power-law model can reproduce satisfactorily the X-ray spectrum of the source in most observations, in some epochs the quality of the fit is significantly improved by the addition of a thermal component and/or a layer of intrinsic absorption (in excess of the Galactic one). We modelled the thermal component with a simple blackbody profile, with temperatures of about $kT\sim110$~eV. This thermal component is mostly detected in observations taken before 2012, as in the latest \chandra\ observations, the loss of effective area in the soft band \citep[below $\approx1$~keV; see][]{Plucinsky2018,Plucinsky2022} prevents us from constraining its presence. The layer of intrinsic neutral absorption improves the quality of the fit in the most recent observations (from 2012 and in 2020), and it is characterised by a column density in the $(0.1-2)\times10^{22}$~cm$^{-2}$ range. The photon index $\Gamma$ varies significantly too, spanning a wide range of values, from $\Gamma=0.8$ to $\Gamma=2$.

This spectral evolution is associated with significant long-term variability in the source's flux. As reported in Fig. \ref{fig:lc}, the source's flux varies by a factor of about 3 in time scales of days, both in the soft (0.3--2~keV) and hard (2--10~keV) band. We inspected the correlation between the spectral shape and the flux state of the source and, as highlighted in Fig. \ref{fig:fg1}, \src\ appears to exhibit two distinct behaviours. When the source's flux is below $\approx3\times10^{-13}$~erg/s/cm$^2$, it shows the full range of spectral steepness described above. When instead the source flux is above the $\approx3\times10^{-13}$~erg/s/cm$^2$ threshold, it seems to exhibit a softer-when-brighter behaviour. However, the heterogeneity of the available dataset (e.g., the limited effective area in the soft X-ray regime of {\em Chandra} in the more recent observations, the presence of a poorly constrained local absorber, and the possibility that some spectral changes may be linked to the short-term variability), limits a more in-depth analysis of this feature.

The X-ray spectral properties of the source are listed in Tab.~\ref{tab:spec}. An exemplary spectrum (rebinned for graphical purposes) is shown in Fig. \ref{fig:xspec}, and the rest of the spectra are reported in Appendix \ref{appendix:spectra}. A visual inspection of the reported spectra highlights the extreme variability in the spectral shape of \src, corroborating the conclusion that the variations in the photon index of the power-law component, reported in Tab.~\ref{tab:spec}, are intrinsic and not driven by external factors, such as unmodeled absorption components.

\begin{table*}
    \centering
    \caption{X-ray spectral parameters of the source in the different epochs.}
    \begin{tabular}{ccccccccc}
    \hline
    \hline
     Epoch & Instrument & $C/\nu^a$ & $N_\textup{H}^b$ & $\Gamma$ & $kT^c$ & $F_\textup{X,s}^d$ & $F_\textup{X,h}^e$ & $L_\textup{X}^f$\\ 
    \hline
    2000 Jun 21 & XMM & 332.08/316 & --                     & $1.60\pm0.05$          & $0.11\pm0.01$    & $0.95\pm0.02$       & $1.17\pm0.08$          & $2.09\pm0.08$\\
    2006 Jun 11 & XMM & 192.28/182 & --                     & $1.20\pm0.09$          & $0.032\pm0.001$ & $0.57\pm0.06$       & $1.2\pm0.1$            & $1.7\pm0.1$\\
    2006 Jun 14 & XMM & 200.53/189 & --                     & $0.8\pm0.1$            & $0.12\pm0.04$    & $0.32\pm0.03$       & $1.5\pm0.2$            & $1.8\pm0.2$\\
    2010 Jun 24 & XMM & 108.89/102 & --                     & $1.77_{-0.08}^{+0.12}$ & $0.95\pm0.02$    & $1.95\pm0.09$       & $1.9\pm0.2$            & $3.7\pm0.2$\\
    2010 Nov 11 & CXO & 165.98/176 & --                     & $1.4\pm0.1$            & --               & $0.49\pm0.04$       & $1.1\pm0.1$            & $1.6\pm0.1$\\
    2010 Dec 03 & XMM & 193.69/215 & --                     & $1.72\pm0.07$          & $0.13\pm0.02$    & $1.99\pm0.09$       & $1.8\pm0.2$            & $3.7\pm0.2$\\
    2010 Dec 05 & XMM & 179.80/191 & --                     & $1.98\pm0.06$          & $0.13\pm0.02$    & $4.6\pm0.2$         & $3.0\pm0.3$            & $7.5_{-0.3}^{+0.4}$\\
    2010 Dec 11 & XMM & 76.39/69   & --                     & $2.0\pm0.2$            & $0.06\pm0.02$    & $3.2\pm0.03$        & $2.3_{-0.4}^{+0.5}$    & $5.4_{-0.5}^{+0.6}$\\
    2012 Jun 02 & XMM & 238.15/229 & --                     & $1.99\pm0.06$          & $0.09\pm0.01$    & $1.55\pm0.04$       & $1.07_{-0.07}^{+0.08}$ & $2.57\pm0.08$\\
    2012 Jun 04 & XMM & 217.47/198 & $0.07\pm0.01$          & $1.9\pm0.1$            & $0.084\pm0.005$  & $1.7\pm0.1$         & $0.81_{-0.07}^{+0.08}$ & $2.4\pm0.1$\\
    2020 Mar 03 & CXO & 241.84/248 & $0.4\pm0.3$            & $1.3\pm0.2$            &  --              & $1.0_{-0.2}^{+0.3}$ & $2.9\pm0.2$            & $3.9_{-0.3}^{+0.4}$\\
    2020 Mar 04 & CXO & 221.13/245 & $0.2\pm0.2$            & $1.1\pm0.2$            &  --              & $0.8\pm0.2$         & $3.2\pm0.3$            & $3.9_{-0.3}^{+0.4}$\\
    2020 Mar 04 & CXO & 235.21/264 & $0.2_{-0.2}^{+0.3}$    & $1.2\pm0.2$            &  --              & $1.1_{-0.2}^{+0.3}$ & $3.6\pm0.3$            & $4.6_{-0.4}^{+0.5}$\\
    2020 Aug 11 & CXO & 300.77/276 & $0.04_{-0.04}^{+0.13}$ & $1.8\pm0.1$            &  --              & $2.2_{-0.2}^{+0.4}$ & $2.8\pm0.2$            & $4.9_{-0.3}^{+0.5}$\\
    2020 Nov 02 & CXO & 46.57/63   & $1.7_{-1.0}^{+1.2}$    & $1.3\pm0.6$            &  --              & $0.5_{-0.3}^{+0.8}$ & $1.5\pm0.3$            & $2.0_{-0.4}^{+0.9}$\\ 
    2020 Nov 03 & CXO & 209.59/204 & $1.2\pm0.04$           & $1.7\pm0.2$            &  --              & $1.7_{-0.6}^{+0.8}$ & $2.7_{-0.2}^{+0.3}$    & $4.3_{-0.5}^{+0.8}$\\
    2020 Nov 08 & CXO & 132.27/125 &  --                    & $1.8\pm0.1$            &  --              & $2.6_{-0.4}^{+0.5}$ & $2.9\pm0.4$            & $5.4_{-0.5}^{+0.6}$\\
    \hline
    \hline
    \end{tabular}
    \tablefoot{
    \tablefoottext{a}{Value of the statistic divided by the degrees of freedom.}
    \tablefoottext{b}{Intrinsic absorption in units of $10^{22}$~cm$^{-2}$.}
    \tablefoottext{c}{Temperature of the blackbody component in keV.}
    \tablefoottext{d}{Unabsorbed flux in the soft 0.3--2 keV band, in units of $10^{-13}$~erg/s/cm$^2$.}
    \tablefoottext{e}{Unabsorbed flux in the hard 2--10 keV band, in units of $10^{-13}$~erg/s/cm$^2$.}
    \tablefoottext{f}{Unabsorbed luminosity in the full 0.3--10 keV band, in units of $10^{41}$~erg/s.}
 }
    \label{tab:spec}
\end{table*}

\begin{figure}
	\includegraphics[width=\columnwidth]{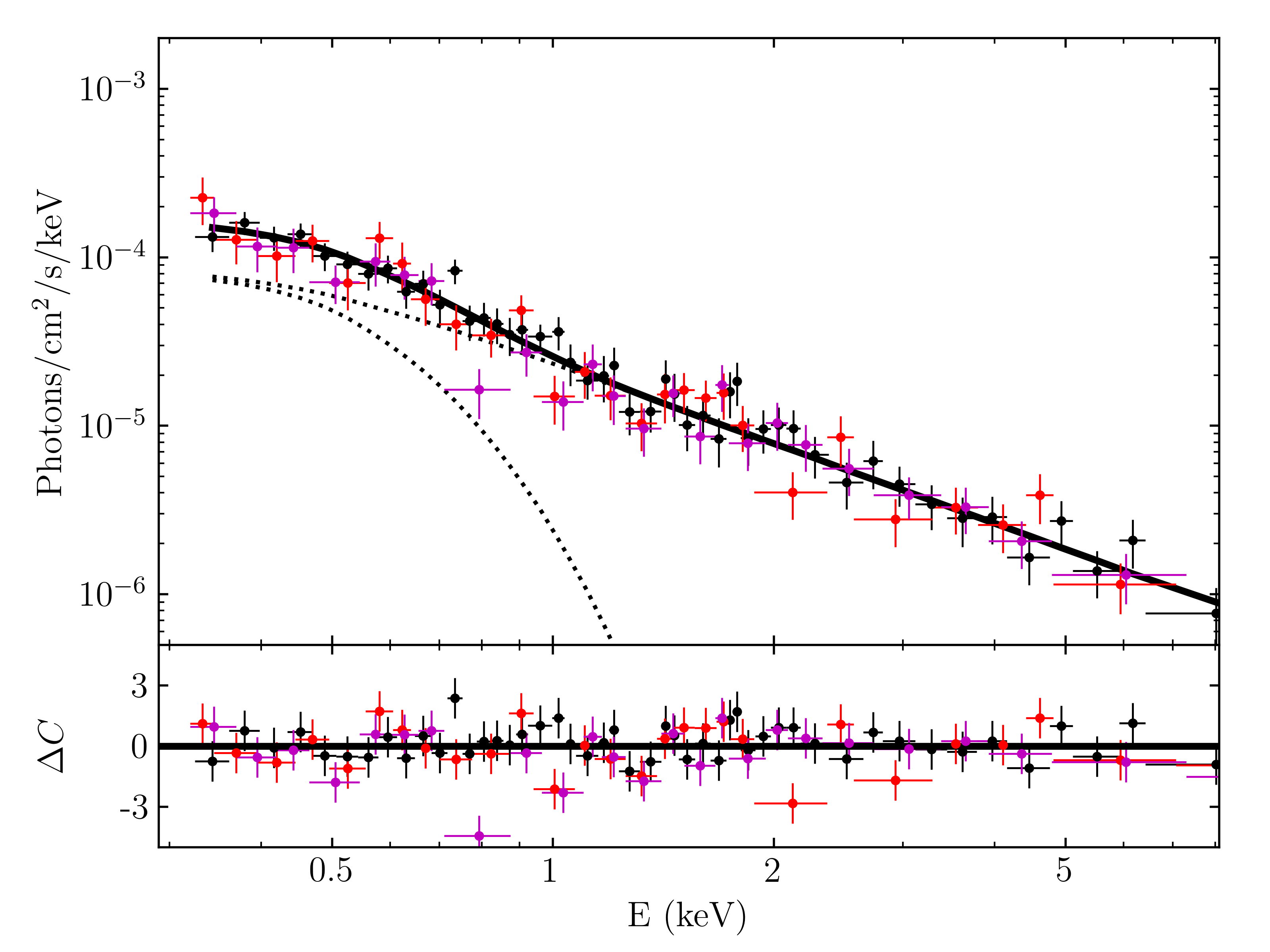}
    \caption{X-ray spectrum (upper panel) and residuals (lower panel) of the J1257, taken with \xmm\ in June 2000 (ObsID 0124710101). In black EPIC/pn, in red EPIC/MOS1, and in magenta EPIC/MOS2 data. The best-fitting model is shown by a solid line, the two components, black body and power-law, are shown by the dotted black lines. 
    \label{fig:xspec}}
\end{figure}

\begin{figure}
	\includegraphics[width=\columnwidth]{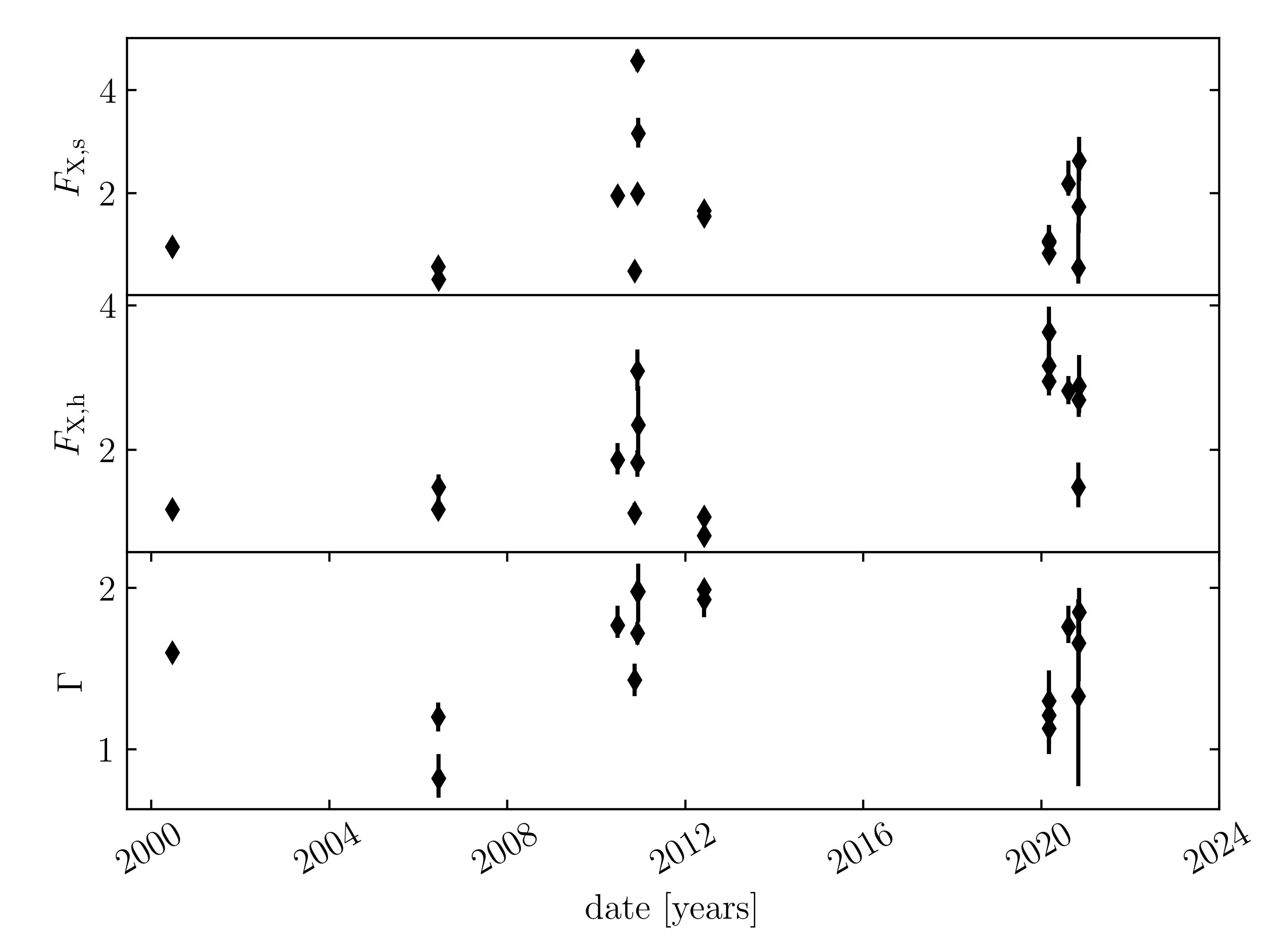}
    \caption{X-ray unabsorbed flux in the soft 0.3--2\,keV band (upper panel), in the hard 2--10\,keV band  (mid panel), and photon index ($\Gamma$) in each epoch. 
    \label{fig:lc}}
\end{figure}

\begin{figure}
	\includegraphics[width=\columnwidth]{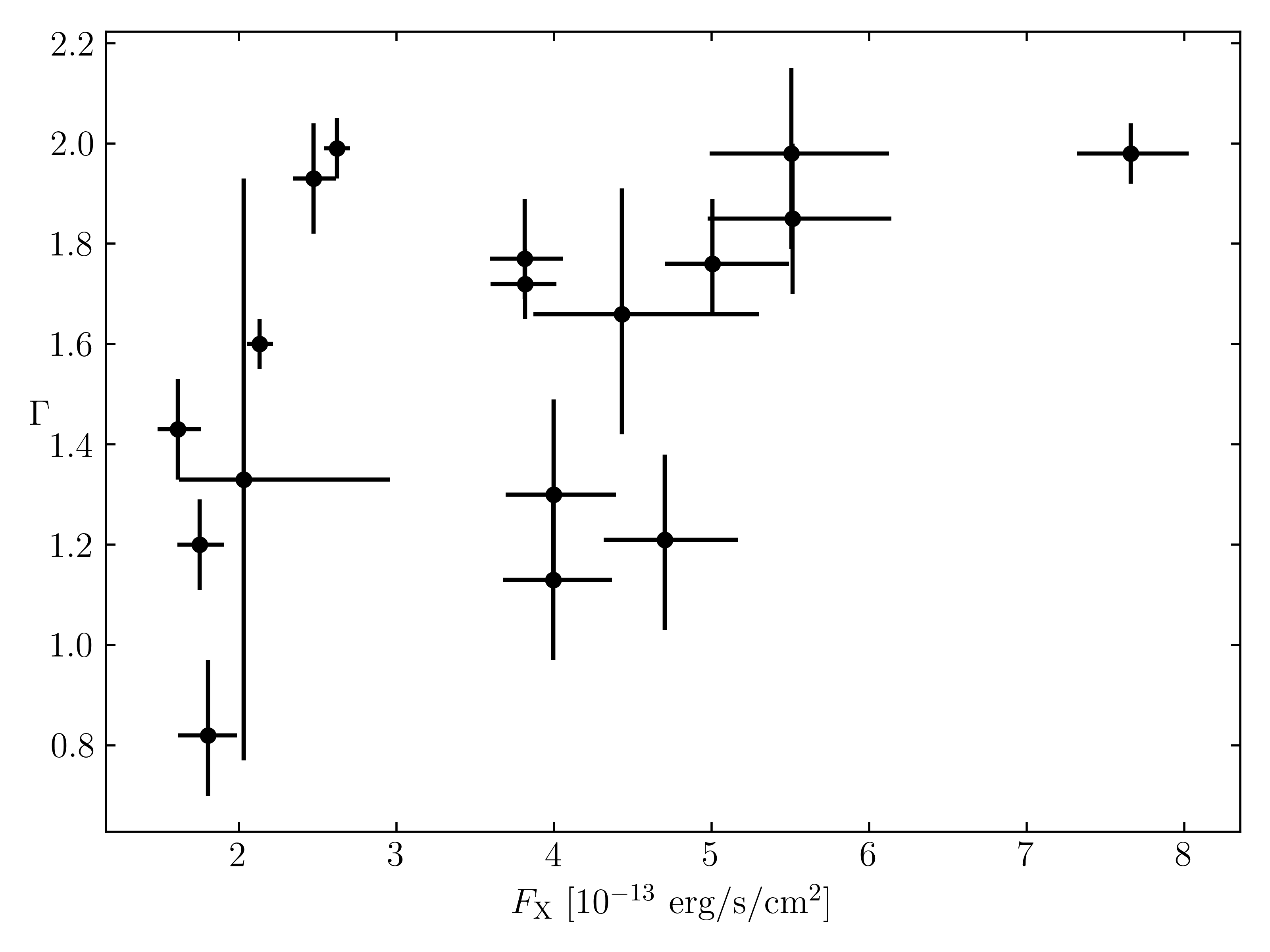}
    \caption{Spectral photon index ($\Gamma$) versus X-ray unabsorbed flux in the 0.3--10~keV band. 
    \label{fig:fg1}}
\end{figure}

\section{Discussion}\label{sec:discussion}

To exclude the possibility that we had detected an extra-nuclear source, we cross-matched \src\ coordinates, as derived by the CIAO source detection tool \texttt{wavdetect}\footnote{\url{https://cxc.cfa.harvard.edu/ciao/ahelp/wavdetect.html}}, with those of the central SMBH as reported in the 2MASS catalogue \citep{Skrutskie2006}. Our analysis confirms that the source showing the possible modulation is the central SMBH. Optical observations of \src\ reveal the presence of a broad (FWHM $\sim2000-3000$\,km\,s$^{-1}$) H$\alpha$ emission line and a H$\beta$ emission line with FWHM $\sim1000$\,km\,s$^{-1}$, supporting a classification as an intermediate-type Seyfert galaxy \citep{VeronCetty2010,Toba2014,Liu2021,Negus2024}. 

Our analysis shows that \src\ possesses peculiar X-ray variability. A repeated modulation, and possibly a flare-like behaviour, is present on the timescales of $\simeq20-30$\,ks. Unfortunately, although \src\ has been observed multiple times by both \chandra\ and \xmm, most observations do not last long enough for a definitive classification of this variability. The long-term light curve shows that the source flux can change up to a factor of 3 in a few days, while the spectral analysis reveals that \src\ shows hints of a softer-when-brighter behaviour. 

In this Section, we discuss the most plausible origins of this repeated variability in light of our findings. 

\subsection{Quasi-periodic modulation candidate}\label{sec:qpo}

Looking at Fig.~\ref{fig:3BestChandraLC} and Fig.~\ref{fig:discoveryPDSs}, it is tempting to classify the modulation as a QPO. The modulation seems to be present throughout 2020, as shown by the phase-folded light curve in Fig.~\ref{fig:2020ChandraEfold} and the light curves of ObsIDs 22930, 23361, 24853, and 24854 shown in Appendix~\ref{appendix:timing}.
Although \src\ has been observed multiple times by both \chandra\ and \xmm, most observations do not last long enough to appreciate a modulation on the timescales of 20--30\,ks. Even in the best dataset (C1), we only detect a few cycles of the putative modulation, which is not enough for a firm detection. Therefore, we will limit our discussion to a simple scenario\footnote{For a list of possible, more complicated models explaining QPOs in AGNs, see e.g. Table~1 of \cite{Pasham2024}.}. Given the uncertainties on the timescale of the modulation, in this section we will consider a frequency range of $33-50$\,\textmu Hz, which corresponds to the aforementioned 20--30\,ks range.

If the variability is indeed caused by a QPO, its centroid frequency can be associated with a fundamental frequency of the accretion disk at the innermost stable circular orbit (ISCO), that is the minimal radius at which stable circular motion is still possible \citep[see e.g.][]{Jefremov2015}. In this case, the QPO is either related to the Keplerian orbital frequency $\nu_\phi$, the vertical epicyclic frequency $\nu_\theta$, or the Lense-Thirring frequency $\nu_\mathrm{LT}$ of the matter falling onto the SMBH \citep[see e.g.][]{Stella1998,Stella1999}. The three frequencies can be linked to the mass $M_\mathrm{BH}$, the radius of the ISCO $R_\mathrm{ISCO}$, and the dimensionless spin parameter $a_*$ of \src\ using the expressions derived by \cite{Kato1990}:
\begin{equation}\label{eq:keplernu}
    \nu_\phi=\frac{c^3}{2\pi GM_\mathrm{BH}}\left[\frac{1}{R_\mathrm{ISCO}^{3/2}+a_*}\right]
\end{equation}
\begin{equation}\label{eq:verticalnu}
    \nu_\theta = \nu_\phi\left[1-\frac{4a_*}{R_\mathrm{ISCO}^{3/2}}+\frac{3a_*^2}{R_\mathrm{ISCO}^2}\right]^{1/2}
\end{equation}
\begin{equation}\label{eq:ltnu}
    \nu_\mathrm{LT}=\nu_\phi-\nu_\theta
\end{equation}
where $c$ is the speed of light and $G$ is the gravitational constant.

As already mentioned in Sect.~\ref{sec:introduction}, \cite{Liu2021} derived a mass $\sim10^{6.3}\solarM$ for \src, in line with the other SMBHs showing (candidate) QPOs \citep[see e.g. Figure~5 of][]{Yan2024}. Using this mass estimate and the equations above, testing for $-1\leq a_*\leq 1$, we find that in this scenario the observed frequency range is consistent only with the Lense-Thirring frequency of eq.~\eqref{eq:ltnu}. However, when compared to other SMBHs showing (candidate) QPOs, it is hard to reconcile the mass of \src\ with the observed timescales of the modulation. By looking again at Figure~5 of \cite{Yan2024}, QPO frequencies of $\simeq33-50$\,\textmu Hz would be extremely low for a SMBH with a mass $\sim10^{6.3}\solarM$ such as \src.

\subsection{J1257 as a possible QPE candidate}\label{sec:qpe_g}

The \chandra\ light curve in Fig.~\ref{fig:3BestChandraLC} comprises three main X-ray flares that are close to symmetric, separated by similar time intervals of $51.3$~ks and $50.0$~ks, and followed by a shorter-duration and higher amplitude burst towards the end of the third observation. The interpretation in terms of a sinusoidal modulation with period of $\simeq 25.4$\,ks (as derived from our fit in Figure~\ref{fig:3BestChandraLC}) in which some of the peaks fall within the gaps is not unique. Although the nature of the light curve prevents us from firmly assessing the significance of the putative periodicity, it is worth briefly exploring further possible interpretations. Here, we consider the possibility that the observed variability might be associated with QPEs. This suggestion must be considered as speculative at this stage and would need to be confronted against future, uninterrupted X-ray exposures.  

\begin{figure}
    \centering    \includegraphics[width=0.9\columnwidth]{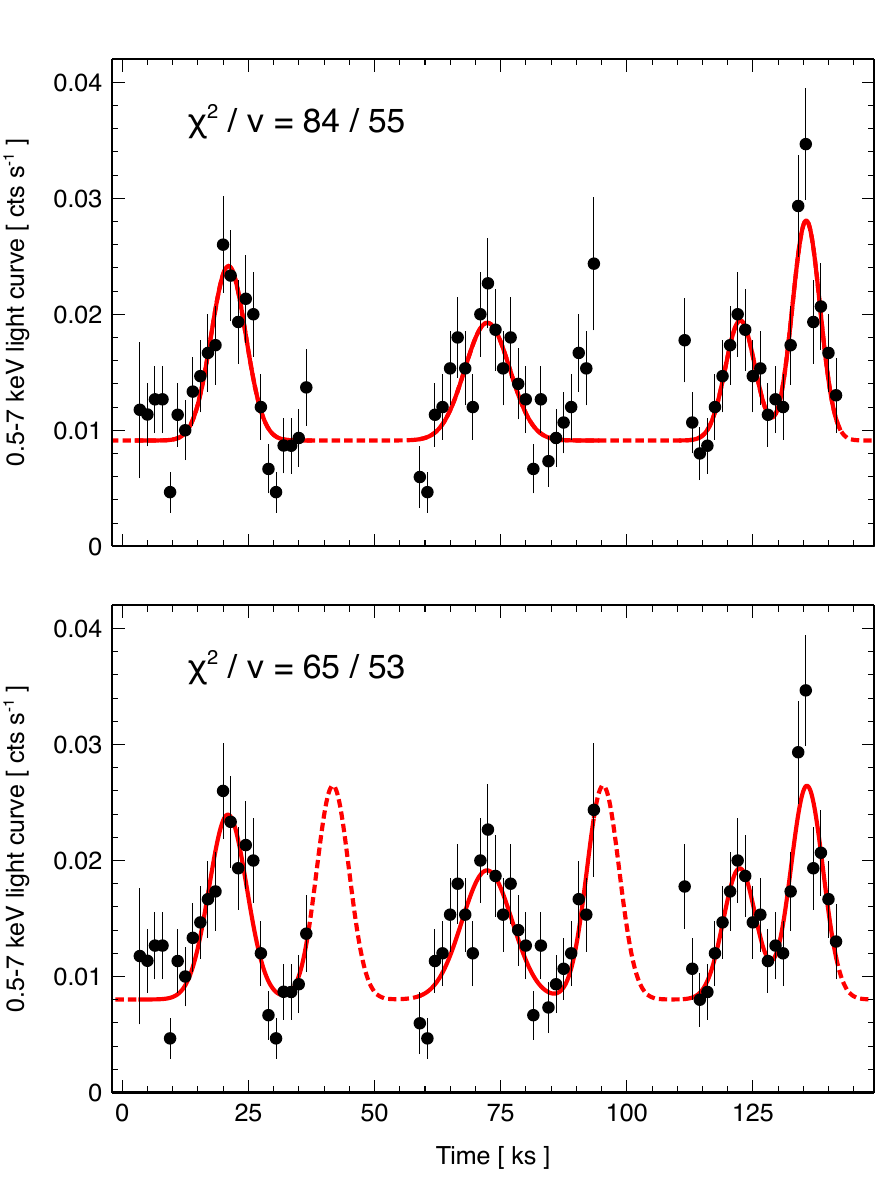}
    \caption{\chandra\ light curve in the 0.5--7\,keV band of ObsIDs 22648, 22649, and 23182, as shown in Fig.~\ref{fig:3BestChandraLC}. The upper and lower panels show in red the best-fitting model~1 and model~2, respectively (see text for details). The model extrapolation within the gaps is shown as a dotted line.}
    \label{fig:qpe-like}
\end{figure}

A model in which the sinusoidal modulation is replaced by three Gaussian functions, and the last shorter duration flare is also described with the same function representing QPEs, is shown in the upper panel of Fig.~\ref{fig:qpe-like}. The double QPE structure seen in the third observation is possibly repeated in the first two as well, as hinted by the rapid rise in count rate following the main flares. In fact, adding two Gaussian functions with all parameters fixed to those of the last flare (except centroid) to account for the rise towards the end of the first and second observation returns a statistical result of $\chi^2/{\rm dof} = 65/53$ as shown in the lower panel of the Fig.~\ref{fig:qpe-like}. In this framework, the X-ray light curve could be consistent with the typical QPE one, in which longer and shorter recurrence times (as well as stronger and weaker QPEs) alternate \citep[for one representative example see][]{Miniutti2023b}.

\begin{figure}
    \centering    \includegraphics[width=0.95\columnwidth]{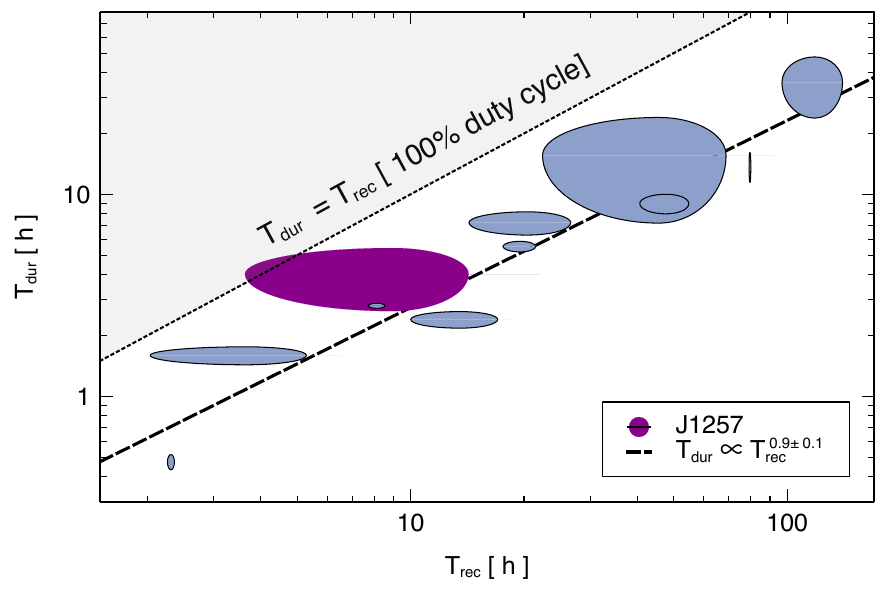}
    \caption{Location of J1257 (purple region) on the $T_{\rm dur}$-$T_{\rm rec}$ plane defined by the current QPE population (light blue regions). The grey area represents events with $T_{\rm dur} \geq T_{\rm rec}$ that cannot be classified as well-isolated QPEs. We point out that QPEs with long recurrence time and short duration, which would populate the lower-right corner of the plane, are extremely difficult to detect, so that the clear correlation seen in the QPE timing plane could be affected, at least partly, by an observational bias. QPE data are taken from \citet{Arcodia2025}. The dashed line is a best-fitting relation of the form $T_{\rm dur} \propto T_{\rm rec}^{p}$ resulting in $p=0.9\pm 0.1$, that is in a relation consistent with being linear.}
    \label{fig:Tdur-Trec}
\end{figure}

An interpretation of the X-ray variability in terms of QPEs appears therefore speculative - mostly due to the limited number of observed cycles - but plausible. It is then interesting to consider whether the derived timing parameters (recurrence time and QPE duration) are in line with the overall observed QPE population. To be conservative, we considered the largest possible range in recurrence time between peaks ($T_{\rm rec}$) and burst duration ($T_{\rm dur}$) derived from the single- and double-peak QPE models fitted on the C1 data set. We infer $T_{\rm rec} = 13.1$-$51.3$\,ks and $T_{\rm dur} = 9.5$-$19.5$\,ks in J1257. These estimates allow us to place J1257 onto the $T_{\rm dur}$-$T_{\rm rec}$ plane of the current QPE population \citep[see e.g. Figure~8 of][]{Arcodia2025}. The location of J1257 in the QPE $T_{\rm dur}$--$T_{\rm rec}$ plane is shown in Fig.~\ref{fig:Tdur-Trec} as a purple area and appears to be consistent with J1257 being part of the QPE family. We note that the $T_{\rm dur}$--$T_{\rm rec}$ linear relation (see the best-fitting model in Figure~\ref{fig:Tdur-Trec}, dashed line) is empirical only and suggests a roughly constant duty cycle for QPEs. The relation has only been studied in the context of EMRI-based QPE models and suggests that QPE emission is dominated by the interaction between the accretion disk and debris that are ablated from an orbiting star at each star-disk collision rather than by the star-disk collision itself \citep{Linial2025,Mummery2025}. Here, we simply consider it as an indication that the timing properties of the X-ray flares in J1257 are roughly consistent with the overall QPE population.

On the other hand, the spectral properties of J1257 are very distinct from those of X-ray QPEs. The latter are thermal-like, soft X-ray flares with typical temperatures of $kT\sim100-200$\,eV, therefore contributing very marginally above 2\,keV. All sources observed thus far exhibit a clear counter-clockwise pattern in the $L-T$ plane during the QPE evolution \citep[see e.g. Figure~18 in][]{Miniutti2023b}. Conversely, the X-ray flaring activity in J1257 appears to be almost achromatic with only a slightly higher relative amplitude in the soft X-ray band, as shown in Fig.~\ref{fig:2020ChandraEfold}.

Besides, and this is the most relevant difference, QPEs are detected above a quiescent X-ray emission that is fully consistent with a post-TDE, compact accretion disk with little to no contribution above $\sim2$\,keV. Instead, the X-ray spectrum of J1257 is clearly that of a standard unobscured type-1 AGN comprising a hard X-ray power law continuum and a soft excess (although the spectral variability of the target is extreme for AGN, see Figure~\ref{fig:lc}). The detection of broad optical emission lines confirms the presence of an active nucleus capable of sustaining a mature broad-line region, in contrast with the TDE case. Such differences imply a different accretion flow structure and geometry, which might have an impact on the QPE spectral shape and evolution. No specific model for QPEs in AGN has been presented thus far, so it is not possible to assess whether the distinct spectral evolution of the flares in J1257 could be consistent with QPEs occurring in AGN (or following a TDE in an AGN) which might be different from that of typical post-TDE QPEs. 

We conclude that the X-ray variability pattern in J1257 is reminiscent of QPEs in terms of timing properties but that the flares' X-ray spectral shape and evolution are not consistent with the current QPE population, preventing us from classifying J1257 as a bona-fide QPE source. Theoretical studies of the observational properties of QPE emission in AGN for the different QPE models proposed so far are needed to make progress, along with uninterrupted X-ray observations that will deliver higher-quality time-resolved spectral data and shed light on the recurrence pattern, if any.

\section{Conclusions}\label{sec:Conclusions}

We have reported on the timing and spectral analysis of \chandra\ and \xmm\ data of \src, an intermediate-type Seyfert galaxy at $z=0.02$. Prompted by the detection of a repeated modulation at timescales of $\simeq$20--30\,ks, we found that \src\ also shows long-term X-ray variability coupled with a complex spectral evolution in the last 20 years, reported here for the first time. 
Whatever the origin, the results we showed in this paper suggest that \src\ is a new specimen of the growing class of SMBHs showing repeated X-ray variability. In particular, it could be either a QPO candidate at a particularly long period or a QPE candidate with an unusually hard spectrum, making \src\ a particularly interesting source. If confirmed, given the possible connection of these phenomena with EMRIs, \src\ could also represent a source of interest for future missions such as LISA, for a deeper comprehension of the SMBH phenomenology. Due to the limited number of observed cycles, the modulation's statistical significance cannot be estimated robustly, and both the confirmation of its presence and the detailed characterisation of its spectral-timing properties must await longer, uninterrupted future X-ray observations.

\begin{acknowledgements}
    We acknowledge the use of the following software: \texttt{scipy} \citep{Virtanen:2020}, \texttt{SAS} \citep{Gabriel2004}, \texttt{CIAO} \citep{Fruscione2006}, \texttt{Stingray} \citep{Huppenkothen2019,Huppenkothen2019a,Bachetti2024}, \texttt{hendrics} \citep{Bachetti2018}.
    MI is supported by the AASS Ph.D. joint research programme between the University of Rome "Sapienza" and the University of Rome "Tor Vergata", with the collaboration of the National Institute of Astrophysics (INAF). GM thanks the Spanish MICIU/AEI/10.13039/501100011033 grants n.\ PID2020-115325GB-C31 and n.\ PID2023-147338NB-C21 for support. RA acknowledges financial support from INAF through the grant ``INAF-Astronomy Fellowships in Italy 2022 - (GOG)''.
\end{acknowledgements}

%
\bibliographystyle{aa} 
\bibliography{Bibliography/silmarillion} 

@ARTICLE{Giustini2020,
       author = {{Giustini}, Margherita and {Miniutti}, Giovanni and {Saxton}, Richard D.},
        title = "{X-ray quasi-periodic eruptions from the galactic nucleus of RX J1301.9+2747}",
      journal = {\aap},
         year = 2020,
        month = apr,
       volume = {636},
          eid = {L2},
        pages = {L2},
          doi = {10.1051/0004-6361/202037610},
archivePrefix = {arXiv},
       eprint = {2002.08967},
 primaryClass = {astro-ph.HE},
       adsurl = {https://ui.adsabs.harvard.edu/abs/2020A&A...636L...2G}
}

@ARTICLE{Chakraborty2025,
       author = {{Chakraborty}, Joheen and {Kara}, Erin and {Arcodia}, Riccardo and {Buchner}, Johannes and {Giustini}, Margherita and {Hern{\'a}ndez-Garc{\'\i}a}, Lorena and {Linial}, Itai and {Masterson}, Megan and {Miniutti}, Giovanni and {Mummery}, Andrew and {Panagiotou}, Christos and {Quintin}, Erwan and {S{\'a}nchez-S{\'a}ez}, Paula},
        title = "{Discovery of Quasiperiodic Eruptions in the Tidal Disruption Event and Extreme Coronal Line Emitter AT2022upj: Implications for the QPE/TDE Fraction and a Connection to ECLEs}",
      journal = {\apjl},
         year = 2025,
        month = apr,
       volume = {983},
       number = {2},
          eid = {L39},
        pages = {L39},
          doi = {10.3847/2041-8213/adc2f8},
archivePrefix = {arXiv},
       eprint = {2503.19013},
 primaryClass = {astro-ph.HE},
       adsurl = {https://ui.adsabs.harvard.edu/abs/2025ApJ...983L..39C}
}

@ARTICLE{Lyu2015,
       author = {{Lyu}, Ming and {M{\'e}ndez}, Mariano and {Zhang}, Guobao and {Keek}, L.},
        title = "{Spectral and timing analysis of the mHz QPOs in the neutron-star low-mass X-ray binary 4U 1636-53}",
      journal = {\mnras},
         year = 2015,
        month = nov,
       volume = {454},
       number = {1},
        pages = {541-549},
          doi = {10.1093/mnras/stv1971},
archivePrefix = {arXiv},
       eprint = {1508.06256},
 primaryClass = {astro-ph.HE},
       adsurl = {https://ui.adsabs.harvard.edu/abs/2015MNRAS.454..541L}
}

@ARTICLE{Motta2018,
       author = {{Motta}, S.~E. and {Franchini}, A. and {Lodato}, G. and {Mastroserio}, G.},
        title = "{On the different flavours of Lense-Thirring precession around accreting stellar mass black holes}",
      journal = {\mnras},
         year = 2018,
        month = jan,
       volume = {473},
       number = {1},
        pages = {431-439},
          doi = {10.1093/mnras/stx2358},
archivePrefix = {arXiv},
       eprint = {1709.02608},
 primaryClass = {astro-ph.HE},
       adsurl = {https://ui.adsabs.harvard.edu/abs/2018MNRAS.473..431M}
}

@ARTICLE{Nicholl2024,
       author = {{Nicholl}, M. and {Pasham}, D.~R. and {Mummery}, A. and {Guolo}, M. and {Gendreau}, K. and {Dewangan}, G.~C. and {Ferrara}, E.~C. and {Remillard}, R. and {Bonnerot}, C. and {Chakraborty}, J. and {Hajela}, A. and {Dhillon}, V.~S. and {Gillan}, A.~F. and {Greenwood}, J. and {Huber}, M.~E. and {Janiuk}, A. and {Salvesen}, G. and {van Velzen}, S. and {Aamer}, A. and {Alexander}, K.~D. and {Angus}, C.~R. and {Arzoumanian}, Z. and {Auchettl}, K. and {Berger}, E. and {de Boer}, T. and {Cendes}, Y. and {Chambers}, K.~C. and {Chen}, T. -W. and {Chornock}, R. and {Fulton}, M.~D. and {Gao}, H. and {Gillanders}, J.~H. and {Gomez}, S. and {Gompertz}, B.~P. and {Fabian}, A.~C. and {Herman}, J. and {Ingram}, A. and {Kara}, E. and {Laskar}, T. and {Lawrence}, A. and {Lin}, C. -C. and {Lowe}, T.~B. and {Magnier}, E.~A. and {Margutti}, R. and {McGee}, S.~L. and {Minguez}, P. and {Moore}, T. and {Nathan}, E. and {Oates}, S.~R. and {Patra}, K.~C. and {Ramsden}, P. and {Ravi}, V. and {Ridley}, E.~J. and {Sheng}, X. and {Smartt}, S.~J. and {Smith}, K.~W. and {Srivastav}, S. and {Stein}, R. and {Stevance}, H.~F. and {Turner}, S.~G.~D. and {Wainscoat}, R.~J. and {Weston}, J. and {Wevers}, T. and {Young}, D.~R.},
        title = "{Quasi-periodic X-ray eruptions years after a nearby tidal disruption event}",
      journal = {\nat},
         year = 2024,
        month = oct,
       volume = {634},
       number = {8035},
        pages = {804-808},
          doi = {10.1038/s41586-024-08023-6},
archivePrefix = {arXiv},
       eprint = {2409.02181},
 primaryClass = {astro-ph.HE},
       adsurl = {https://ui.adsabs.harvard.edu/abs/2024Natur.634..804N}
}

@ARTICLE{Kejriwal2024,
       author = {{Kejriwal}, Shubham and {Witzany}, Vojt{\v{e}}ch and {Zaja{\v{c}}ek}, Michal and {Pasham}, Dheeraj R. and {Chua}, Alvin J.~K.},
        title = "{Repeating nuclear transients as candidate electromagnetic counterparts of LISA extreme mass ratio inspirals}",
      journal = {\mnras},
         year = 2024,
        month = aug,
       volume = {532},
       number = {2},
        pages = {2143-2158},
          doi = {10.1093/mnras/stae1599},
archivePrefix = {arXiv},
       eprint = {2404.00941},
 primaryClass = {astro-ph.HE},
       adsurl = {https://ui.adsabs.harvard.edu/abs/2024MNRAS.532.2143K}
}

@ARTICLE{Smith2021,
       author = {{Smith}, Krista Lynne and {Tandon}, Celia R. and {Wagoner}, Robert V.},
        title = "{Confrontation of Observation and Theory: High-frequency QPOs in X-Ray Binaries, Tidal Disruption Events, and Active Galactic Nuclei}",
      journal = {\apj},
         year = 2021,
        month = jan,
       volume = {906},
       number = {2},
          eid = {92},
        pages = {92},
          doi = {10.3847/1538-4357/abc9b7},
archivePrefix = {arXiv},
       eprint = {2011.05346},
 primaryClass = {astro-ph.HE},
       adsurl = {https://ui.adsabs.harvard.edu/abs/2021ApJ...906...92S}
}

@ARTICLE{Yan2024,
       author = {{Yan}, Y.~K. and {Zhang}, P. and {Liu}, Q.~Z. and {Chang}, Z. and {Liu}, G.~C. and {Yan}, J.~Z. and {Zeng}, X.~Y.},
        title = "{An X-ray high-frequency quasi-periodic oscillation in NGC 1365}",
      journal = {\aap},
         year = 2024,
        month = nov,
       volume = {691},
          eid = {A7},
        pages = {A7},
          doi = {10.1051/0004-6361/202450875},
       adsurl = {https://ui.adsabs.harvard.edu/abs/2024A&A...691A...7Y}
}

@ARTICLE{Chakraborty2021,
       author = {{Chakraborty}, Joheen and {Kara}, Erin and {Masterson}, Megan and {Giustini}, Margherita and {Miniutti}, Giovanni and {Saxton}, Richard},
        title = "{Possible X-Ray Quasi-periodic Eruptions in a Tidal Disruption Event Candidate}",
      journal = {\apjl},
         year = 2021,
        month = nov,
       volume = {921},
       number = {2},
          eid = {L40},
        pages = {L40},
          doi = {10.3847/2041-8213/ac313b},
archivePrefix = {arXiv},
       eprint = {2110.10786},
 primaryClass = {astro-ph.HE},
       adsurl = {https://ui.adsabs.harvard.edu/abs/2021ApJ...921L..40C}
}

@ARTICLE{Shu2018,
       author = {{Shu}, X.~W. and {Wang}, S.~S. and {Dou}, L.~M. and {Jiang}, N. and {Wang}, J.~X. and {Wang}, T.~G.},
        title = "{A Long Decay of X-Ray Flux and Spectral Evolution in the Supersoft Active Galactic Nucleus GSN 069}",
      journal = {\apjl},
         year = 2018,
        month = apr,
       volume = {857},
       number = {2},
          eid = {L16},
        pages = {L16},
          doi = {10.3847/2041-8213/aaba17},
archivePrefix = {arXiv},
       eprint = {1809.00319},
 primaryClass = {astro-ph.HE},
       adsurl = {https://ui.adsabs.harvard.edu/abs/2018ApJ...857L..16S}
}

@ARTICLE{Komossa2015,
       author = {{Komossa}, S.},
        title = "{Tidal disruption of stars by supermassive black holes: Status of observations}",
      journal = {Journal of High Energy Astrophysics},
         year = 2015,
        month = sep,
       volume = {7},
        pages = {148-157},
          doi = {10.1016/j.jheap.2015.04.006},
archivePrefix = {arXiv},
       eprint = {1505.01093},
 primaryClass = {astro-ph.HE},
       adsurl = {https://ui.adsabs.harvard.edu/abs/2015JHEAp...7..148K}
}

@ARTICLE{Arcodia2021,
       author = {{Arcodia}, R. and {Merloni}, A. and {Nandra}, K. and {Buchner}, J. and {Salvato}, M. and {Pasham}, D. and {Remillard}, R. and {Comparat}, J. and {Lamer}, G. and {Ponti}, G. and {Malyali}, A. and {Wolf}, J. and {Arzoumanian}, Z. and {Bogensberger}, D. and {Buckley}, D.~A.~H. and {Gendreau}, K. and {Gromadzki}, M. and {Kara}, E. and {Krumpe}, M. and {Markwardt}, C. and {Ramos-Ceja}, M.~E. and {Rau}, A. and {Schramm}, M. and {Schwope}, A.},
        title = "{X-ray quasi-periodic eruptions from two previously quiescent galaxies}",
      journal = {\nat},
         year = 2021,
        month = apr,
       volume = {592},
       number = {7856},
        pages = {704-707},
          doi = {10.1038/s41586-021-03394-6},
archivePrefix = {arXiv},
       eprint = {2104.13388},
 primaryClass = {astro-ph.HE},
       adsurl = {https://ui.adsabs.harvard.edu/abs/2021Natur.592..704A}
}

@ARTICLE{Masterson2022,
       author = {{Masterson}, Megan and {Kara}, Erin and {Ricci}, Claudio and {Garc{\'\i}a}, Javier A. and {Fabian}, Andrew C. and {Pinto}, Ciro and {Kosec}, Peter and {Remillard}, Ronald A. and {Loewenstein}, Michael and {Trakhtenbrot}, Benny and {Arcavi}, Iair},
        title = "{Evolution of a Relativistic Outflow and X-Ray Corona in the Extreme Changing-look AGN 1ES 1927+654}",
      journal = {\apj},
         year = 2022,
        month = jul,
       volume = {934},
       number = {1},
          eid = {35},
        pages = {35},
          doi = {10.3847/1538-4357/ac76c0},
archivePrefix = {arXiv},
       eprint = {2206.05140},
 primaryClass = {astro-ph.HE},
       adsurl = {https://ui.adsabs.harvard.edu/abs/2022ApJ...934...35M}
}

@ARTICLE{Masterson2025,
       author = {{Masterson}, Megan and {Kara}, Erin and {Panagiotou}, Christos and {Alston}, William N. and {Chakraborty}, Joheen and {Burdge}, Kevin and {Ricci}, Claudio and {Laha}, Sibasish and {Arcavi}, Iair and {Arcodia}, Riccardo and {Cenko}, S. Bradley and {Fabian}, Andrew C. and {Garc{\'\i}a}, Javier A. and {Giustini}, Margherita and {Ingram}, Adam and {Kosec}, Peter and {Loewenstein}, Michael and {Meyer}, Eileen T. and {Miniutti}, Giovanni and {Pinto}, Ciro and {Remillard}, Ronald A. and {Sadaula}, Dev R. and {Shuvo}, Onic I. and {Trakhtenbrot}, Benny and {Wang}, Jingyi},
        title = "{Millihertz oscillations near the innermost orbit of a supermassive black hole}",
      journal = {\nat},
         year = 2025,
        month = feb,
       volume = {638},
       number = {8050},
        pages = {370-375},
          doi = {10.1038/s41586-024-08385-x},
archivePrefix = {arXiv},
       eprint = {2501.01581},
 primaryClass = {astro-ph.HE},
       adsurl = {https://ui.adsabs.harvard.edu/abs/2025Natur.638..370M}
}

@article{Abramowicz1988,
 adsurl = {https://ui.adsabs.harvard.edu/abs/1988ApJ...332..646A},
 author = {{Abramowicz}, M.~A. and {Czerny}, B. and {Lasota}, J.~P. and {Szuszkiewicz}, E.},
 doi = {10.1086/166683},
 journal = {\apj},
 month = sep,
 pages = {646},
 title = {{Slim Accretion Disks}},
 volume = {332},
 year = {1988}
}

@article{Xia2025,
      title={Strong Amplitude Modulation of Hard-band X-ray QPO with Soft-band Flux in RE J1034+396}, 
      author={Ruisong Xia and Hao Liu and Yongquan Xue},
      year={2025},
      eprint={2503.05158},
      archivePrefix={arXiv},
      primaryClass={astro-ph.HE},
      url={https://arxiv.org/abs/2503.05158},
      doi={doi.org/10.48550/arXiv.2503.05158},
}

@ARTICLE{Xia2024,
       author = {{Xia}, Ruisong and {Liu}, Hao and {Xue}, Yongquan},
        title = "{First Observational Evidence for an Interconnected Evolution between Time Lag and QPO Frequency among AGNs}",
      journal = {\apjl},
         year = 2024,
        month = feb,
       volume = {961},
       number = {2},
          eid = {L32},
        pages = {L32},
          doi = {10.3847/2041-8213/ad1bf2},
archivePrefix = {arXiv},
       eprint = {2401.04926},
 primaryClass = {astro-ph.HE},
       adsurl = {https://ui.adsabs.harvard.edu/abs/2024ApJ...961L..32X}
}

@article{Arcodia2024,
 adsurl = {https://ui.adsabs.harvard.edu/abs/2024A&A...684A..64A},
 archiveprefix = {arXiv},
 author = {{Arcodia}, R. and {Liu}, Z. and {Merloni}, A. and {Malyali}, A. and {Rau}, A. and {Chakraborty}, J. and {Goodwin}, A. and {Buckley}, D. and {Brink}, J. and {Gromadzki}, M. and {Arzoumanian}, Z. and {Buchner}, J. and {Kara}, E. and {Nandra}, K. and {Ponti}, G. and {Salvato}, M. and {Anderson}, G. and {Baldini}, P. and {Grotova}, I. and {Krumpe}, M. and {Maitra}, C. and {Miller-Jones}, J.~C.~A. and {Ramos-Ceja}, M.~E.},
 doi = {10.1051/0004-6361/202348881},
 eid = {A64},
 eprint = {2401.17275},
 journal = {\aap},
 month = apr,
 pages = {A64},
 primaryclass = {astro-ph.HE},
 title = {{The more the merrier: SRG/eROSITA discovers two further galaxies showing X-ray quasi-periodic eruptions}},
 volume = {684},
 year = {2024}
}

@inproceedings{Arnaud1996,
 adsurl = {https://ui.adsabs.harvard.edu/abs/1996ASPC..101...17A},
 author = {{Arnaud}, K.~A.},
 booktitle = {Astronomical Data Analysis Software and Systems V},
 editor = {{Jacoby}, George H. and {Barnes}, Jeannette},
 month = jan,
 pages = {17},
 series = {Astronomical Society of the Pacific Conference Series},
 title = {{XSPEC: The First Ten Years}},
 volume = {101},
 year = {1996}
}

@software{Bachetti2018,
 adsurl = {https://ui.adsabs.harvard.edu/abs/2018ascl.soft05019B},
 author = {{Bachetti}, Matteo},
 eid = {ascl:1805.019},
 howpublished = {Astrophysics Source Code Library, record ascl:1805.019},
 month = may,
 title = {{HENDRICS: High ENergy Data Reduction Interface from the Command Shell}},
 year = {2018}
}

@article{Bartlett2017,
 adsurl = {https://ui.adsabs.harvard.edu/abs/2017MNRAS.466.4659B},
 archiveprefix = {arXiv},
 author = {{Bartlett}, E.~S. and {Coe}, M.~J. and {Israel}, G.~L. and {Clark}, J.~S. and {Esposito}, P. and {D'Elia}, V. and {Udalski}, A.},
 doi = {10.1093/mnras/stx032},
 eprint = {1701.01729},
 journal = {\mnras},
 month = apr,
 number = {4},
 pages = {4659-4671},
 primaryclass = {astro-ph.HE},
 title = {{SXP 7.92: a recently rediscovered Be/X-ray binary in the Small Magellanic Cloud, viewed edge on}},
 volume = {466},
 year = {2017}
}

@article{Burrows2005,
 adsurl = {https://ui.adsabs.harvard.edu/abs/2005SSRv..120..165B},
 archiveprefix = {arXiv},
 author = {{Burrows}, David N. and {Hill}, J.~E. and {Nousek}, J.~A. and {Kennea}, J.~A. and {Wells}, A. and {Osborne}, J.~P. and {Abbey}, A.~F. and {Beardmore}, A. and {Mukerjee}, K. and {Short}, A.~D.~T. and {Chincarini}, G. and {Campana}, S. and {Citterio}, O. and {Moretti}, A. and {Pagani}, C. and {Tagliaferri}, G. and {Giommi}, P. and {Capalbi}, M. and {Tamburelli}, F. and {Angelini}, L. and {Cusumano}, G. and {Br{\"a}uninger}, H.~W. and {Burkert}, W. and {Hartner}, G.~D.},
 doi = {10.1007/s11214-005-5097-2},
 eprint = {astro-ph/0508071},
 journal = {\ssr},
 month = oct,
 number = {3-4},
 pages = {165-195},
 primaryclass = {astro-ph},
 title = {{The Swift X-Ray Telescope}},
 volume = {120},
 year = {2005}
}

@article{Casella2008,
 adsurl = {https://ui.adsabs.harvard.edu/abs/2008MNRAS.387.1707C},
 archiveprefix = {arXiv},
 author = {{Casella}, P. and {Ponti}, G. and {Patruno}, A. and {Belloni}, T. and {Miniutti}, G. and {Zampieri}, L.},
 doi = {10.1111/j.1365-2966.2008.13372.x},
 eprint = {0804.3378},
 journal = {\mnras},
 month = jul,
 number = {4},
 pages = {1707-1711},
 primaryclass = {astro-ph},
 title = {{Weighing the black holes in ultraluminous X-ray sources through timing}},
 volume = {387},
 year = {2008}
}

@article{Cash1979,
 adsurl = {https://ui.adsabs.harvard.edu/abs/1979ApJ...228..939C},
 author = {{Cash}, W.},
 doi = {10.1086/156922},
 journal = {\apj},
 month = mar,
 pages = {939-947},
 title = {{Parameter estimation in astronomy through application of the likelihood ratio.}},
 volume = {228},
 year = {1979}
}

@article{Esposito2013,
 adsurl = {https://ui.adsabs.harvard.edu/abs/2013MNRAS.433.3464E},
 archiveprefix = {arXiv},
 author = {{Esposito}, P. and {Israel}, G.~L. and {Sidoli}, L. and {Rodr{\'\i}guez Castillo}, G.~A. and {Masetti}, N. and {D'Avanzo}, P. and {Campana}, S.},
 doi = {10.1093/mnras/stt1010},
 eprint = {1306.1106},
 journal = {\mnras},
 month = aug,
 number = {4},
 pages = {3464-3471},
 primaryclass = {astro-ph.HE},
 title = {{CXOU J005047.9-731817: a 292-s X-ray binary pulsar in the Small Magellanic Cloud}},
 volume = {433},
 year = {2013}
}

@article{Esposito2013a,
 adsurl = {https://ui.adsabs.harvard.edu/abs/2013MNRAS.436.3380E},
 archiveprefix = {arXiv},
 author = {{Esposito}, P. and {Israel}, G.~L. and {Sidoli}, L. and {Mapelli}, M. and {Zampieri}, L. and {Motta}, S.~E.},
 doi = {10.1093/mnras/stt1819},
 eprint = {1309.6328},
 journal = {\mnras},
 month = dec,
 number = {4},
 pages = {3380-3387},
 primaryclass = {astro-ph.HE},
 title = {{Discovery of a 6.4 h black hole binary in NGC 4490}},
 volume = {436},
 year = {2013}
}

@article{Esposito2015,
 adsurl = {https://ui.adsabs.harvard.edu/abs/2015MNRAS.450.1705E},
 archiveprefix = {arXiv},
 author = {{Esposito}, P. and {Israel}, G.~L. and {de Martino}, D. and {D'Avanzo}, P. and {Testa}, V. and {Sidoli}, L. and {Di Stefano}, R. and {Belfiore}, A. and {Mapelli}, M. and {Piranomonte}, S. and {Rodr{\'\i}guez Castillo}, G.~A. and {Moretti}, A. and {D'Elia}, V. and {Verrecchia}, F. and {Campana}, S. and {Rea}, N.},
 doi = {10.1093/mnras/stv724},
 eprint = {1503.08830},
 journal = {\mnras},
 month = jun,
 number = {2},
 pages = {1705-1715},
 primaryclass = {astro-ph.HE},
 title = {{Swift J201424.9+152930: discovery of a new deeply eclipsing binary with 491-s and 3.4-h modulations}},
 volume = {450},
 year = {2015}
}

@inproceedings{Fruscione2006,
 adsurl = {https://ui.adsabs.harvard.edu/abs/2006SPIE.6270E..1VF},
 author = {{Fruscione}, Antonella and {McDowell}, Jonathan C. and {Allen}, Glenn E. and {Brickhouse}, Nancy S. and {Burke}, Douglas J. and {Davis}, John E. and {Durham}, Nick and {Elvis}, Martin and {Galle}, Elizabeth C. and {Harris}, Daniel E. and {Huenemoerder}, David P. and {Houck}, John C. and {Ishibashi}, Bish and {Karovska}, Margarita and {Nicastro}, Fabrizio and {Noble}, Michael S. and {Nowak}, Michael A. and {Primini}, Frank A. and {Siemiginowska}, Aneta and {Smith}, Randall K. and {Wise}, Michael},
 booktitle = {Observatory Operations: Strategies, Processes, and Systems},
 doi = {10.1117/12.671760},
 editor = {{Silva}, David R. and {Doxsey}, Rodger E.},
 eid = {62701V},
 month = jun,
 pages = {62701V},
 series = {Society of Photo-Optical Instrumentation Engineers (SPIE) Conference Series},
 title = {{CIAO: Chandra's data analysis system}},
 volume = {6270},
 year = {2006}
}

@inproceedings{Gabriel2004,
 adsurl = {https://ui.adsabs.harvard.edu/abs/2004ASPC..314..759G},
 author = {{Gabriel}, C. and {Denby}, M. and {Fyfe}, D.~J. and {Hoar}, J. and {Ibarra}, A. and {Ojero}, E. and {Osborne}, J. and {Saxton}, R.~D. and {Lammers}, U. and {Vacanti}, G.},
 booktitle = {Astronomical Data Analysis Software and Systems (ADASS) XIII},
 editor = {{Ochsenbein}, Francois and {Allen}, Mark G. and {Egret}, Daniel},
 month = jul,
 pages = {759},
 series = {Astronomical Society of the Pacific Conference Series},
 title = {{The XMM-Newton SAS - Distributed Development and Maintenance of a Large Science Analysis System: A Critical Analysis}},
 volume = {314},
 year = {2004}
}

@ARTICLE{Zhang2023,
       author = {{Zhang}, Haoyang and {Yang}, Shenbang and {Dai}, Benzhong},
        title = "{Search for X-Ray Quasiperiodicity of Six AGNs Using the Gaussian Process Method}",
      journal = {\apj},
         year = 2023,
        month = mar,
       volume = {946},
       number = {1},
          eid = {52},
        pages = {52},
          doi = {10.3847/1538-4357/acbe37},
archivePrefix = {arXiv},
       eprint = {2304.08044},
 primaryClass = {astro-ph.HE},
       adsurl = {https://ui.adsabs.harvard.edu/abs/2023ApJ...946...52Z}
}

@article{Gierlinski2008,
 adsurl = {https://ui.adsabs.harvard.edu/abs/2008Natur.455..369G},
 author = {{Gierli{\'n}ski}, Marek and {Middleton}, Matthew and {Ward}, Martin and {Done}, Chris},
 doi = {10.1038/nature07277},
 journal = {\nat},
 month = sep,
 number = {7211},
 pages = {369-371},
 title = {{A periodicity of \raisebox{-0.5ex}\textasciitilde1hour in X-ray emission from the active galaxy RE J1034+396}},
 volume = {455},
 year = {2008}
}

@ARTICLE{Ahn2012,
       author = {{Ahn}, Christopher P. and {Alexandroff}, Rachael and {Allende Prieto}, Carlos and {Anderson}, Scott F. and {Anderton}, Timothy and {Andrews}, Brett H. and {Aubourg}, {\'E}ric and {Bailey}, Stephen and {Balbinot}, Eduardo and {Barnes}, Rory and {Bautista}, Julian and {Beers}, Timothy C. and {Beifiori}, Alessandra and {Berlind}, Andreas A. and {Bhardwaj}, Vaishali and {Bizyaev}, Dmitry and {Blake}, Cullen H. and {Blanton}, Michael R. and {Blomqvist}, Michael and {Bochanski}, John J. and {Bolton}, Adam S. and {Borde}, Arnaud and {Bovy}, Jo and {Brandt}, W.~N. and {Brinkmann}, J. and {Brown}, Peter J. and {Brownstein}, Joel R. and {Bundy}, Kevin and {Busca}, N.~G. and {Carithers}, William and {Carnero}, Aurelio R. and {Carr}, Michael A. and {Casetti-Dinescu}, Dana I. and {Chen}, Yanmei and {Chiappini}, Cristina and {Comparat}, Johan and {Connolly}, Natalia and {Crepp}, Justin R. and {Cristiani}, Stefano and {Croft}, Rupert A.~C. and {Cuesta}, Antonio J. and {da Costa}, Luiz N. and {Davenport}, James R.~A. and {Dawson}, Kyle S. and {de Putter}, Roland and {De Lee}, Nathan and {Delubac}, Timoth{\'e}e and {Dhital}, Saurav and {Ealet}, Anne and {Ebelke}, Garrett L. and {Edmondson}, Edward M. and {Eisenstein}, Daniel J. and {Escoffier}, S. and {Esposito}, Massimiliano and {Evans}, Michael L. and {Fan}, Xiaohui and {Femen{\'\i}a Castell{\'a}}, Bruno and {Fern{\'a}ndez Alvar}, Emma and {Ferreira}, Leticia D. and {Filiz Ak}, N. and {Finley}, Hayley and {Fleming}, Scott W. and {Font-Ribera}, Andreu and {Frinchaboy}, Peter M. and {Garc{\'\i}a-Hern{\'a}ndez}, D.~A. and {Garc{\'\i}a P{\'e}rez}, A.~E. and {Ge}, Jian and {G{\'e}nova-Santos}, R. and {Gillespie}, Bruce A. and {Girardi}, L{\'e}o and {Gonz{\'a}lez Hern{\'a}ndez}, Jonay I. and {Grebel}, Eva K. and {Gunn}, James E. and {Guo}, Hong and {Haggard}, Daryl and {Hamilton}, Jean-Christophe and {Harris}, David W. and {Hawley}, Suzanne L. and {Hearty}, Frederick R. and {Ho}, Shirley and {Hogg}, David W. and {Holtzman}, Jon A. and {Honscheid}, Klaus and {Huehnerhoff}, J. and {Ivans}, Inese I. and {Ivezi{\'c}}, {\v{Z}}eljko and {Jacobson}, Heather R. and {Jiang}, Linhua and {Johansson}, Jonas and {Johnson}, Jennifer A. and {Kauffmann}, Guinevere and {Kirkby}, David and {Kirkpatrick}, Jessica A. and {Klaene}, Mark A. and {Knapp}, Gillian R. and {Kneib}, Jean-Paul and {Le Goff}, Jean-Marc and {Leauthaud}, Alexie and {Lee}, Khee-Gan and {Lee}, Young Sun and {Long}, Daniel C. and {Loomis}, Craig P. and {Lucatello}, Sara and {Lundgren}, Britt and {Lupton}, Robert H. and {Ma}, Bo and {Ma}, Zhibo and {MacDonald}, Nicholas and {Mack}, Claude E. and {Mahadevan}, Suvrath and {Maia}, Marcio A.~G. and {Majewski}, Steven R. and {Makler}, Martin and {Malanushenko}, Elena and {Malanushenko}, Viktor and {Manchado}, A. and {Mandelbaum}, Rachel and {Manera}, Marc and {Maraston}, Claudia and {Margala}, Daniel and {Martell}, Sarah L. and {McBride}, Cameron K. and {McGreer}, Ian D. and {McMahon}, Richard G. and {M{\'e}nard}, Brice and {Meszaros}, Sz. and {Miralda-Escud{\'e}}, Jordi and {Montero-Dorta}, Antonio D. and {Montesano}, Francesco and {Morrison}, Heather L. and {Muna}, Demitri and {Munn}, Jeffrey A. and {Murayama}, Hitoshi and {Myers}, Adam D. and {Neto}, A.~F. and {Nguyen}, Duy Cuong and {Nichol}, Robert C. and {Nidever}, David L. and {Noterdaeme}, Pasquier and {Nuza}, Sebasti{\'a}n E. and {Ogando}, Ricardo L.~C. and {Olmstead}, Matthew D. and {Oravetz}, Daniel J. and {Owen}, Russell and {Padmanabhan}, Nikhil and {Palanque-Delabrouille}, Nathalie and {Pan}, Kaike and {Parejko}, John K. and {Parihar}, Prachi and {P{\^a}ris}, Isabelle and {Pattarakijwanich}, Petchara and {Pepper}, Joshua and {Percival}, Will J. and {P{\'e}rez-Fournon}, Ismael and {P{\'e}rez-R{\`a}fols}, Ignasi and {Petitjean}, Patrick and {Pforr}, Janine and {Pieri}, Matthew M. and {Pinsonneault}, Marc H. and {Porto de Mello}, G.~F. and {Prada}, Francisco and {Price-Whelan}, Adrian M. and {Raddick}, M. Jordan and {Rebolo}, Rafael and {Rich}, James and {Richards}, Gordon T. and {Robin}, Annie C. and {Rocha-Pinto}, Helio J. and {Rockosi}, Constance M. and {Roe}, Natalie A. and {Ross}, Ashley J. and {Ross}, Nicholas P. and {Rossi}, Graziano and {Rubi{\~n}o-Martin}, J.~A. and {Samushia}, Lado and {Sanchez Almeida}, J. and {S{\'a}nchez}, Ariel G. and {Santiago}, Bas{\'\i}lio and {Sayres}, Conor and {Schlegel}, David J. and {Schlesinger}, Katharine J. and {Schmidt}, Sarah J. and {Schneider}, Donald P. and {Schultheis}, Mathias and {Schwope}, Axel D. and {Sc{\'o}ccola}, C.~G. and {Seljak}, Uros and {Sheldon}, Erin and {Shen}, Yue and {Shu}, Yiping and {Simmerer}, Jennifer and {Simmons}, Audrey E. and {Skibba}, Ramin A. and {Skrutskie}, M.~F. and {Slosar}, A. and {Sobreira}, Flavia and {Sobeck}, Jennifer S. and {Stassun}, Keivan G. and {Steele}, Oliver and {Steinmetz}, Matthias},
        title = "{The Ninth Data Release of the Sloan Digital Sky Survey: First Spectroscopic Data from the SDSS-III Baryon Oscillation Spectroscopic Survey}",
      journal = {\apjs},
         year = 2012,
        month = dec,
       volume = {203},
       number = {2},
          eid = {21},
        pages = {21},
          doi = {10.1088/0067-0049/203/2/21},
archivePrefix = {arXiv},
       eprint = {1207.7137},
 primaryClass = {astro-ph.IM},
       adsurl = {https://ui.adsabs.harvard.edu/abs/2012ApJS..203...21A}
}

@ARTICLE{Hernandez-Garcia2025,
       author = {{Hern{\'a}ndez-Garc{\'\i}a}, Lorena and {Chakraborty}, Joheen and {S{\'a}nchez-S{\'a}ez}, Paula and {Ricci}, Claudio and {Cuadra}, Jorge and {McKernan}, Barry and {Ford}, K.~E. Saavik and {Ar{\'e}valo}, Patricia and {Rau}, Arne and {Arcodia}, Riccardo and {Kara}, Erin and {Liu}, Zhu and {Merloni}, Andrea and {Bruni}, Gabriele and {Goodwin}, Adelle and {Arzoumanian}, Zaven and {Assef}, Roberto J. and {Baldini}, Pietro and {Bayo}, Amelia and {Bauer}, Franz E. and {Bernal}, Santiago and {Brightman}, Murray and {Calistro Rivera}, Gabriela and {Gendreau}, Keith and {Homan}, David and {Krumpe}, Mirko and {Lira}, Paulina and {Mart{\'\i}nez-Aldama}, Mary Loli and {Salvato}, Mara and {Sotomayor}, Bel{\'e}n},
        title = "{Discovery of extreme quasi-periodic eruptions in a newly accreting massive black hole}",
      journal = {Nature Astronomy},
         year = 2025,
        month = jun,
       volume = {9},
        pages = {895-906},
          doi = {10.1038/s41550-025-02523-9},
archivePrefix = {arXiv},
       eprint = {2504.07169},
 primaryClass = {astro-ph.HE},
       adsurl = {https://ui.adsabs.harvard.edu/abs/2025NatAs...9..895H}
}

@article{HI4PICollaboration2016,
 adsurl = {https://ui.adsabs.harvard.edu/abs/2016A&A...594A.116H},
 archiveprefix = {arXiv},
 author = {{HI4PI Collaboration} and {Ben Bekhti}, N. and {Fl{\"o}er}, L. and {Keller}, R. and {Kerp}, J. and {Lenz}, D. and {Winkel}, B. and {Bailin}, J. and {Calabretta}, M.~R. and {Dedes}, L. and {Ford}, H.~A. and {Gibson}, B.~K. and {Haud}, U. and {Janowiecki}, S. and {Kalberla}, P.~M.~W. and {Lockman}, F.~J. and {McClure-Griffiths}, N.~M. and {Murphy}, T. and {Nakanishi}, H. and {Pisano}, D.~J. and {Staveley-Smith}, L.},
 doi = {10.1051/0004-6361/201629178},
 eid = {A116},
 eprint = {1610.06175},
 journal = {\aap},
 month = oct,
 pages = {A116},
 primaryclass = {astro-ph.GA},
 title = {{HI4PI: A full-sky H I survey based on EBHIS and GASS}},
 volume = {594},
 year = {2016}
}

@article{Huppenkothen2019,
 adsurl = {https://ui.adsabs.harvard.edu/abs/2019ApJ...881...39H},
 archiveprefix = {arXiv},
 author = {{Huppenkothen}, Daniela and {Bachetti}, Matteo and {Stevens}, Abigail L. and {Migliari}, Simone and {Balm}, Paul and {Hammad}, Omar and {Khan}, Usman Mahmood and {Mishra}, Himanshu and {Rashid}, Haroon and {Sharma}, Swapnil and {Martinez Ribeiro}, Evandro and {Valles Blanco}, Ricardo},
 doi = {10.3847/1538-4357/ab258d},
 eid = {39},
 eprint = {1901.07681},
 journal = {\apj},
 month = aug,
 number = {1},
 pages = {39},
 primaryclass = {astro-ph.IM},
 title = {{Stingray: A Modern Python Library for Spectral Timing}},
 volume = {881},
 year = {2019}
}

@ARTICLE{Kato1990,
       author = {{Kato}, Shoji},
        title = "{Trapped One-Armed Corrugation Waves and QPO's}",
      journal = {\pasj},
         year = 1990,
        month = feb,
       volume = {42},
        pages = {99-113},
       adsurl = {https://ui.adsabs.harvard.edu/abs/1990PASJ...42...99K}
}

@ARTICLE{Skrutskie2006,
       author = {{Skrutskie}, M.~F. and {Cutri}, R.~M. and {Stiening}, R. and {Weinberg}, M.~D. and {Schneider}, S. and {Carpenter}, J.~M. and {Beichman}, C. and {Capps}, R. and {Chester}, T. and {Elias}, J. and {Huchra}, J. and {Liebert}, J. and {Lonsdale}, C. and {Monet}, D.~G. and {Price}, S. and {Seitzer}, P. and {Jarrett}, T. and {Kirkpatrick}, J.~D. and {Gizis}, J.~E. and {Howard}, E. and {Evans}, T. and {Fowler}, J. and {Fullmer}, L. and {Hurt}, R. and {Light}, R. and {Kopan}, E.~L. and {Marsh}, K.~A. and {McCallon}, H.~L. and {Tam}, R. and {Van Dyk}, S. and {Wheelock}, S.},
        title = "{The Two Micron All Sky Survey (2MASS)}",
      journal = {\aj},
         year = 2006,
        month = feb,
       volume = {131},
       number = {2},
        pages = {1163-1183},
          doi = {10.1086/498708},
       adsurl = {https://ui.adsabs.harvard.edu/abs/2006AJ....131.1163S}
}

@ARTICLE{Bachetti2024,
       author = {{Bachetti}, Matteo and {Huppenkothen}, Daniela and {Stevens}, Abigail and {Swinbank}, John and {Mastroserio}, Guglielmo and {Lucchini}, Matteo and {Lai}, Eleonora and {Buchner}, Johannes and {Desai}, Amogh and {Joshi}, Gaurav and {Pisanu}, Francesco and {Pisupati}, Sri and {Sharma}, Swapnil and {Tripathi}, Mihir and {Vats}, Dhruv},
        title = "{Stingray 2: A fast and modern Python library for spectral timing}",
      journal = {The Journal of Open Source Software},
         year = 2024,
        month = oct,
       volume = {9},
       number = {102},
          eid = {7389},
        pages = {7389},
          doi = {10.21105/joss.07389},
       adsurl = {https://ui.adsabs.harvard.edu/abs/2024JOSS....9.7389B}
}

@article{Huppenkothen2019a,
 adsurl = {https://ui.adsabs.harvard.edu/abs/2019JOSS....4.1393H},
 author = {{Huppenkothen}, Daniela and {Bachetti}, Matteo and {Stevens}, Abigail and {Migliari}, Simone and {Balm}, Paul and {Hammad}, Omar and {Khan}, Usman and {Mishra}, Himanshu and {Rashid}, Haroon and {Sharma}, Swapnil and {Ribeiro}, Evandro and {Blanco}, Ricardo},
 doi = {10.21105/joss.01393},
 eid = {1393},
 journal = {The Journal of Open Source Software},
 month = jun,
 number = {38},
 pages = {1393},
 title = {{stingray: A modern Python library for spectral timing}},
 volume = {4},
 year = {2019}
}

@article{Imbrogno2024,
 adsurl = {https://ui.adsabs.harvard.edu/abs/2024A&A...689A.284I},
 archiveprefix = {arXiv},
 author = {{Imbrogno}, Matteo and {Motta}, Sara Elisa and {Amato}, Roberta and {Israel}, Gian Luca and {Rodr{\'\i}guez Castillo}, Guillermo Andres and {Brightman}, Murray and {Casella}, Piergiorgio and {Bachetti}, Matteo and {F{\"u}rst}, Felix and {Stella}, Luigi and {Pinto}, Ciro and {Pintore}, Fabio and {Tombesi}, Francesco and {G{\'u}rpide}, Andr{\'e}s and {Middleton}, Matthew J. and {Salvaggio}, Chiara and {Tiengo}, Andrea and {Belfiore}, Andrea and {De Luca}, Andrea and {Esposito}, Paolo and {Wolter}, Anna and {Earnshaw}, Hannah P. and {Walton}, Dominic J. and {Roberts}, Timothy P. and {Zampieri}, Luca and {Marelli}, Martino and {Salvaterra}, Ruben},
 doi = {10.1051/0004-6361/202450432},
 eid = {A284},
 eprint = {2407.09240},
 journal = {\aap},
 month = sep,
 pages = {A284},
 primaryclass = {astro-ph.HE},
 title = {{Skipping a beat: Discovery of persistent quasi-periodic oscillations associated with pulsed fraction drop of the spin signal in M51 ULX-7}},
 volume = {689},
 year = {2024}
}

@article{Ingram2009,
 adsurl = {https://ui.adsabs.harvard.edu/abs/2009MNRAS.397L.101I},
 archiveprefix = {arXiv},
 author = {{Ingram}, Adam and {Done}, Chris and {Fragile}, P. Chris},
 doi = {10.1111/j.1745-3933.2009.00693.x},
 eprint = {0901.1238},
 journal = {\mnras},
 month = jul,
 number = {1},
 pages = {L101-L105},
 primaryclass = {astro-ph.SR},
 title = {{Low-frequency quasi-periodic oscillations spectra and Lense-Thirring precession}},
 volume = {397},
 year = {2009}
}

@article{Ingram2019,
 adsurl = {https://ui.adsabs.harvard.edu/abs/2019NewAR..8501524I},
 archiveprefix = {arXiv},
 author = {{Ingram}, Adam R. and {Motta}, Sara E.},
 doi = {10.1016/j.newar.2020.101524},
 eid = {101524},
 eprint = {2001.08758},
 journal = {\nar},
 month = sep,
 pages = {101524},
 primaryclass = {astro-ph.HE},
 title = {{A review of quasi-periodic oscillations from black hole X-ray binaries: Observation and theory}},
 volume = {85},
 year = {2019}
}

@article{Israel1996,
 adsurl = {https://ui.adsabs.harvard.edu/abs/1996ApJ...468..369I},
 archiveprefix = {arXiv},
 author = {{Israel}, G.~L. and {Stella}, L.},
 doi = {10.1086/177697},
 eprint = {astro-ph/9603038},
 journal = {\apj},
 month = sep,
 pages = {369},
 primaryclass = {astro-ph},
 title = {{A New Technique for the Detection of Periodic Signals in ``Colored'' Power Spectra}},
 volume = {468},
 year = {1996}
}

@ARTICLE{Lui2025arXiv,
       author = {{Lui}, Leif and {Torres-Orjuela}, Alejandro and {Chowdhury}, Rudrani Kar and {Dai}, Lixin},
        title = "{Gravitational Wave Signatures of Quasi-Periodic Eruptions: LISA Detection Prospects for RX J1301.9+2747}",
      journal = {arXiv e-prints},
         year = 2025,
        month = aug,
          eid = {arXiv:2508.07961},
        pages = {arXiv:2508.07961},
          doi = {10.48550/arXiv.2508.07961},
archivePrefix = {arXiv},
       eprint = {2508.07961},
 primaryClass = {astro-ph.HE},
       adsurl = {https://ui.adsabs.harvard.edu/abs/2025arXiv250807961L}
}

@ARTICLE{Quintin2023,
       author = {{Quintin}, E. and {Webb}, N.~A. and {Guillot}, S. and {Miniutti}, G. and {Kammoun}, E.~S. and {Giustini}, M. and {Arcodia}, R. and {Soucail}, G. and {Clerc}, N. and {Amato}, R. and {Markwardt}, C.~B.},
        title = "{Tormund's return: Hints of quasi-periodic eruption features from a recent optical tidal disruption event}",
      journal = {\aap},
         year = 2023,
        month = jul,
       volume = {675},
          eid = {A152},
        pages = {A152},
          doi = {10.1051/0004-6361/202346440},
archivePrefix = {arXiv},
       eprint = {2306.00438},
 primaryClass = {astro-ph.HE},
       adsurl = {https://ui.adsabs.harvard.edu/abs/2023A&A...675A.152Q}
}

@article{Israel2016a,
 adsurl = {https://ui.adsabs.harvard.edu/abs/2016MNRAS.462.4371I},
 archiveprefix = {arXiv},
 author = {{Israel}, G.~L. and {Esposito}, P. and {Rodr{\'\i}guez Castillo}, G.~A. and {Sidoli}, L.},
 doi = {10.1093/mnras/stw1897},
 eprint = {1608.00077},
 journal = {\mnras},
 month = nov,
 number = {4},
 pages = {4371-4385},
 primaryclass = {astro-ph.HE},
 title = {{The Chandra ACIS Timing Survey Project: glimpsing a sample of faint X-ray pulsators}},
 volume = {462},
 year = {2016}
}

@ARTICLE{Ricci2021,
       author = {{Ricci}, C. and {Loewenstein}, M. and {Kara}, E. and {Remillard}, R. and {Trakhtenbrot}, B. and {Arcavi}, I. and {Gendreau}, K.~C. and {Arzoumanian}, Z. and {Fabian}, A.~C. and {Li}, R. and {Ho}, L.~C. and {MacLeod}, C.~L. and {Cackett}, E. and {Altamirano}, D. and {Gandhi}, P. and {Kosec}, P. and {Pasham}, D. and {Steiner}, J. and {Chan}, C. -H.},
        title = "{The 450 Day X-Ray Monitoring of the Changing-look AGN 1ES 1927+654}",
      journal = {\apjs},
         year = 2021,
        month = jul,
       volume = {255},
       number = {1},
          eid = {7},
        pages = {7},
          doi = {10.3847/1538-4365/abe94b},
archivePrefix = {arXiv},
       eprint = {2102.05666},
 primaryClass = {astro-ph.HE},
       adsurl = {https://ui.adsabs.harvard.edu/abs/2021ApJS..255....7R}
}

@ARTICLE{Wevers2024,
       author = {{Wevers}, T. and {French}, K.~D. and {Zabludoff}, A.~I. and {Fischer}, T.~C. and {Rowlands}, K. and {Guolo}, M. and {Dalla Barba}, B. and {Arcodia}, R. and {Berton}, M. and {Bian}, F. and {Linial}, I. and {Miniutti}, G. and {Pasham}, D.~R.},
        title = "{X-Ray Quasi-periodic Eruptions and Tidal Disruption Events Prefer Similar Host Galaxies}",
      journal = {\apjl},
         year = 2024,
        month = jul,
       volume = {970},
       number = {1},
          eid = {L23},
        pages = {L23},
          doi = {10.3847/2041-8213/ad5f1b},
archivePrefix = {arXiv},
       eprint = {2406.02678},
 primaryClass = {astro-ph.HE},
       adsurl = {https://ui.adsabs.harvard.edu/abs/2024ApJ...970L..23W}
}

@ARTICLE{Toba2014,
       author = {{Toba}, Y. and {Oyabu}, S. and {Matsuhara}, H. and {Malkan}, M.~A. and {Gandhi}, P. and {Nakagawa}, T. and {Isobe}, N. and {Shirahata}, M. and {Oi}, N. and {Ohyama}, Y. and {Takita}, S. and {Yamauchi}, C. and {Yano}, K.},
        title = "{Luminosity and Redshift Dependence of the Covering Factor of Active Galactic Nuclei viewed with WISE and Sloan Digital Sky Survey}",
      journal = {\apj},
         year = 2014,
        month = jun,
       volume = {788},
       number = {1},
          eid = {45},
        pages = {45},
          doi = {10.1088/0004-637X/788/1/45},
archivePrefix = {arXiv},
       eprint = {1404.4937},
 primaryClass = {astro-ph.GA},
       adsurl = {https://ui.adsabs.harvard.edu/abs/2014ApJ...788...45T}
}

@ARTICLE{VeronCetty2010,
       author = {{V{\'e}ron-Cetty}, M. -P. and {V{\'e}ron}, P.},
        title = "{A catalogue of quasars and active nuclei: 13th edition}",
      journal = {\aap},
         year = 2010,
        month = jul,
       volume = {518},
          eid = {A10},
        pages = {A10},
          doi = {10.1051/0004-6361/201014188},
       adsurl = {https://ui.adsabs.harvard.edu/abs/2010A&A...518A..10V}
}

@ARTICLE{Negus2024,
       author = {{Negus}, James and {Comerford}, Julia M. and {S{\'a}nchez}, Francisco M{\"u}ller},
        title = "{A Catalog of Broad H{\ensuremath{\alpha}} and H{\ensuremath{\beta}} Active Galactic Nuclei in MaNGA}",
      journal = {\apj},
         year = 2024,
        month = aug,
       volume = {971},
       number = {1},
          eid = {92},
        pages = {92},
          doi = {10.3847/1538-4357/ad4c68},
archivePrefix = {arXiv},
       eprint = {2405.12873},
 primaryClass = {astro-ph.GA},
       adsurl = {https://ui.adsabs.harvard.edu/abs/2024ApJ...971...92N}
}

@article{Jansen2001,
 adsurl = {https://ui.adsabs.harvard.edu/abs/2001A&A...365L...1J},
 author = {{Jansen}, F. and {Lumb}, D. and {Altieri}, B. and {Clavel}, J. and {Ehle}, M. and {Erd}, C. and {Gabriel}, C. and {Guainazzi}, M. and {Gondoin}, P. and {Much}, R. and {Munoz}, R. and {Santos}, M. and {Schartel}, N. and {Texier}, D. and {Vacanti}, G.},
 doi = {10.1051/0004-6361:20000036},
 journal = {\aap},
 month = jan,
 pages = {L1-L6},
 title = {{XMM-Newton observatory. I. The spacecraft and operations}},
 volume = {365},
 year = {2001}
}

@article{Lomb1976,
 adsurl = {https://ui.adsabs.harvard.edu/abs/1976Ap&SS..39..447L},
 author = {{Lomb}, N.~R.},
 doi = {10.1007/BF00648343},
 journal = {\apss},
 month = feb,
 number = {2},
 pages = {447-462},
 title = {{Least-Squares Frequency Analysis of Unequally Spaced Data}},
 volume = {39},
 year = {1976}
}

@article{Miniutti2019,
 adsurl = {https://ui.adsabs.harvard.edu/abs/2019Natur.573..381M},
 archiveprefix = {arXiv},
 author = {{Miniutti}, G. and {Saxton}, R.~D. and {Giustini}, M. and {Alexander}, K.~D. and {Fender}, R.~P. and {Heywood}, I. and {Monageng}, I. and {Coriat}, M. and {Tzioumis}, A.~K. and {Read}, A.~M. and {Knigge}, C. and {Gandhi}, P. and {Pretorius}, M.~L. and {Ag{\'\i}s-Gonz{\'a}lez}, B.},
 doi = {10.1038/s41586-019-1556-x},
 eprint = {1909.04693},
 journal = {\nat},
 month = sep,
 number = {7774},
 pages = {381-384},
 primaryclass = {astro-ph.HE},
 title = {{Nine-hour X-ray quasi-periodic eruptions from a low-mass black hole galactic nucleus}},
 volume = {573},
 year = {2019}
}

@ARTICLE{Miniutti2023,
       author = {{Miniutti}, G. and {Giustini}, M. and {Arcodia}, R. and {Saxton}, R.~D. and {Chakraborty}, J. and {Read}, A.~M. and {Kara}, E.},
        title = "{Alive and kicking: A new QPE phase in GSN 069 revealing a quiescent luminosity threshold for QPEs}",
      journal = {\aap},
         year = 2023,
        month = jun,
       volume = {674},
          eid = {L1},
        pages = {L1},
          doi = {10.1051/0004-6361/202346653},
archivePrefix = {arXiv},
       eprint = {2305.09717},
 primaryClass = {astro-ph.HE},
       adsurl = {https://ui.adsabs.harvard.edu/abs/2023A&A...674L...1M}
}

@article{Morgan1997,
 adsurl = {https://ui.adsabs.harvard.edu/abs/1997ApJ...482..993M},
 author = {{Morgan}, E.~H. and {Remillard}, R.~A. and {Greiner}, J.},
 doi = {10.1086/304191},
 journal = {\apj},
 month = jun,
 number = {2},
 pages = {993-1010},
 title = {{RXTE Observations of QPOs in the Black Hole Candidate GRS 1915+105}},
 volume = {482},
 year = {1997}
}

@article{Motch1983,
 adsurl = {https://ui.adsabs.harvard.edu/abs/1983A&A...119..171M},
 author = {{Motch}, C. and {Ricketts}, M.~J. and {Page}, C.~G. and {Ilovaisky}, S.~A. and {Chevalier}, C.},
 journal = {\aap},
 month = mar,
 pages = {171-176},
 title = {{Simultaneous X-ray/optical observations of GX 339-4 during the May 1981 optically bright state.}},
 volume = {119},
 year = {1983}
}

@ARTICLE{Pasham2024,
       author = {{Pasham}, Dheeraj R. and {Tombesi}, Francesco and {Sukov{\'a}}, Petra and {Zaja{\v{c}}ek}, Michal and {Rakshit}, Suvendu and {Coughlin}, Eric and {Kosec}, Peter and {Karas}, Vladim{\'\i}r and {Masterson}, Megan and {Mummery}, Andrew and {Holoien}, Thomas W. -S. and {Guolo}, Muryel and {Hinkle}, Jason and {Ripperda}, Bart and {Witzany}, Vojt{\v{e}}ch and {Shappee}, Ben and {Kara}, Erin and {Horesh}, Assaf and {van Velzen}, Sjoert and {Sfaradi}, Itai and {Kaplan}, David and {Burger}, Noam and {Murphy}, Tara and {Remillard}, Ronald and {Steiner}, James F. and {Wevers}, Thomas and {Arcodia}, Riccardo and {Buchner}, Johannes and {Merloni}, Andrea and {Malyali}, Adam and {Fabian}, Andy and {Fausnaugh}, Michael and {Daylan}, Tansu and {Altamirano}, Diego and {Payne}, Anna and {Ferraraa}, Elizabeth C.},
        title = "{A case for a binary black hole system revealed via quasi-periodic outflows}",
      journal = {Science Advances},
         year = 2024,
        month = mar,
       volume = {10},
       number = {13},
          eid = {eadj8898},
        pages = {eadj8898},
          doi = {10.1126/sciadv.adj8898},
archivePrefix = {arXiv},
       eprint = {2402.10140},
 primaryClass = {astro-ph.HE},
       adsurl = {https://ui.adsabs.harvard.edu/abs/2024SciA...10J8898P}
}

@ARTICLE{Linial2023,
       author = {{Linial}, Itai and {Metzger}, Brian D.},
        title = "{EMRI + TDE = QPE: Periodic X-Ray Flares from Star-Disk Collisions in Galactic Nuclei}",
      journal = {\apj},
         year = 2023,
        month = nov,
       volume = {957},
       number = {1},
          eid = {34},
        pages = {34},
          doi = {10.3847/1538-4357/acf65b},
archivePrefix = {arXiv},
       eprint = {2303.16231},
 primaryClass = {astro-ph.HE},
       adsurl = {https://ui.adsabs.harvard.edu/abs/2023ApJ...957...34L}
}

@dataset{GAIA2020,
       author = {{Gaia Collaboration}},
        title = "{VizieR Online Data Catalog: Gaia EDR3 (Gaia Collaboration, 2020)}",
 howpublished = {VizieR On-line Data Catalog: I/350.  Originally published in: 2021A\&A...649A...1G; doi:10.5270/esa-1ug},
         year = 2020,
        month = nov,
          eid = {I/350},
          doi = {10.26093/cds/vizier.1350},
       adsurl = {https://ui.adsabs.harvard.edu/abs/2020yCat.1350....0G}
}

@ARTICLE{Franchini2023,
       author = {{Franchini}, Alessia and {Bonetti}, Matteo and {Lupi}, Alessandro and {Miniutti}, Giovanni and {Bortolas}, Elisa and {Giustini}, Margherita and {Dotti}, Massimo and {Sesana}, Alberto and {Arcodia}, Riccardo and {Ryu}, Taeho},
        title = "{Quasi-periodic eruptions from impacts between the secondary and a rigidly precessing accretion disc in an extreme mass-ratio inspiral system}",
      journal = {\aap},
         year = 2023,
        month = jul,
       volume = {675},
          eid = {A100},
        pages = {A100},
          doi = {10.1051/0004-6361/202346565},
archivePrefix = {arXiv},
       eprint = {2304.00775},
 primaryClass = {astro-ph.HE},
       adsurl = {https://ui.adsabs.harvard.edu/abs/2023A&A...675A.100F}
}

@article{Remillard2006,
 adsurl = {https://ui.adsabs.harvard.edu/abs/2006ARA&A..44...49R},
 archiveprefix = {arXiv},
 author = {{Remillard}, Ronald A. and {McClintock}, Jeffrey E.},
 doi = {10.1146/annurev.astro.44.051905.092532},
 eprint = {astro-ph/0606352},
 journal = {\araa},
 month = sep,
 number = {1},
 pages = {49-92},
 primaryclass = {astro-ph},
 title = {{X-Ray Properties of Black-Hole Binaries}},
 volume = {44},
 year = {2006}
}

@article{Scargle1982,
 adsurl = {https://ui.adsabs.harvard.edu/abs/1982ApJ...263..835S},
 author = {{Scargle}, J.~D.},
 doi = {10.1086/160554},
 journal = {\apj},
 month = dec,
 pages = {835-853},
 title = {{Studies in astronomical time series analysis. II. Statistical aspects of spectral analysis of unevenly spaced data.}},
 volume = {263},
 year = {1982}
}

@article{Sidoli2016,
 adsurl = {https://ui.adsabs.harvard.edu/abs/2016MNRAS.460.3637S},
 archiveprefix = {arXiv},
 author = {{Sidoli}, L. and {Esposito}, P. and {Motta}, S.~E. and {Israel}, G.~L. and {Rodr{\'\i}guez Castillo}, G.~A.},
 doi = {10.1093/mnras/stw1246},
 eprint = {1605.06356},
 journal = {\mnras},
 month = aug,
 number = {4},
 pages = {3637-3646},
 primaryclass = {astro-ph.HE},
 title = {{XMM-Newton discovery of mHz quasi-periodic oscillations in the high-mass X-ray binary IGR J19140+0951}},
 volume = {460},
 year = {2016}
}

@article{Sidoli2017,
 adsurl = {https://ui.adsabs.harvard.edu/abs/2017MNRAS.469.3056S},
 archiveprefix = {arXiv},
 author = {{Sidoli}, L. and {Israel}, G.~L. and {Esposito}, P. and {Rodr{\'\i}guez Castillo}, G.~A. and {Postnov}, K.},
 doi = {10.1093/mnras/stx1105},
 eprint = {1705.01791},
 journal = {\mnras},
 month = aug,
 number = {3},
 pages = {3056-3061},
 primaryclass = {astro-ph.HE},
 title = {{AX J1910.7+0917: the slowest X-ray pulsar}},
 volume = {469},
 year = {2017}
}

@article{Smith2018,
 adsurl = {https://ui.adsabs.harvard.edu/abs/2018ApJ...860L..10S},
 archiveprefix = {arXiv},
 author = {{Smith}, Krista Lynne and {Mushotzky}, Richard F. and {Boyd}, Patricia T. and {Wagoner}, Robert V.},
 doi = {10.3847/2041-8213/aac88c},
 eid = {L10},
 eprint = {1805.12154},
 journal = {\apjl},
 month = jun,
 number = {1},
 pages = {L10},
 primaryclass = {astro-ph.HE},
 title = {{Evidence for an Optical Low-frequency Quasi-periodic Oscillation in the Kepler Light Curve of an Active Galaxy}},
 volume = {860},
 year = {2018}
}

@article{Stella1998,
 adsurl = {https://ui.adsabs.harvard.edu/abs/1998ApJ...492L..59S},
 archiveprefix = {arXiv},
 author = {{Stella}, Luigi and {Vietri}, Mario},
 doi = {10.1086/311075},
 eprint = {astro-ph/9709085},
 journal = {\apjl},
 month = jan,
 number = {1},
 pages = {L59-L62},
 primaryclass = {astro-ph},
 title = {{Lense-Thirring Precession and Quasi-periodic Oscillations in Low-Mass X-Ray Binaries}},
 volume = {492},
 year = {1998}
}

@article{Stella1999,
 adsurl = {https://ui.adsabs.harvard.edu/abs/1999ApJ...524L..63S},
 archiveprefix = {arXiv},
 author = {{Stella}, Luigi and {Vietri}, Mario and {Morsink}, Sharon M.},
 doi = {10.1086/312291},
 eprint = {astro-ph/9907346},
 journal = {\apjl},
 month = oct,
 number = {1},
 pages = {L63-L66},
 primaryclass = {astro-ph},
 title = {{Correlations in the Quasi-periodic Oscillation Frequencies of Low-Mass X-Ray Binaries and the Relativistic Precession Model}},
 volume = {524},
 year = {1999}
}

@article{Struder2001,
 adsurl = {https://ui.adsabs.harvard.edu/abs/2001A&A...365L..18S},
 author = {{Str{\"u}der}, L. and {Briel}, U. and {Dennerl}, K. and {Hartmann}, R. and {Kendziorra}, E. and {Meidinger}, N. and {Pfeffermann}, E. and {Reppin}, C. and {Aschenbach}, B. and {Bornemann}, W. and {Br{\"a}uninger}, H. and {Burkert}, W. and {Elender}, M. and {Freyberg}, M. and {Haberl}, F. and {Hartner}, G. and {Heuschmann}, F. and {Hippmann}, H. and {Kastelic}, E. and {Kemmer}, S. and {Kettenring}, G. and {Kink}, W. and {Krause}, N. and {M{\"u}ller}, S. and {Oppitz}, A. and {Pietsch}, W. and {Popp}, M. and {Predehl}, P. and {Read}, A. and {Stephan}, K.~H. and {St{\"o}tter}, D. and {Tr{\"u}mper}, J. and {Holl}, P. and {Kemmer}, J. and {Soltau}, H. and {St{\"o}tter}, R. and {Weber}, U. and {Weichert}, U. and {von Zanthier}, C. and {Carathanassis}, D. and {Lutz}, G. and {Richter}, R.~H. and {Solc}, P. and {B{\"o}ttcher}, H. and {Kuster}, M. and {Staubert}, R. and {Abbey}, A. and {Holland}, A. and {Turner}, M. and {Balasini}, M. and {Bignami}, G.~F. and {La Palombara}, N. and {Villa}, G. and {Buttler}, W. and {Gianini}, F. and {Lain{\'e}}, R. and {Lumb}, D. and {Dhez}, P.},
 doi = {10.1051/0004-6361:20000066},
 journal = {\aap},
 month = jan,
 pages = {L18-L26},
 title = {{The European Photon Imaging Camera on XMM-Newton: The pn-CCD camera}},
 volume = {365},
 year = {2001}
}

@article{Tagger1999,
 adsurl = {https://ui.adsabs.harvard.edu/abs/1999A&A...349.1003T},
 archiveprefix = {arXiv},
 author = {{Tagger}, M. and {Pellat}, R.},
 doi = {10.48550/arXiv.astro-ph/9907267},
 eprint = {astro-ph/9907267},
 journal = {\aap},
 month = sep,
 pages = {1003-1016},
 primaryclass = {astro-ph},
 title = {{An accretion-ejection instability in magnetized disks}},
 volume = {349},
 year = {1999}
}

@article{Turner2001,
 adsurl = {https://ui.adsabs.harvard.edu/abs/2001A&A...365L..27T},
 archiveprefix = {arXiv},
 author = {{Turner}, M.~J.~L. and {Abbey}, A. and {Arnaud}, M. and {Balasini}, M. and {Barbera}, M. and {Belsole}, E. and {Bennie}, P.~J. and {Bernard}, J.~P. and {Bignami}, G.~F. and {Boer}, M. and {Briel}, U. and {Butler}, I. and {Cara}, C. and {Chabaud}, C. and {Cole}, R. and {Collura}, A. and {Conte}, M. and {Cros}, A. and {Denby}, M. and {Dhez}, P. and {Di Coco}, G. and {Dowson}, J. and {Ferrando}, P. and {Ghizzardi}, S. and {Gianotti}, F. and {Goodall}, C.~V. and {Gretton}, L. and {Griffiths}, R.~G. and {Hainaut}, O. and {Hochedez}, J.~F. and {Holland}, A.~D. and {Jourdain}, E. and {Kendziorra}, E. and {Lagostina}, A. and {Laine}, R. and {La Palombara}, N. and {Lortholary}, M. and {Lumb}, D. and {Marty}, P. and {Molendi}, S. and {Pigot}, C. and {Poindron}, E. and {Pounds}, K.~A. and {Reeves}, J.~N. and {Reppin}, C. and {Rothenflug}, R. and {Salvetat}, P. and {Sauvageot}, J.~L. and {Schmitt}, D. and {Sembay}, S. and {Short}, A.~D.~T. and {Spragg}, J. and {Stephen}, J. and {Str{\"u}der}, L. and {Tiengo}, A. and {Trifoglio}, M. and {Tr{\"u}mper}, J. and {Vercellone}, S. and {Vigroux}, L. and {Villa}, G. and {Ward}, M.~J. and {Whitehead}, S. and {Zonca}, E.},
 doi = {10.1051/0004-6361:20000087},
 eprint = {astro-ph/0011498},
 journal = {\aap},
 month = jan,
 pages = {L27-L35},
 primaryclass = {astro-ph},
 title = {{The European Photon Imaging Camera on XMM-Newton: The MOS cameras}},
 volume = {365},
 year = {2001}
}

@article{vanderKlis1989,
 adsurl = {https://ui.adsabs.harvard.edu/abs/1989ARA&A..27..517V},
 author = {{van der Klis}, M.},
 doi = {10.1146/annurev.aa.27.090189.002505},
 journal = {\araa},
 month = jan,
 pages = {517-553},
 title = {{Quasi-periodic oscillations and noise in low-mass X-ray binaries.}},
 volume = {27},
 year = {1989}
}

@inproceedings{vanderKlis1989a,
 adsurl = {https://ui.adsabs.harvard.edu/abs/1989ASIC..262...27V},
 author = {{van der Klis}, M.},
 booktitle = {Timing Neutron Stars},
 doi = {10.1007/978-94-009-2273-0_3},
 editor = {{{\"O}gelman}, H. and {van den Heuvel}, E.~P.~J.},
 month = jan,
 pages = {27},
 series = {NATO Advanced Study Institute (ASI) Series C},
 title = {{Fourier techniques in X-ray timing}},
 volume = {262},
 year = {1989}
}

@article{Verner1996,
 adsurl = {https://ui.adsabs.harvard.edu/abs/1996ApJ...465..487V},
 archiveprefix = {arXiv},
 author = {{Verner}, D.~A. and {Ferland}, G.~J. and {Korista}, K.~T. and {Yakovlev}, D.~G.},
 doi = {10.1086/177435},
 eprint = {astro-ph/9601009},
 journal = {\apj},
 month = jul,
 pages = {487},
 primaryclass = {astro-ph},
 title = {{Atomic Data for Astrophysics. II. New Analytic FITS for Photoionization Cross Sections of Atoms and Ions}},
 volume = {465},
 year = {1996}
}

@article{Vikhlinin1994,
 adsurl = {https://ui.adsabs.harvard.edu/abs/1994ApJ...424..395V},
 author = {{Vikhlinin}, A. and {Churazov}, E. and {Gilfanov}, M. and {Sunyaev}, R. and {Dyachkov}, A. and {Khavenson}, N. and {Kremnev}, R. and {Sukhanov}, K. and {Ballet}, J. and {Laurent}, P. and {Salotti}, L. and {Claret}, A. and {Olive}, J.~F. and {Denis}, M. and {Mandrou}, P. and {Roques}, J.~P.},
 doi = {10.1086/173897},
 journal = {\apj},
 month = mar,
 pages = {395},
 title = {{Discovery of a Low-Frequency Broad Quasi-periodic Oscillation Peak in the Power Density Spectrum of Cygnus X-1 with Granat/SIGMA}},
 volume = {424},
 year = {1994}
}

@inproceedings{Weisskopf2000,
 adsurl = {https://ui.adsabs.harvard.edu/abs/2000SPIE.4012....2W},
 archiveprefix = {arXiv},
 author = {{Weisskopf}, Martin C. and {Tananbaum}, Harvey D. and {Van Speybroeck}, Leon P. and {O'Dell}, Stephen L.},
 booktitle = {X-Ray Optics, Instruments, and Missions III},
 doi = {10.1117/12.391545},
 editor = {{Truemper}, Joachim E. and {Aschenbach}, Bernd},
 eprint = {astro-ph/0004127},
 month = jul,
 pages = {2-16},
 primaryclass = {astro-ph},
 series = {Society of Photo-Optical Instrumentation Engineers (SPIE) Conference Series},
 title = {{Chandra X-ray Observatory (CXO): overview}},
 volume = {4012},
 year = {2000}
}

@article{Wilms2000,
 adsurl = {https://ui.adsabs.harvard.edu/abs/2000ApJ...542..914W},
 archiveprefix = {arXiv},
 author = {{Wilms}, J. and {Allen}, A. and {McCray}, R.},
 doi = {10.1086/317016},
 eprint = {astro-ph/0008425},
 journal = {\apj},
 month = oct,
 number = {2},
 pages = {914-924},
 primaryclass = {astro-ph},
 title = {{On the Absorption of X-Rays in the Interstellar Medium}},
 volume = {542},
 year = {2000}
}

@ARTICLE{Xiao2011,
       author = {{Xiao}, Ting and {Barth}, Aaron J. and {Greene}, Jenny E. and {Ho}, Luis C. and {Bentz}, Misty C. and {Ludwig}, Randi R. and {Jiang}, Yanfei},
        title = "{Exploring the Low-mass End of the M $_{BH}$-{\ensuremath{\sigma}}$_{*}$ Relation with Active Galaxies}",
      journal = {\apj},
         year = 2011,
        month = sep,
       volume = {739},
       number = {1},
          eid = {28},
        pages = {28},
          doi = {10.1088/0004-637X/739/1/28},
archivePrefix = {arXiv},
       eprint = {1106.6232},
 primaryClass = {astro-ph.CO},
       adsurl = {https://ui.adsabs.harvard.edu/abs/2011ApJ...739...28X}
}

@ARTICLE{Gezari2021,
       author = {{Gezari}, Suvi},
        title = "{Tidal Disruption Events}",
      journal = {\araa},
         year = 2021,
        month = sep,
       volume = {59},
        pages = {21-58},
          doi = {10.1146/annurev-astro-111720-030029},
archivePrefix = {arXiv},
       eprint = {2104.14580},
 primaryClass = {astro-ph.HE},
       adsurl = {https://ui.adsabs.harvard.edu/abs/2021ARA&A..59...21G}
}

@ARTICLE{Rees1988,
       author = {{Rees}, Martin J.},
        title = "{Tidal disruption of stars by black holes of {}10$^{6}$-{}10$^{8}$ solar masses in nearby galaxies}",
      journal = {\nat},
         year = 1988,
        month = jun,
       volume = {333},
       number = {6173},
        pages = {523-528},
          doi = {10.1038/333523a0},
       adsurl = {https://ui.adsabs.harvard.edu/abs/1988Natur.333..523R}
}

@ARTICLE{Jefremov2015,
       author = {{Jefremov}, Paul I. and {Tsupko}, Oleg Yu. and {Bisnovatyi-Kogan}, Gennady S.},
        title = "{Innermost stable circular orbits of spinning test particles in Schwarzschild and Kerr space-times}",
      journal = {\prd},
         year = 2015,
        month = jun,
       volume = {91},
       number = {12},
          eid = {124030},
        pages = {124030},
          doi = {10.1103/PhysRevD.91.124030},
archivePrefix = {arXiv},
       eprint = {1503.07060},
 primaryClass = {gr-qc},
       adsurl = {https://ui.adsabs.harvard.edu/abs/2015PhRvD..91l4030J}
}

@ARTICLE{AmaroSeoane2017,
       author = {{Amaro-Seoane}, Pau and {Audley}, Heather and {Babak}, Stanislav and {Baker}, John and {Barausse}, Enrico and {Bender}, Peter and {Berti}, Emanuele and {Binetruy}, Pierre and {Born}, Michael and {Bortoluzzi}, Daniele and {Camp}, Jordan and {Caprini}, Chiara and {Cardoso}, Vitor and {Colpi}, Monica and {Conklin}, John and {Cornish}, Neil and {Cutler}, Curt and {Danzmann}, Karsten and {Dolesi}, Rita and {Ferraioli}, Luigi and {Ferroni}, Valerio and {Fitzsimons}, Ewan and {Gair}, Jonathan and {Gesa Bote}, Lluis and {Giardini}, Domenico and {Gibert}, Ferran and {Grimani}, Catia and {Halloin}, Hubert and {Heinzel}, Gerhard and {Hertog}, Thomas and {Hewitson}, Martin and {Holley-Bockelmann}, Kelly and {Hollington}, Daniel and {Hueller}, Mauro and {Inchauspe}, Henri and {Jetzer}, Philippe and {Karnesis}, Nikos and {Killow}, Christian and {Klein}, Antoine and {Klipstein}, Bill and {Korsakova}, Natalia and {Larson}, Shane L and {Livas}, Jeffrey and {Lloro}, Ivan and {Man}, Nary and {Mance}, Davor and {Martino}, Joseph and {Mateos}, Ignacio and {McKenzie}, Kirk and {McWilliams}, Sean T and {Miller}, Cole and {Mueller}, Guido and {Nardini}, Germano and {Nelemans}, Gijs and {Nofrarias}, Miquel and {Petiteau}, Antoine and {Pivato}, Paolo and {Plagnol}, Eric and {Porter}, Ed and {Reiche}, Jens and {Robertson}, David and {Robertson}, Norna and {Rossi}, Elena and {Russano}, Giuliana and {Schutz}, Bernard and {Sesana}, Alberto and {Shoemaker}, David and {Slutsky}, Jacob and {Sopuerta}, Carlos F. and {Sumner}, Tim and {Tamanini}, Nicola and {Thorpe}, Ira and {Troebs}, Michael and {Vallisneri}, Michele and {Vecchio}, Alberto and {Vetrugno}, Daniele and {Vitale}, Stefano and {Volonteri}, Marta and {Wanner}, Gudrun and {Ward}, Harry and {Wass}, Peter and {Weber}, William and {Ziemer}, John and {Zweifel}, Peter},
        title = "{Laser Interferometer Space Antenna}",
      journal = {arXiv e-prints},
         year = 2017,
        month = feb,
          eid = {arXiv:1702.00786},
        pages = {arXiv:1702.00786},
          doi = {10.48550/arXiv.1702.00786},
archivePrefix = {arXiv},
       eprint = {1702.00786},
 primaryClass = {astro-ph.IM},
       adsurl = {https://ui.adsabs.harvard.edu/abs/2017arXiv170200786A}
}

@ARTICLE{AmaroSeoane2018,
       author = {{Amaro-Seoane}, Pau},
        title = "{Relativistic dynamics and extreme mass ratio inspirals}",
      journal = {Living Reviews in Relativity},
         year = 2018,
        month = dec,
       volume = {21},
       number = {1},
          eid = {4},
        pages = {4},
          doi = {10.1007/s41114-018-0013-8},
archivePrefix = {arXiv},
       eprint = {1205.5240},
 primaryClass = {astro-ph.CO},
       adsurl = {https://ui.adsabs.harvard.edu/abs/2018LRR....21....4A}
}

@ARTICLE{Backer1986,
       author = {{Backer}, D.~C. and {Hellings}, R.~W.},
        title = "{Pulsar timing and general relativity.}",
      journal = {\araa},
         year = 1986,
        month = jan,
       volume = {24},
        pages = {537-575},
          doi = {10.1146/annurev.aa.24.090186.002541},
       adsurl = {https://ui.adsabs.harvard.edu/abs/1986ARA&A..24..537B}
}

\begin{appendix} 
 \section{Light curves}\label{appendix:timing}

 \begin{figure}[htbp!]
     \centering
     \includegraphics[width=0.95\columnwidth]{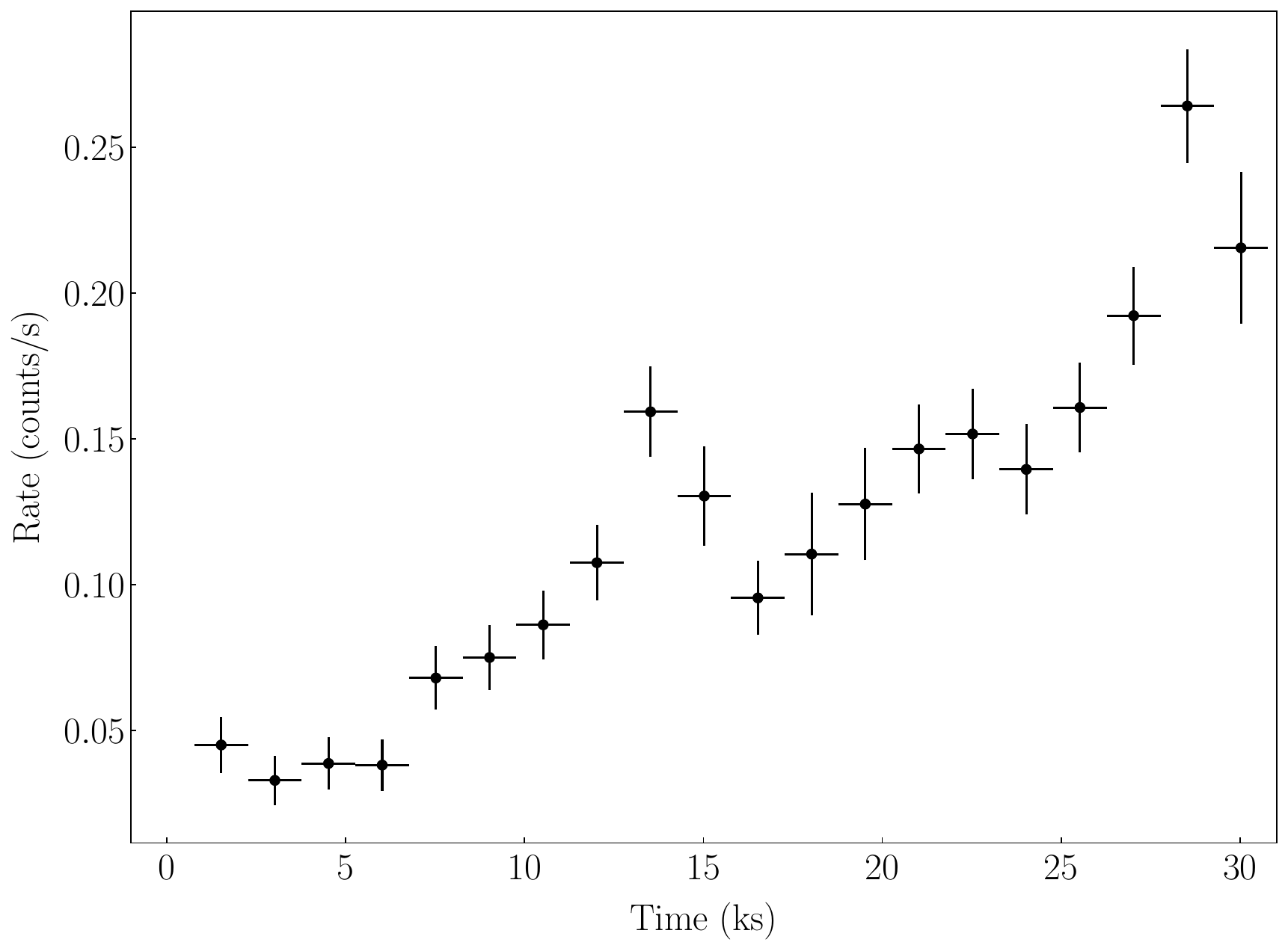}
     \caption{\xmm\ light curve in the 0.3--10\,keV band during ObsID 0124710101.}
     \label{fig:X0124710101}
 \end{figure}

   \begin{figure}[htbp!]
     \centering
     \includegraphics[width=0.95\columnwidth]{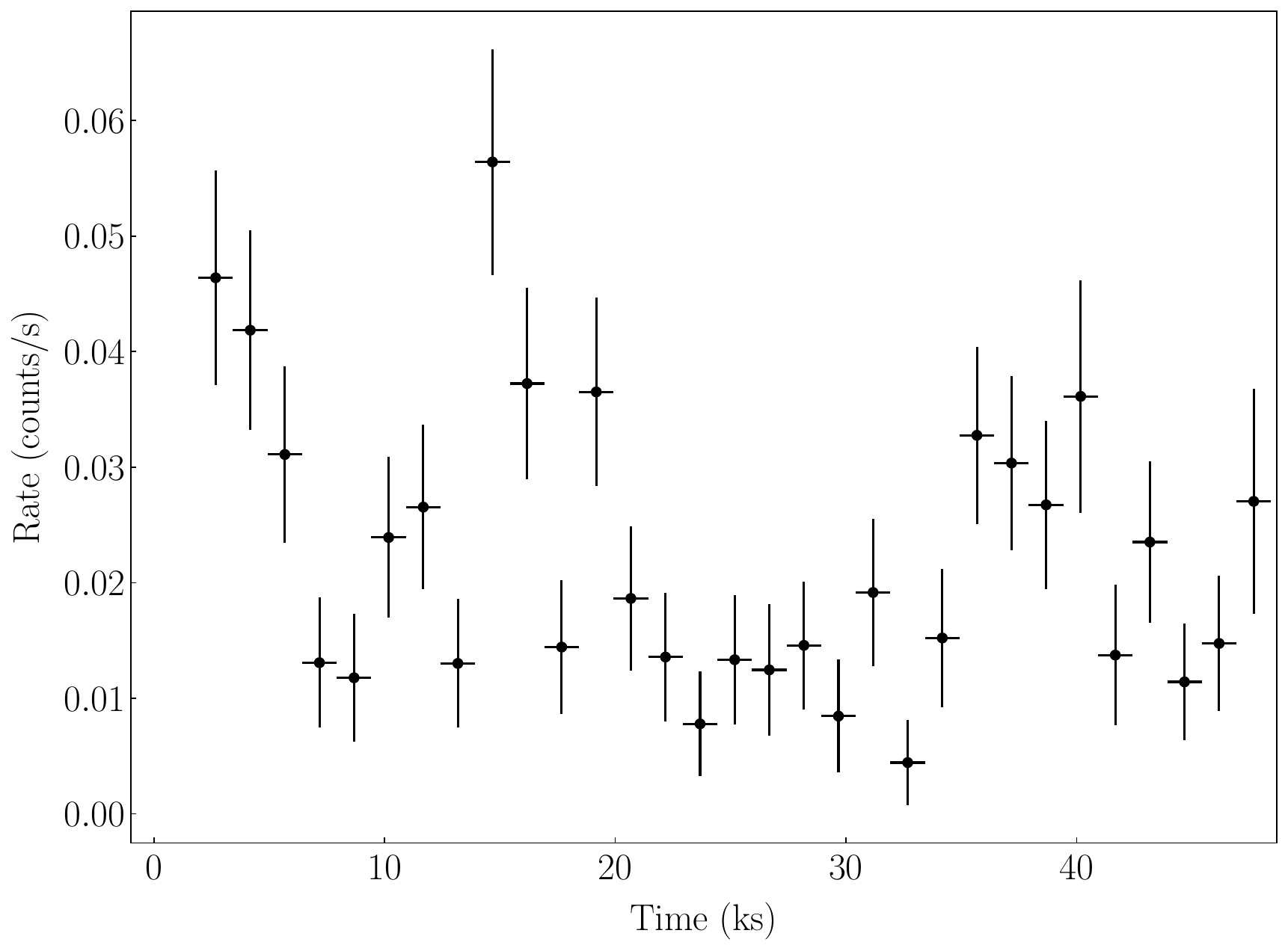}
     \caption{\xmm\ light curve in the 0.3--10\,keV band during ObsID 0403150201.}
     \label{fig:X0403}
 \end{figure}

 \begin{figure}[htbp!]
     \centering
     \includegraphics[width=0.95\columnwidth]{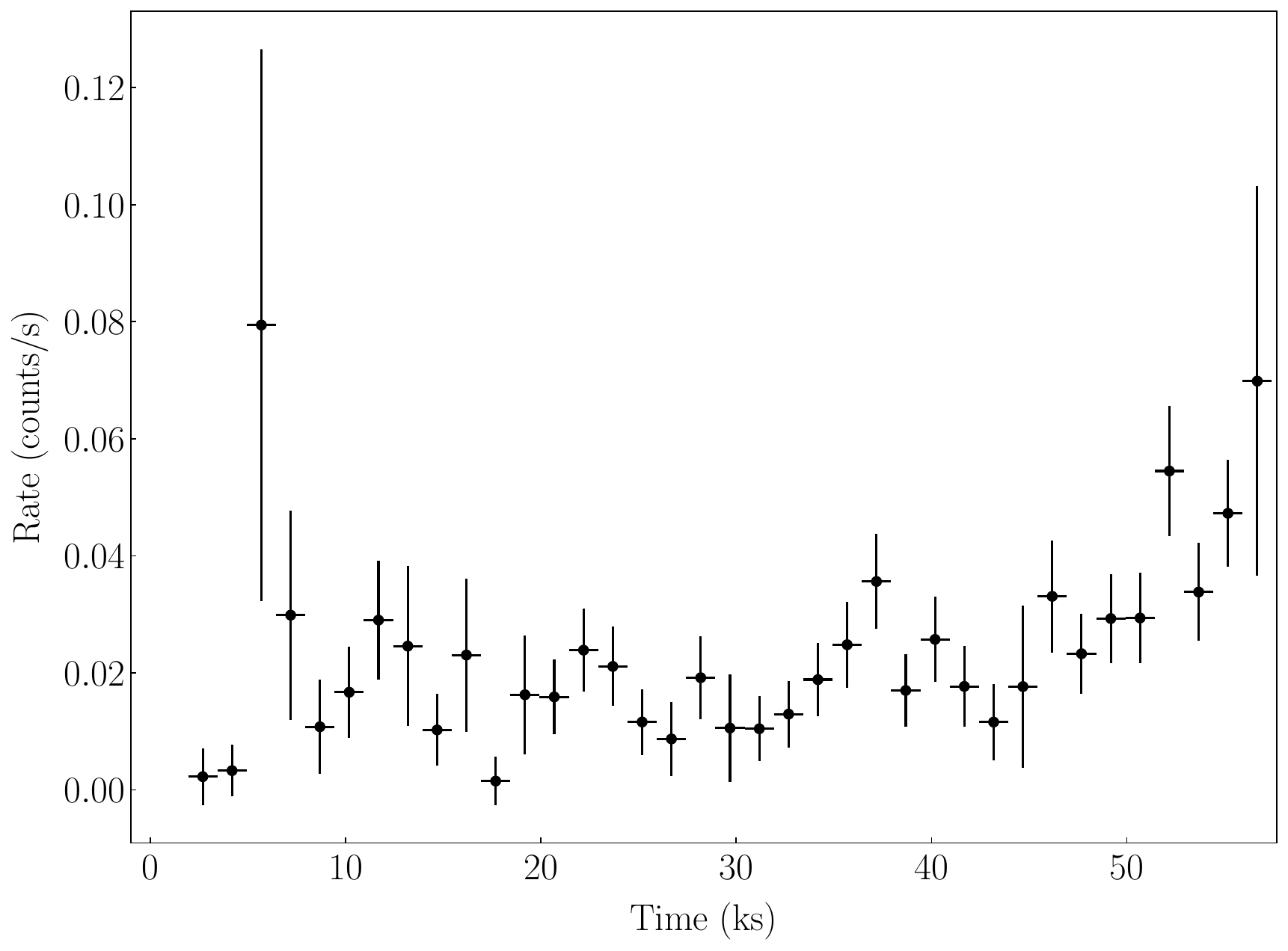}
     \caption{\xmm\ light curve in the 0.3--10\,keV band during ObsID 0403150101.}
     \label{fig:X0403101}
 \end{figure}

 \begin{figure}[htbp!]
     \centering
     \includegraphics[width=0.95\columnwidth]{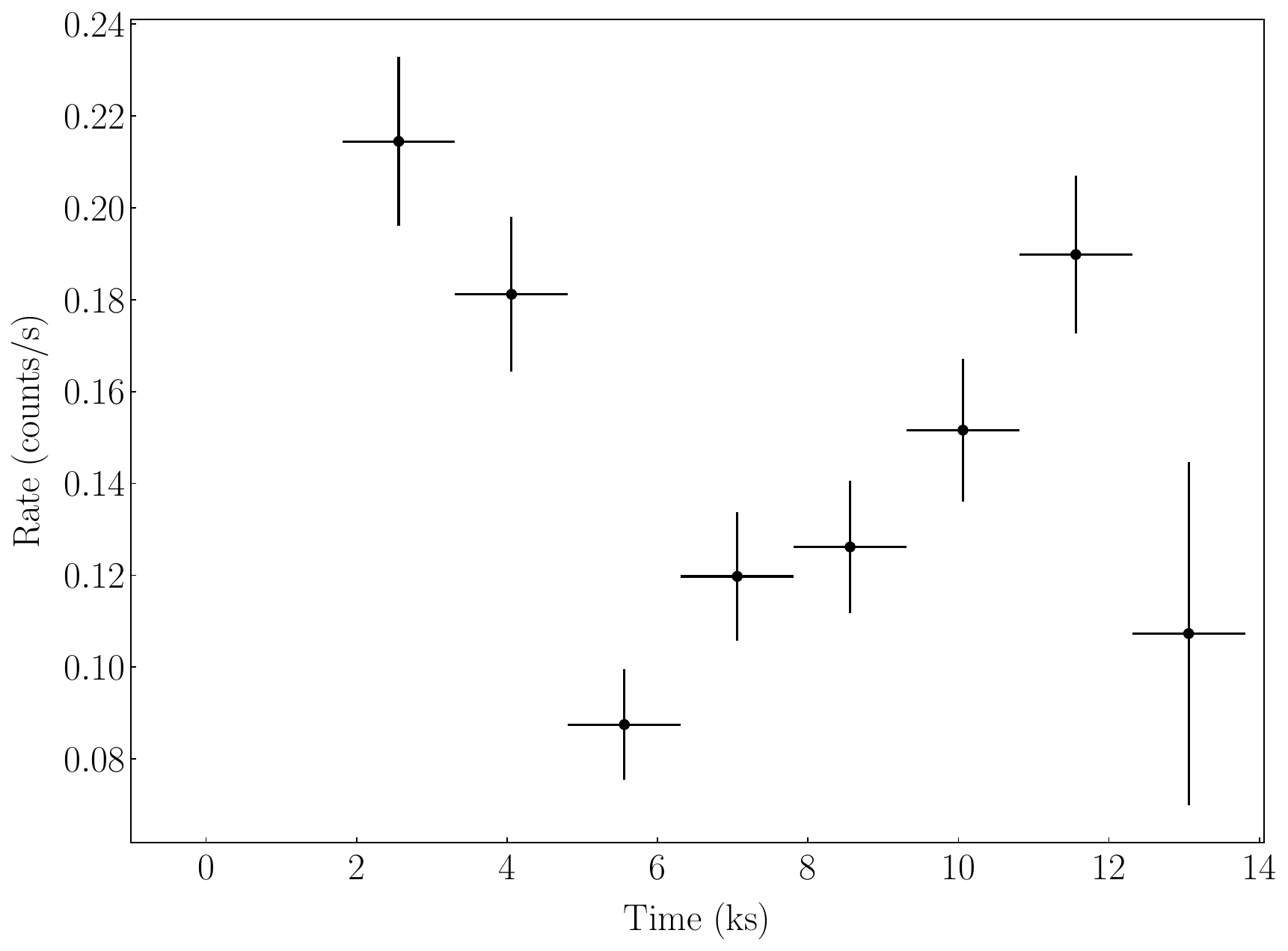}
     \caption{\xmm\ light curve in the 0.3--10\,keV band during ObsID 0652310401.}
     \label{fig:X0652310401}
 \end{figure}

 \begin{figure}[htbp!]
     \centering
     \includegraphics[width=0.95\columnwidth]{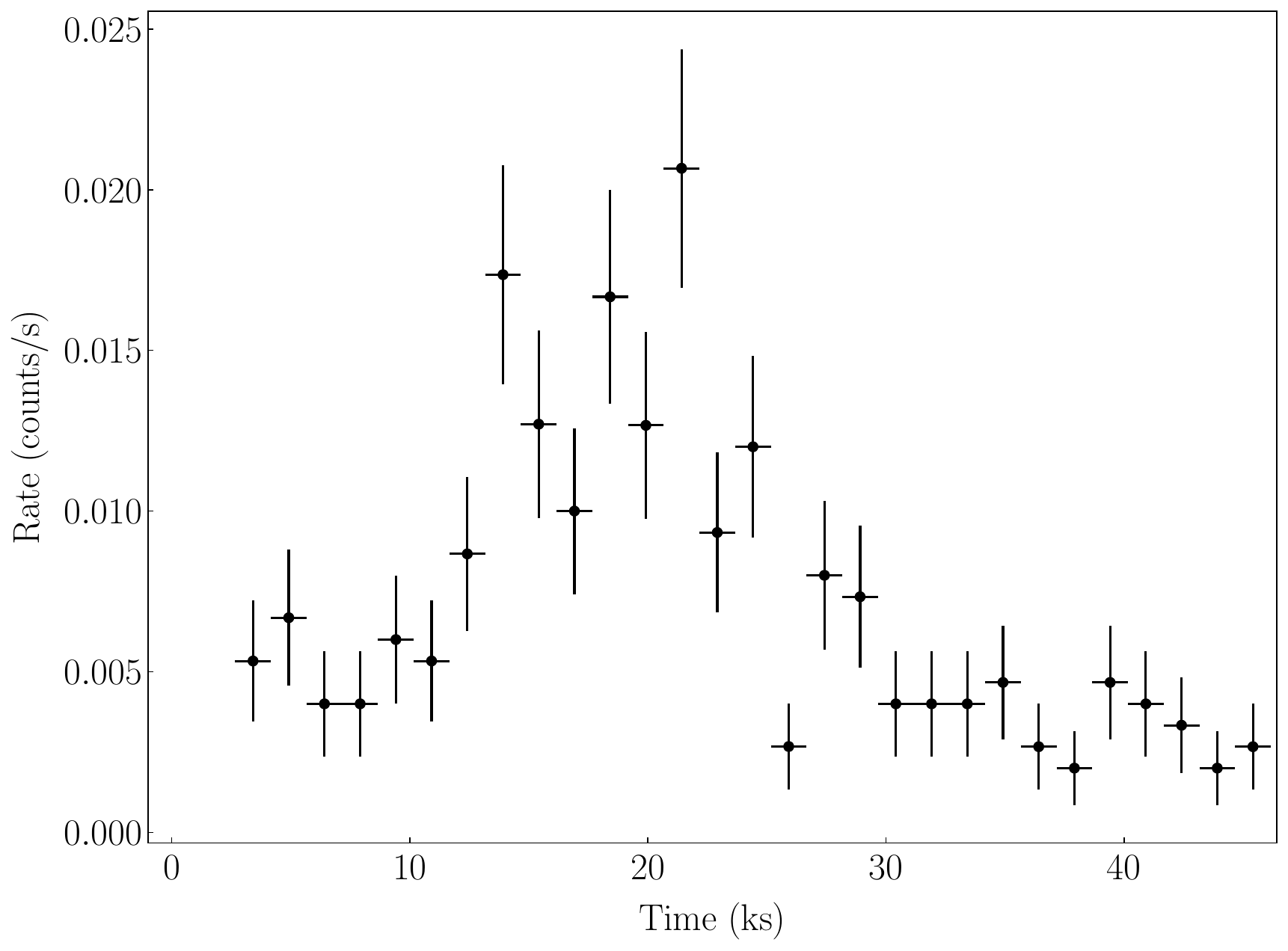}
     \caption{\chandra\ light curve in the 0.5--7\,keV band duèring ObsID 12887.}
     \label{fig:C12887}
 \end{figure} 

 \begin{figure}[htbp!]
     \centering
     \includegraphics[width=0.95\columnwidth]{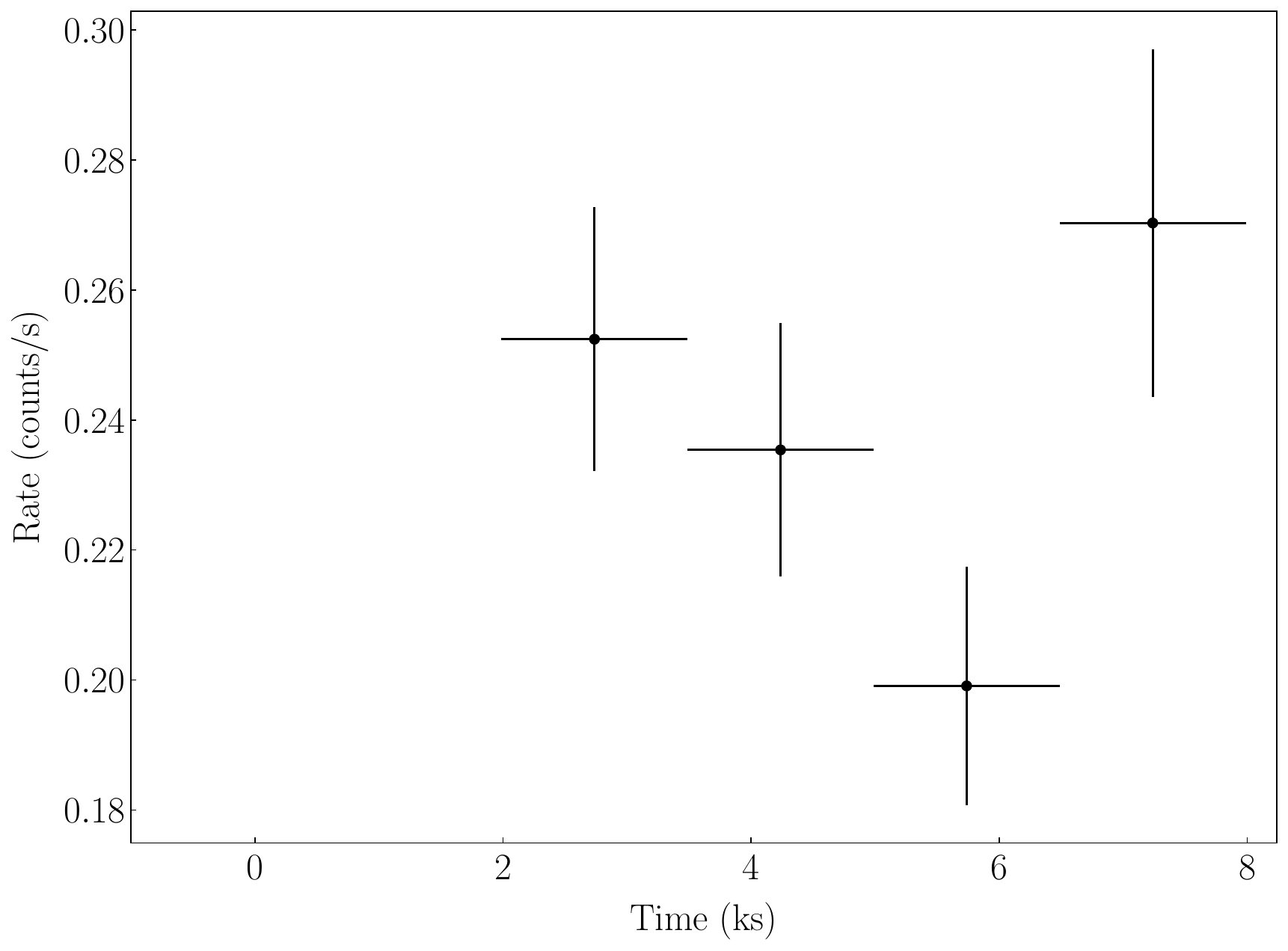}
     \caption{\xmm\ light curve in the 0.3--10\,keV band during ObsID 0652310801.}
     \label{fig:X0652310801}
 \end{figure}

 \begin{figure}[htbp!]
     \centering
     \includegraphics[width=0.95\columnwidth]{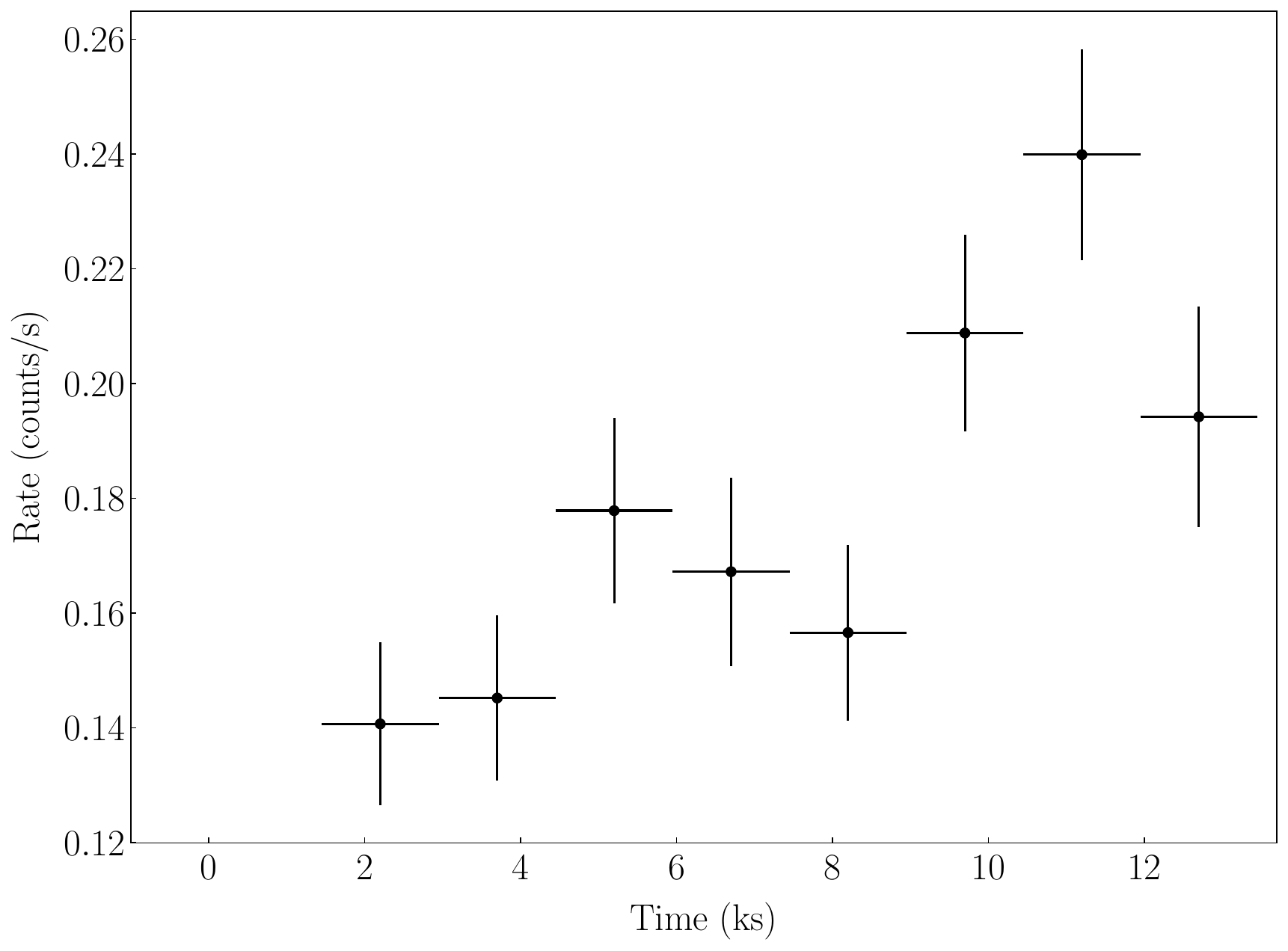}
     \caption{\xmm\ light curve in the 0.3--10\,keV band during ObsID 0652310901.}
     \label{fig:X0652310901}
 \end{figure}

\begin{figure}[htbp!]
     \centering
     \includegraphics[width=0.95\columnwidth]{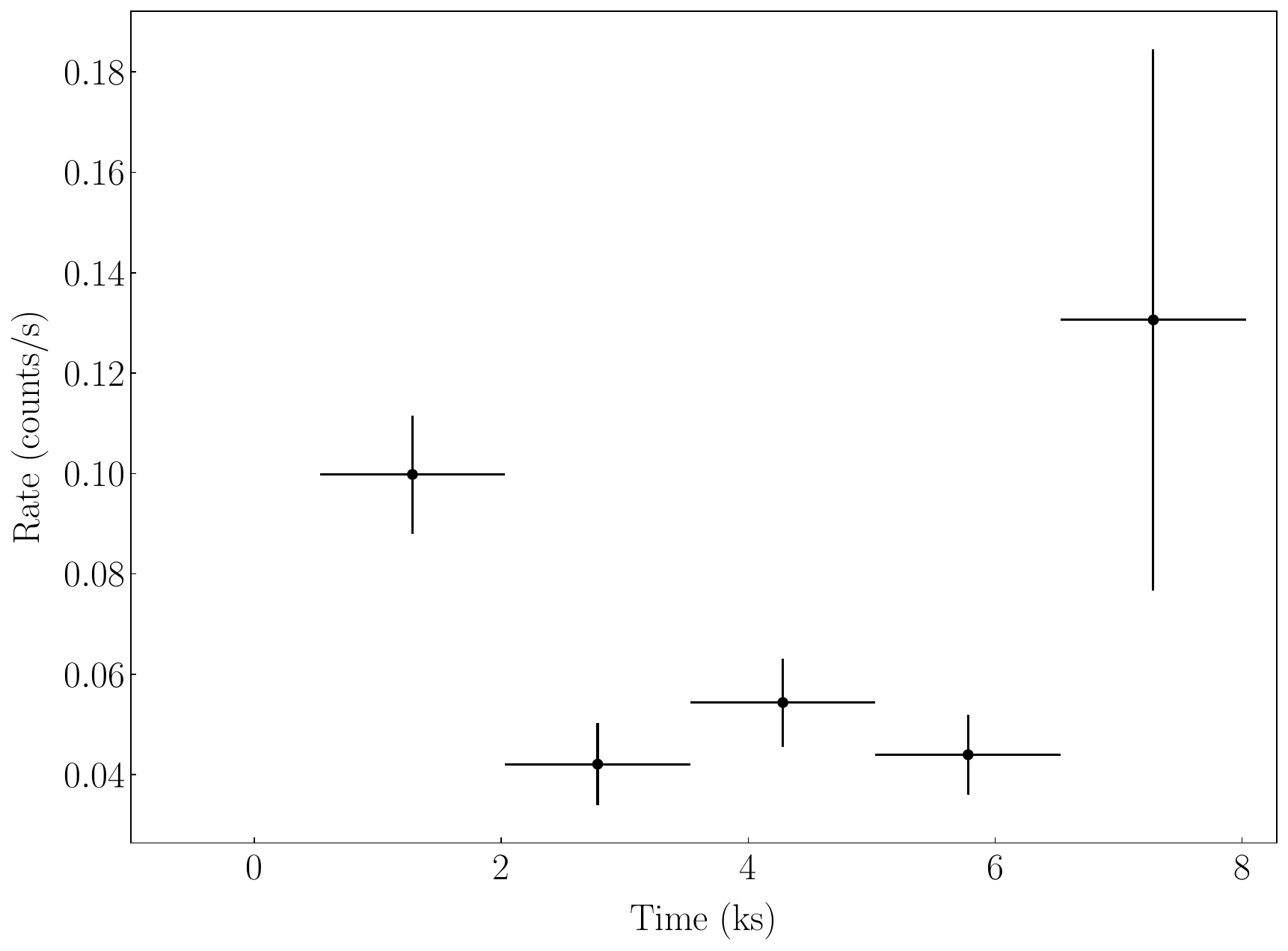}
     \caption{\xmm\ light curve in the 0.3--10\,keV band during ObsID 0652311001.}
     \label{fig:X0652311001}
 \end{figure}

 \begin{figure}[htbp!]
     \centering
     \includegraphics[width=0.95\columnwidth]{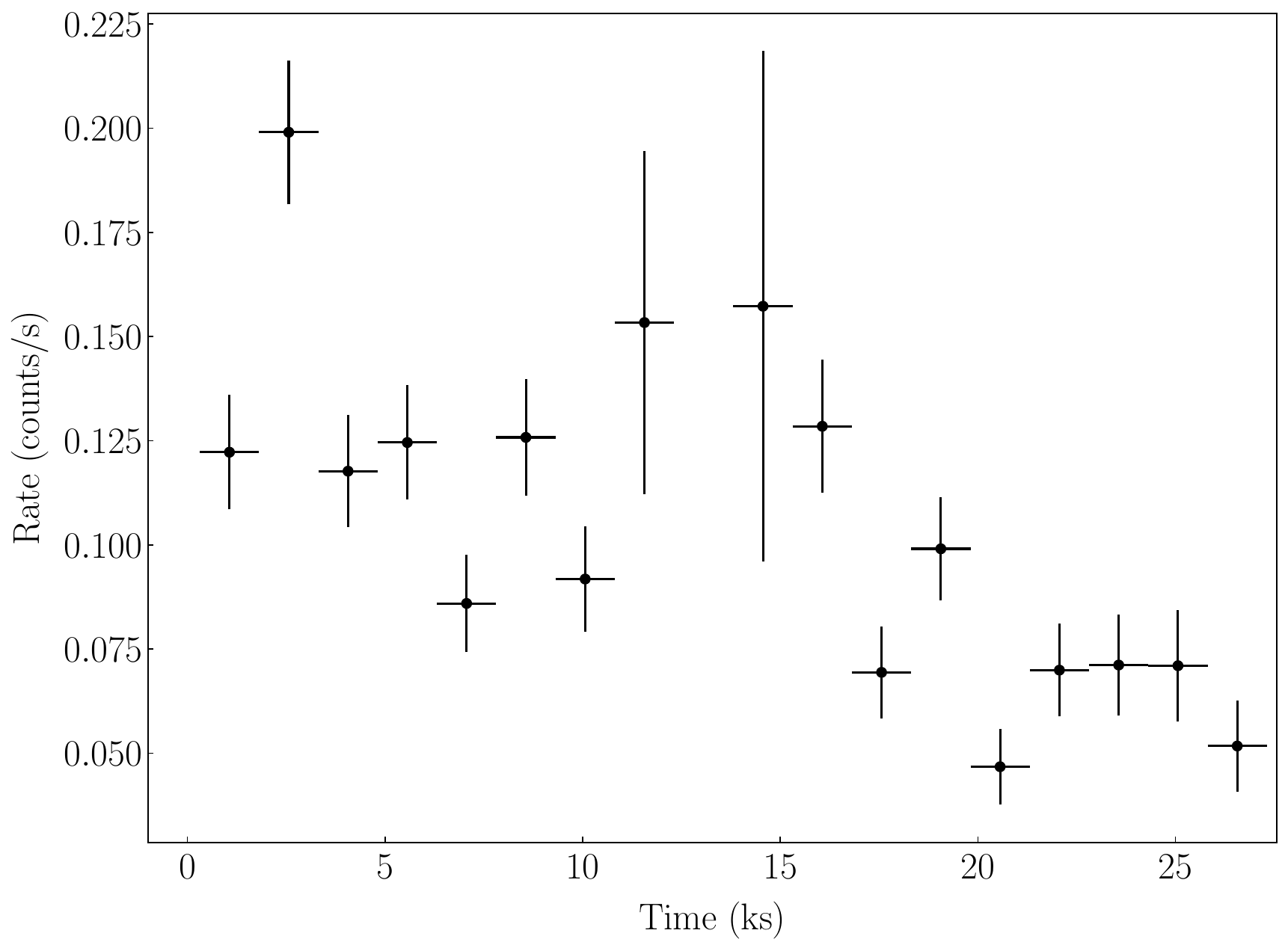}
     \caption{\xmm\ light curve in the 0.3--10\,keV band during ObsID 0691610301.}
     \label{fig:X0691610301}
 \end{figure}

 \begin{figure}[htbp!]
     \centering
     \includegraphics[width=0.95\columnwidth]{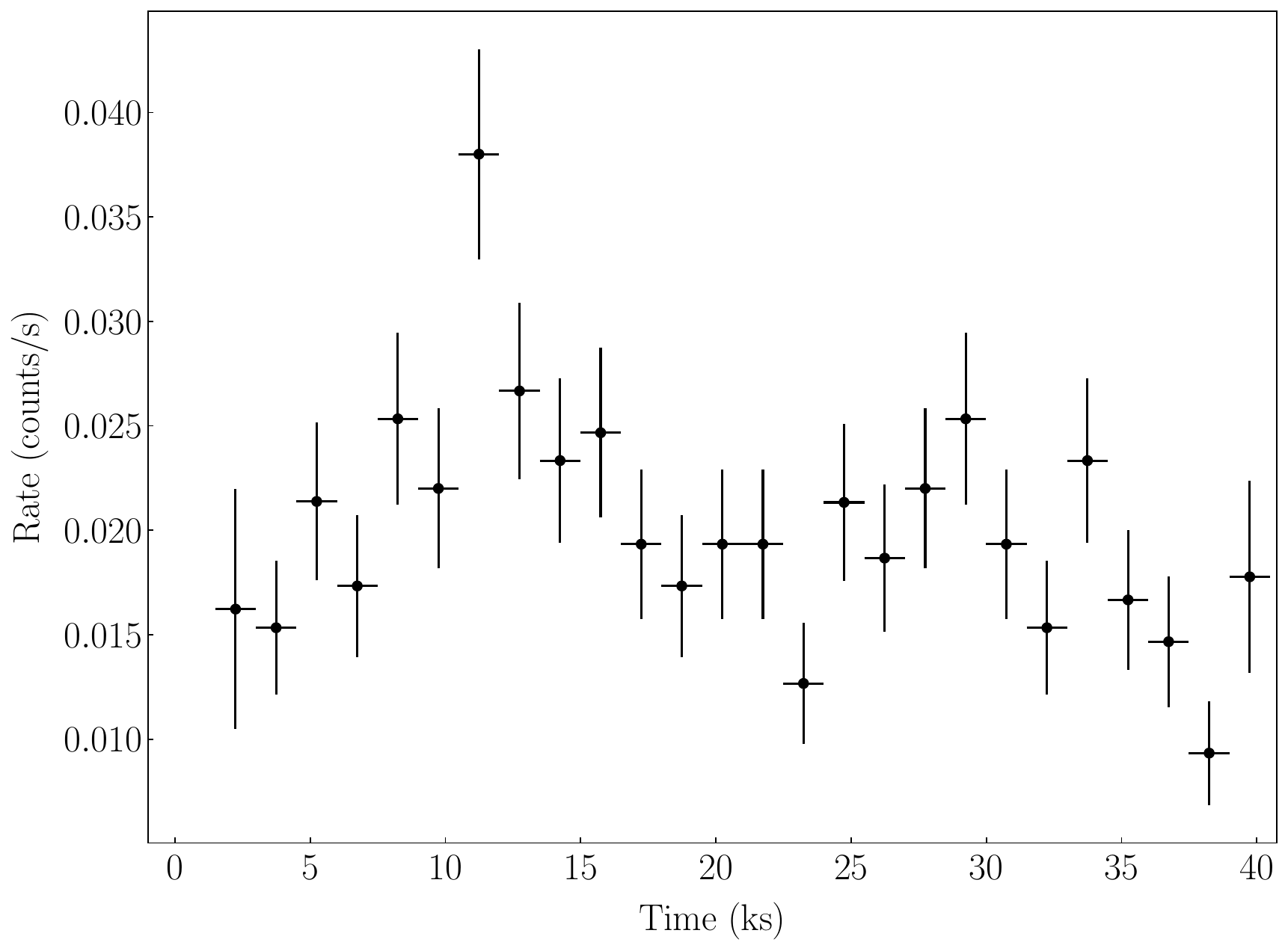}
     \caption{\chandra\ light curve in the 0.5--7\,keV band during ObsID 22930.}
     \label{fig:C22930}
 \end{figure}

  \begin{figure}[htbp!]
     \centering
     \includegraphics[width=0.95\columnwidth]{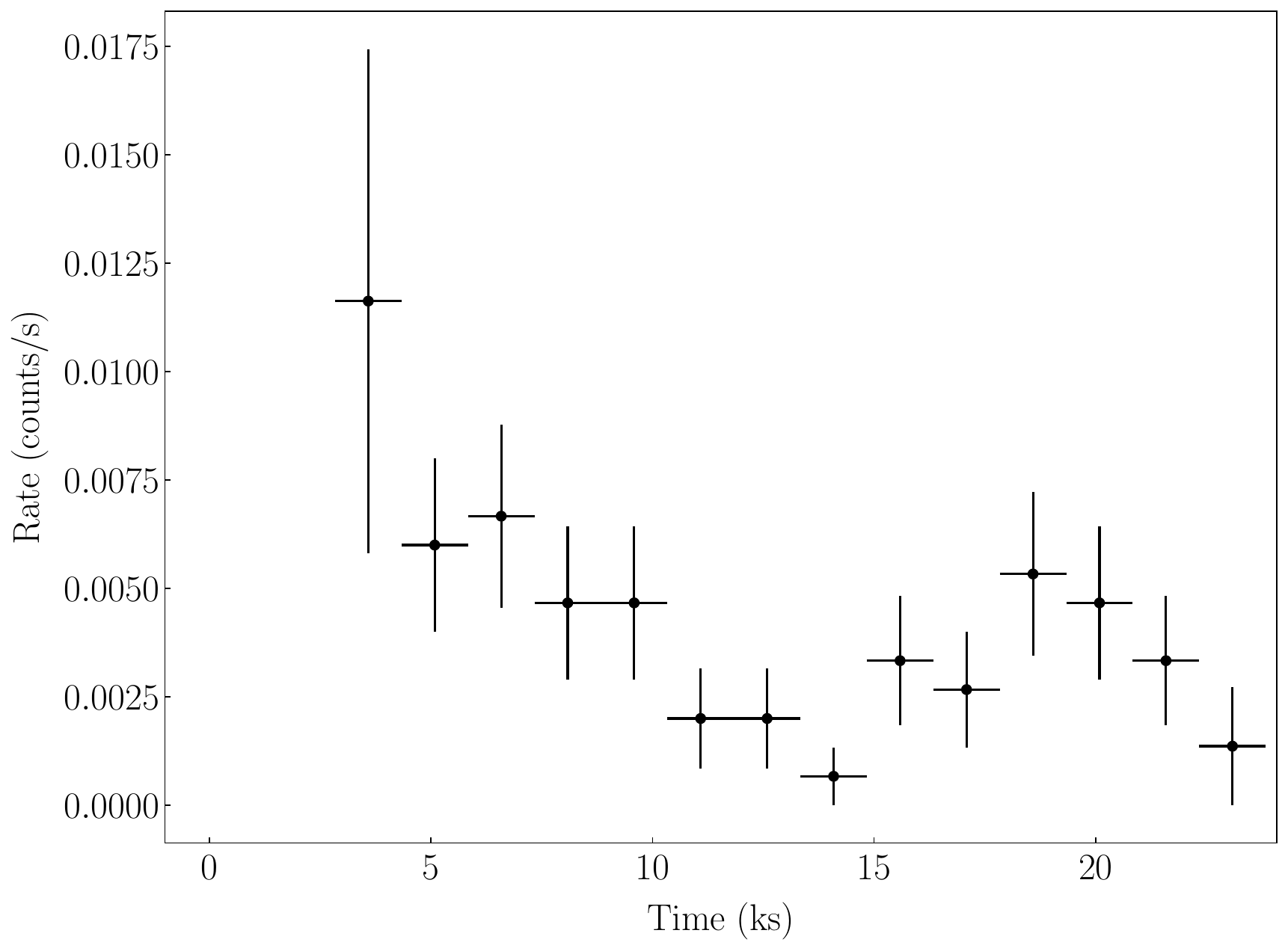}
     \caption{\chandra\ light curve in the 0.5--7\,keV band during ObsID 23361.}
     \label{fig:C23361}
 \end{figure}

  \begin{figure}[htbp!]
     \centering
     \includegraphics[width=0.95\columnwidth]{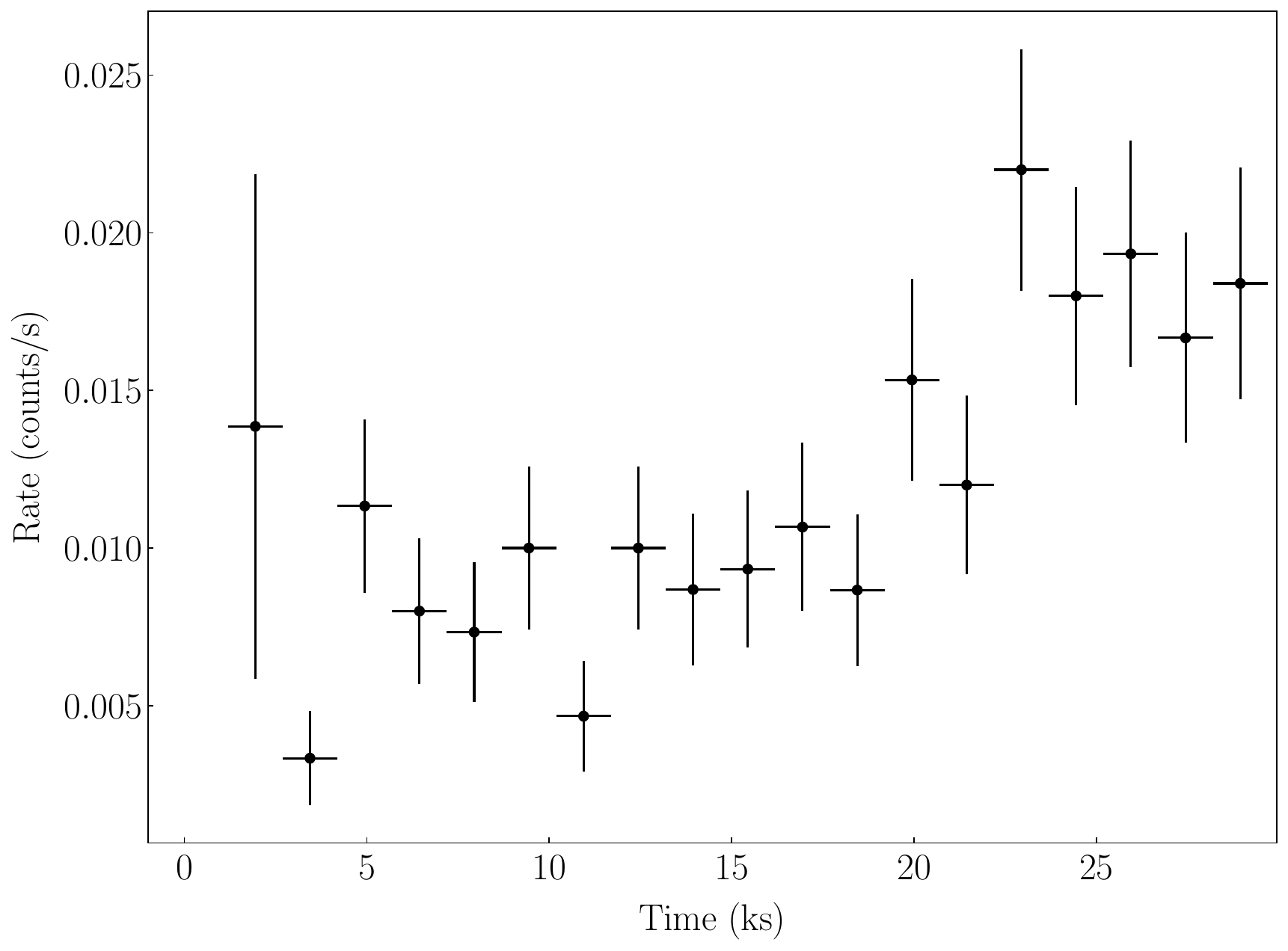}
     \caption{\chandra\ light curve in the 0.5--7\,keV band during ObsID 24853.}
     \label{fig:C24853}
 \end{figure}

  \begin{figure}[htbp!]
     \centering
     \includegraphics[width=0.95\columnwidth]{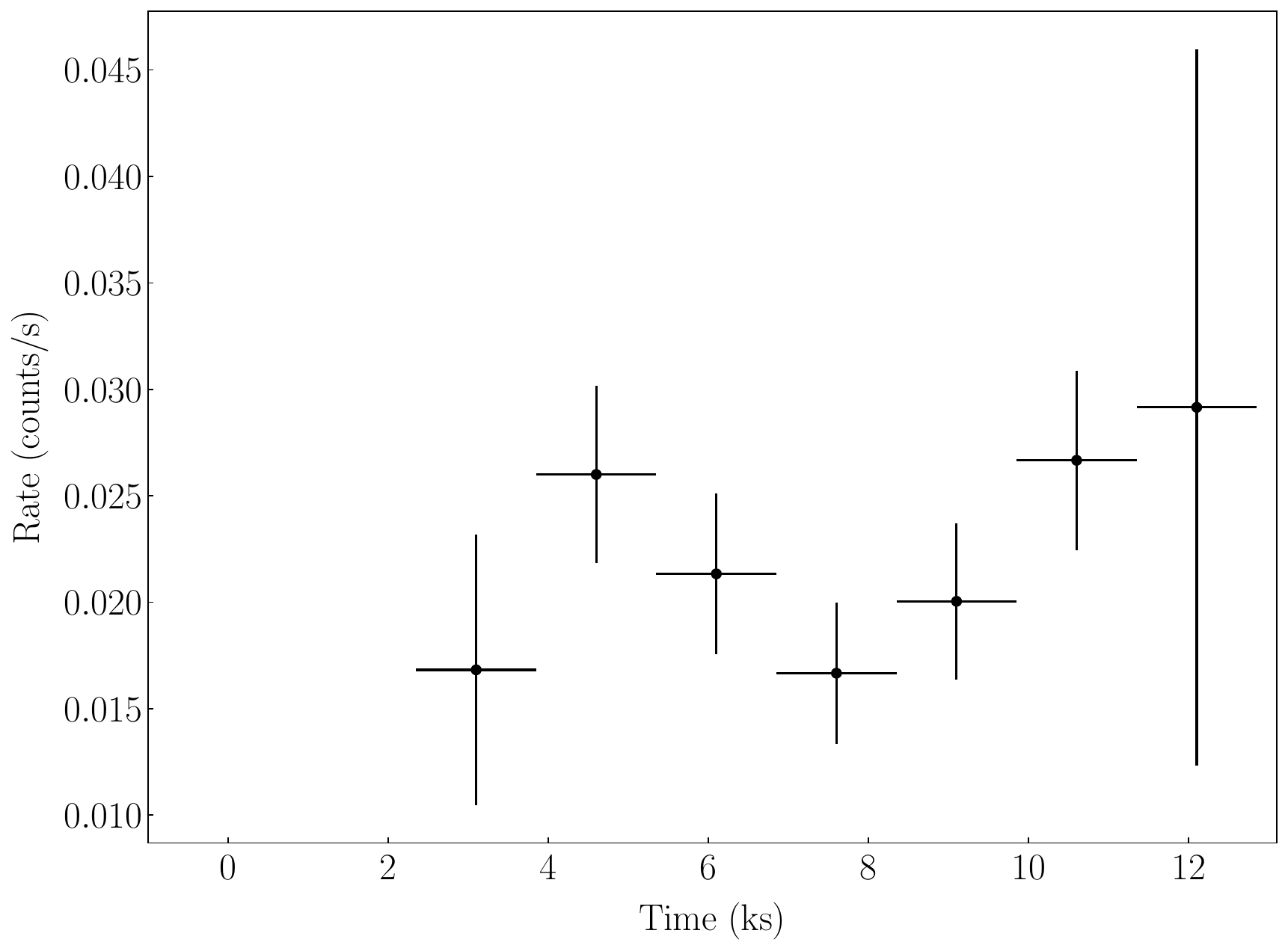}
     \caption{\chandra\ light curve in the 0.5--7\,keV band during ObsID 24854.}
     \label{fig:C24854}
 \end{figure}



\section{Spectra}\label{appendix:spectra} 
 \begin{figure}[htbp!]
     \centering
     \includegraphics[width=0.95\columnwidth]{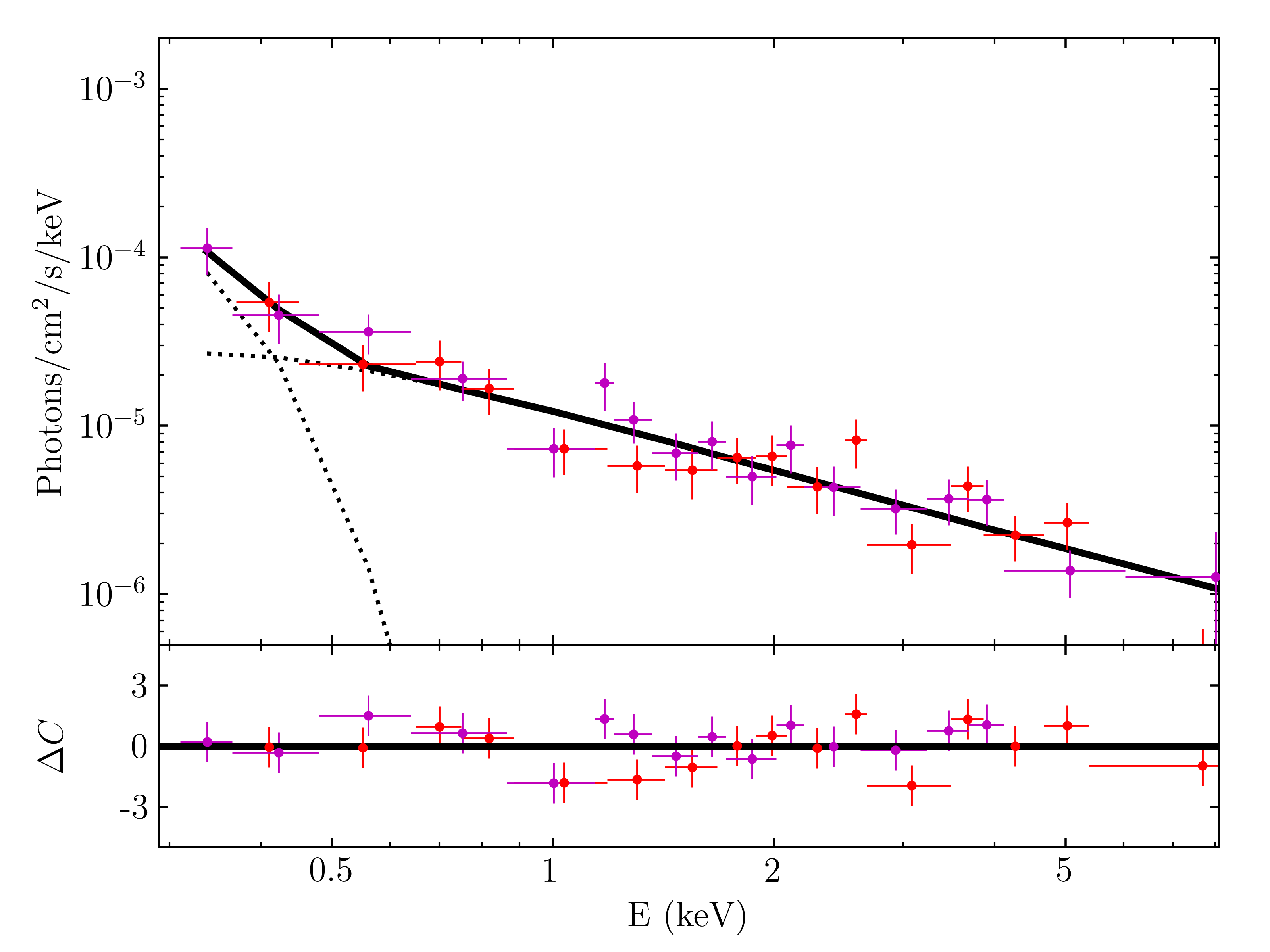}
     \caption{\xmm\ spectrum during ObsID 0403150201.}
     \label{fig:O2_XMM2}
 \end{figure}
 \begin{figure}[htbp!]
     \centering
     \includegraphics[width=0.95\columnwidth]{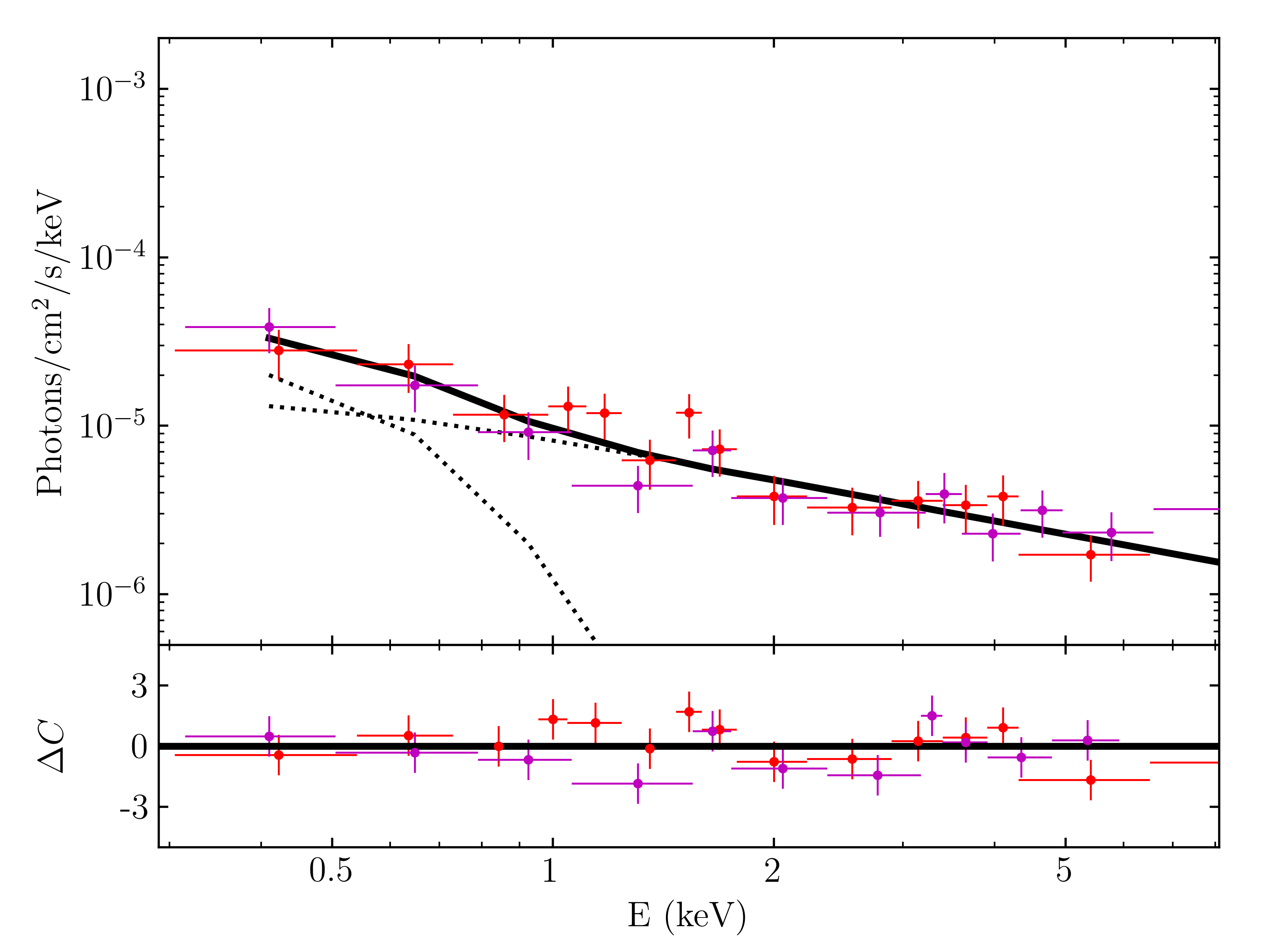}
     \caption{\xmm\ spectrum during ObsID 0403150101.}
     \label{fig:O3_XMM3}
 \end{figure}
 \begin{figure}[htbp!]
     \centering
     \includegraphics[width=0.95\columnwidth]{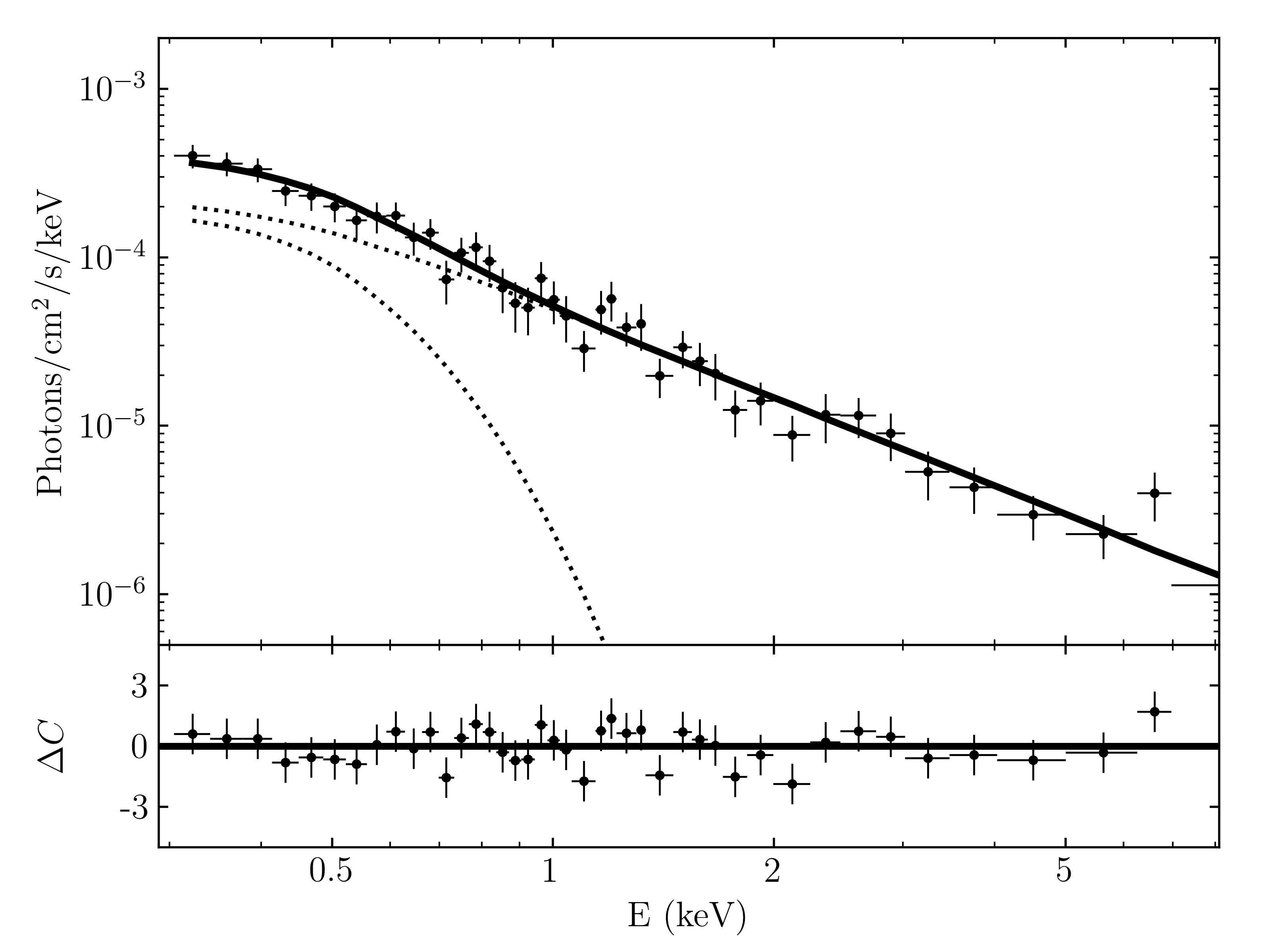}
     \caption{\xmm\ spectrum during ObsID 0652310401.}
     \label{fig:O4_XMM4}
 \end{figure}
 \begin{figure}[htbp!]
     \centering
     \includegraphics[width=0.95\columnwidth]{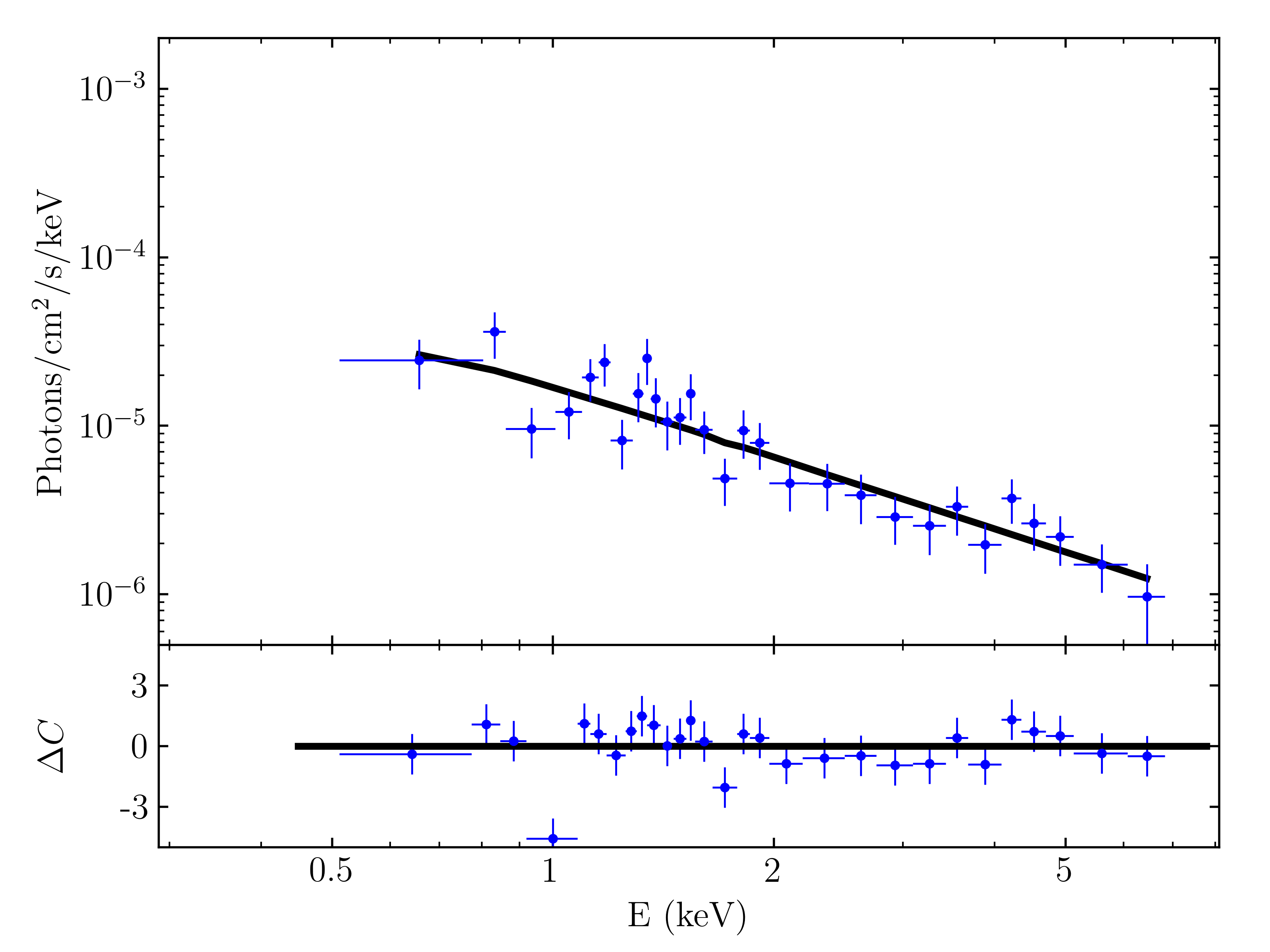}
     \caption{\chandra\ spectrum during ObsID 12887.}
     \label{fig:O5_CXO1}
 \end{figure}
 \begin{figure}[htbp!]
     \centering
     \includegraphics[width=0.95\columnwidth]{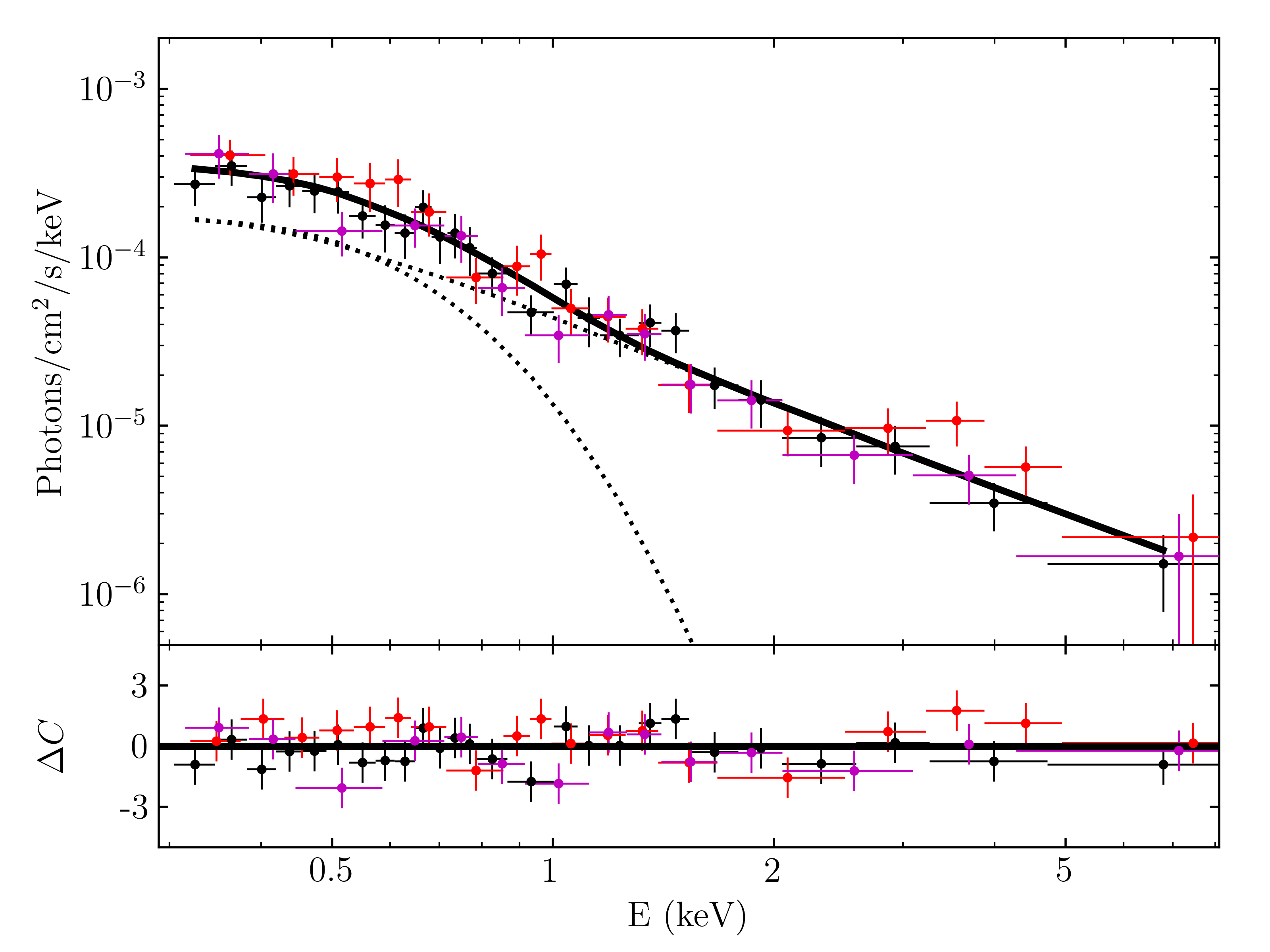}
     \caption{\xmm\ spectrum during ObsID 0652310801.}
     \label{fig:O6_XMM5}
 \end{figure}
 \begin{figure}[htbp!]
     \centering
     \includegraphics[width=0.95\columnwidth]{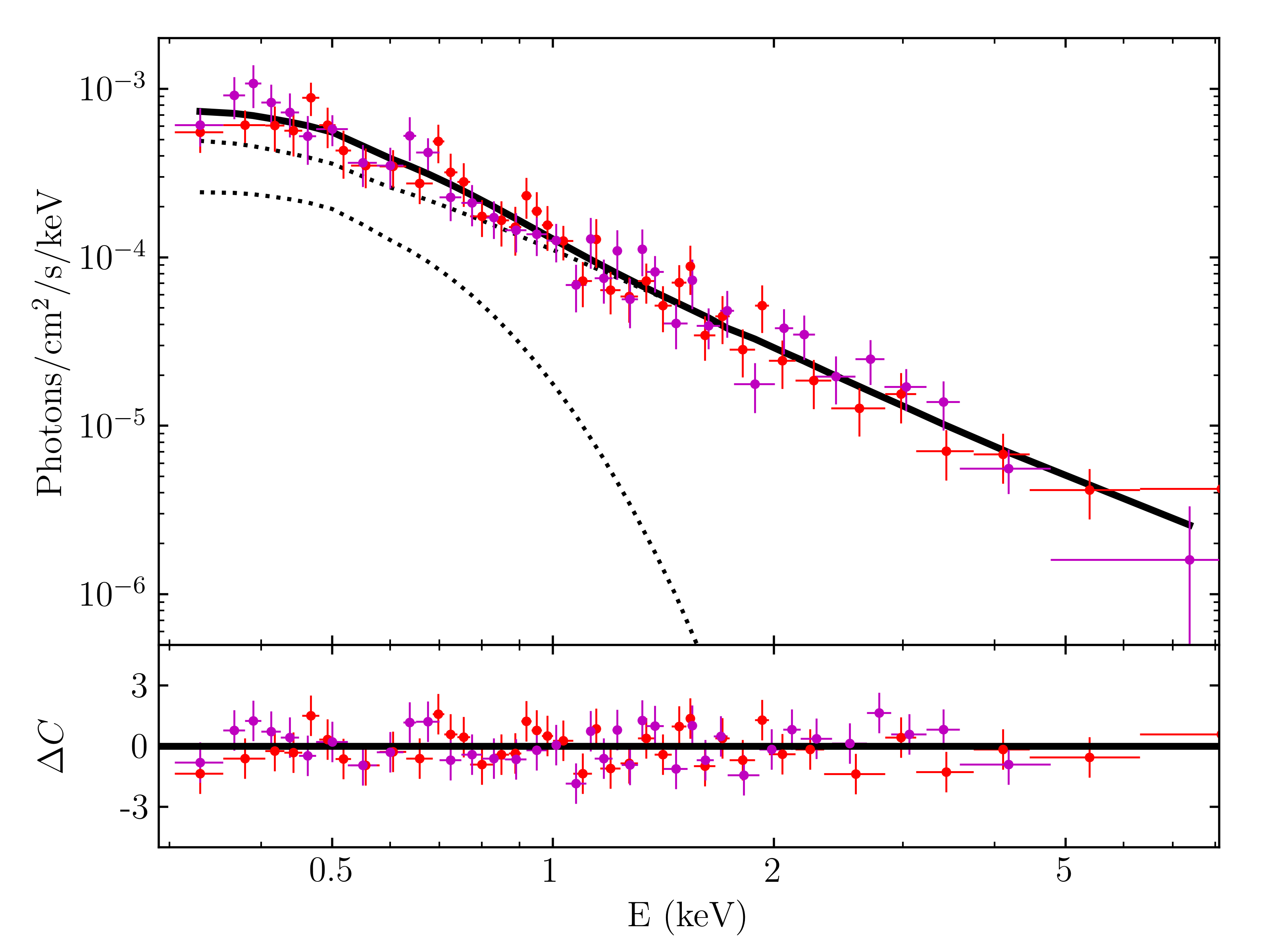}
     \caption{\xmm\ spectrum during ObsID 0652310901.}
     \label{fig:O7_XMM6}
 \end{figure}
 \begin{figure}[htbp!]
     \centering
     \includegraphics[width=0.95\columnwidth]{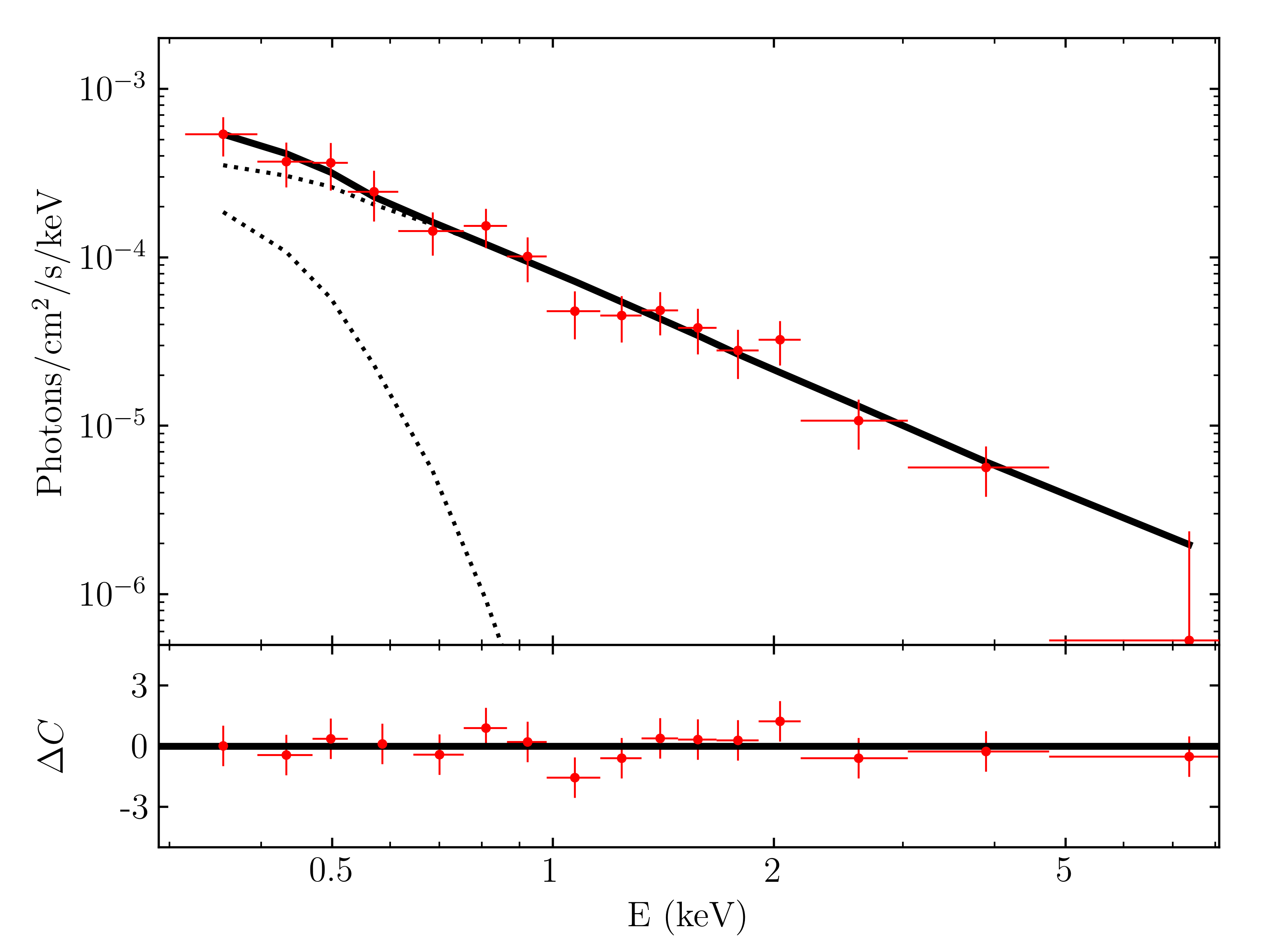}
     \caption{\xmm\ spectrum during ObsID 0652311001.}
     \label{fig:O8_XMM7}
 \end{figure}
 \begin{figure}[htbp!]
     \centering
     \includegraphics[width=0.95\columnwidth]{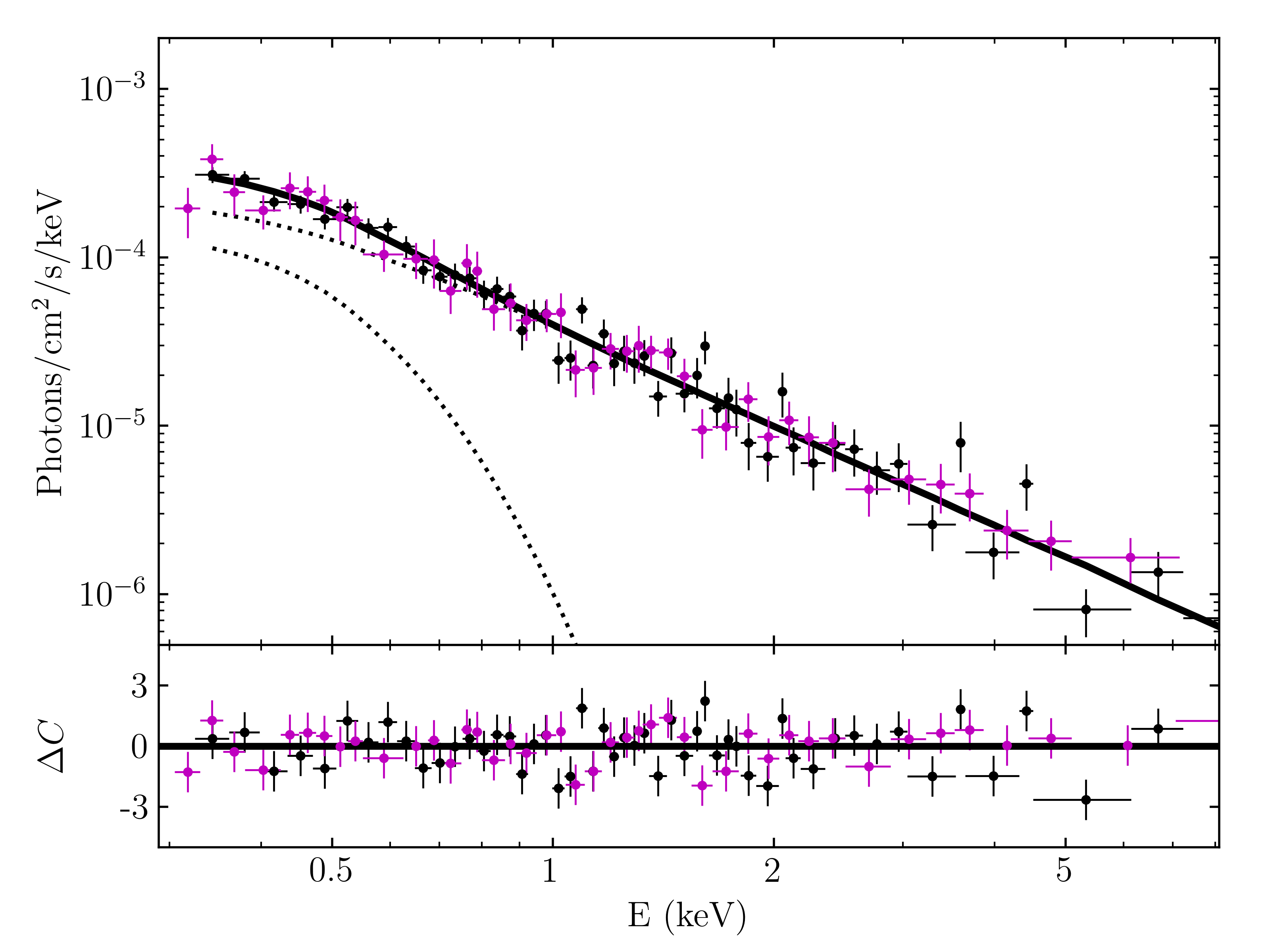}
     \caption{\xmm\ spectrum during ObsID 0691610201.}
     \label{fig:O9_XMM8}
 \end{figure}
 \begin{figure}[htbp!]
     \centering
     \includegraphics[width=0.95\columnwidth]{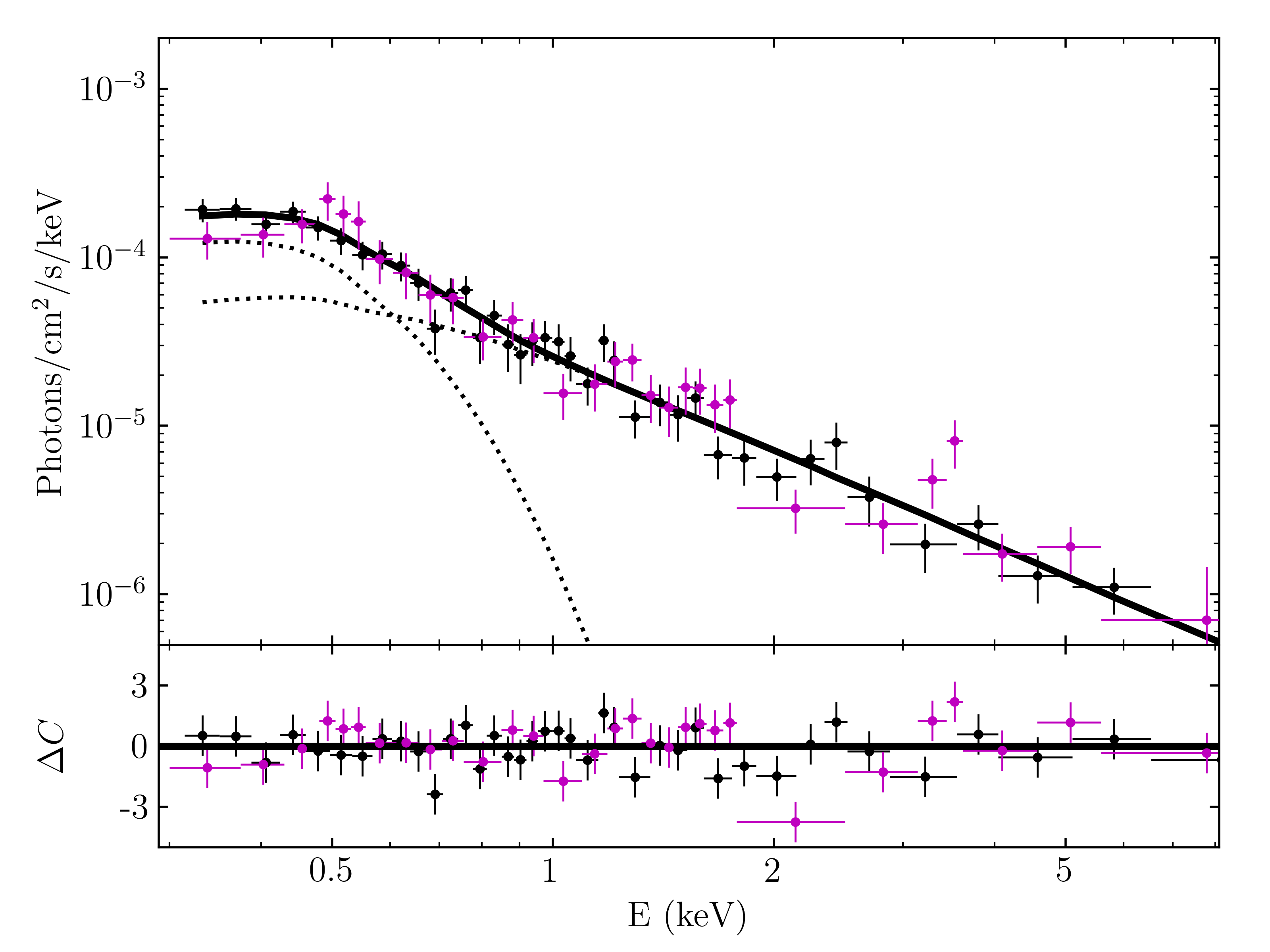}
     \caption{\xmm\ spectrum during ObsID 0691610301.}
     \label{fig:O10_XMM9}
 \end{figure}
 \begin{figure}[htbp!]
     \centering
     \includegraphics[width=0.95\columnwidth]{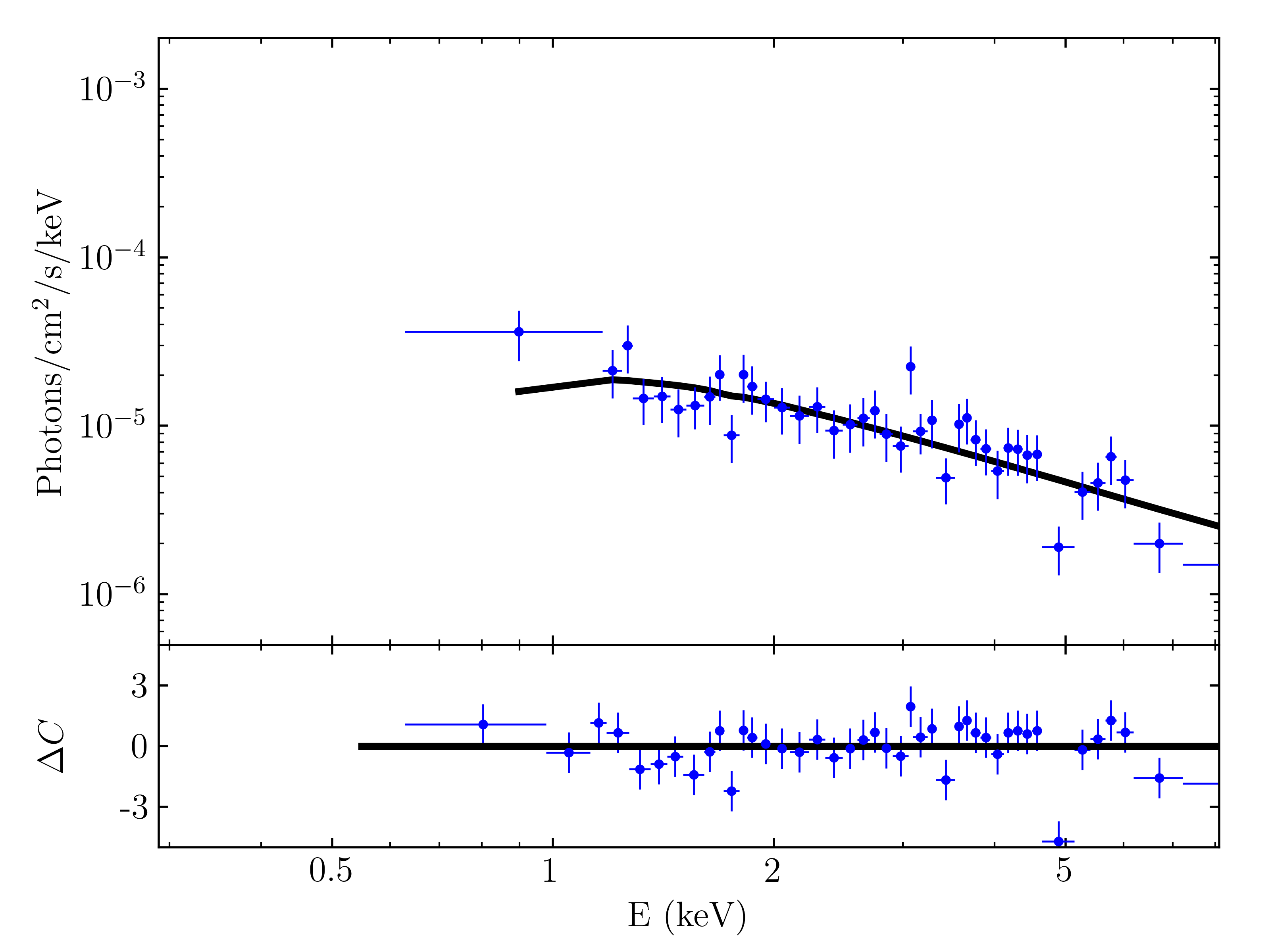}
     \caption{\chandra\ spectrum during ObsID 22648.}
     \label{fig:O11_CXO2}
 \end{figure}
 \begin{figure}[htbp!]
     \centering
     \includegraphics[width=0.95\columnwidth]{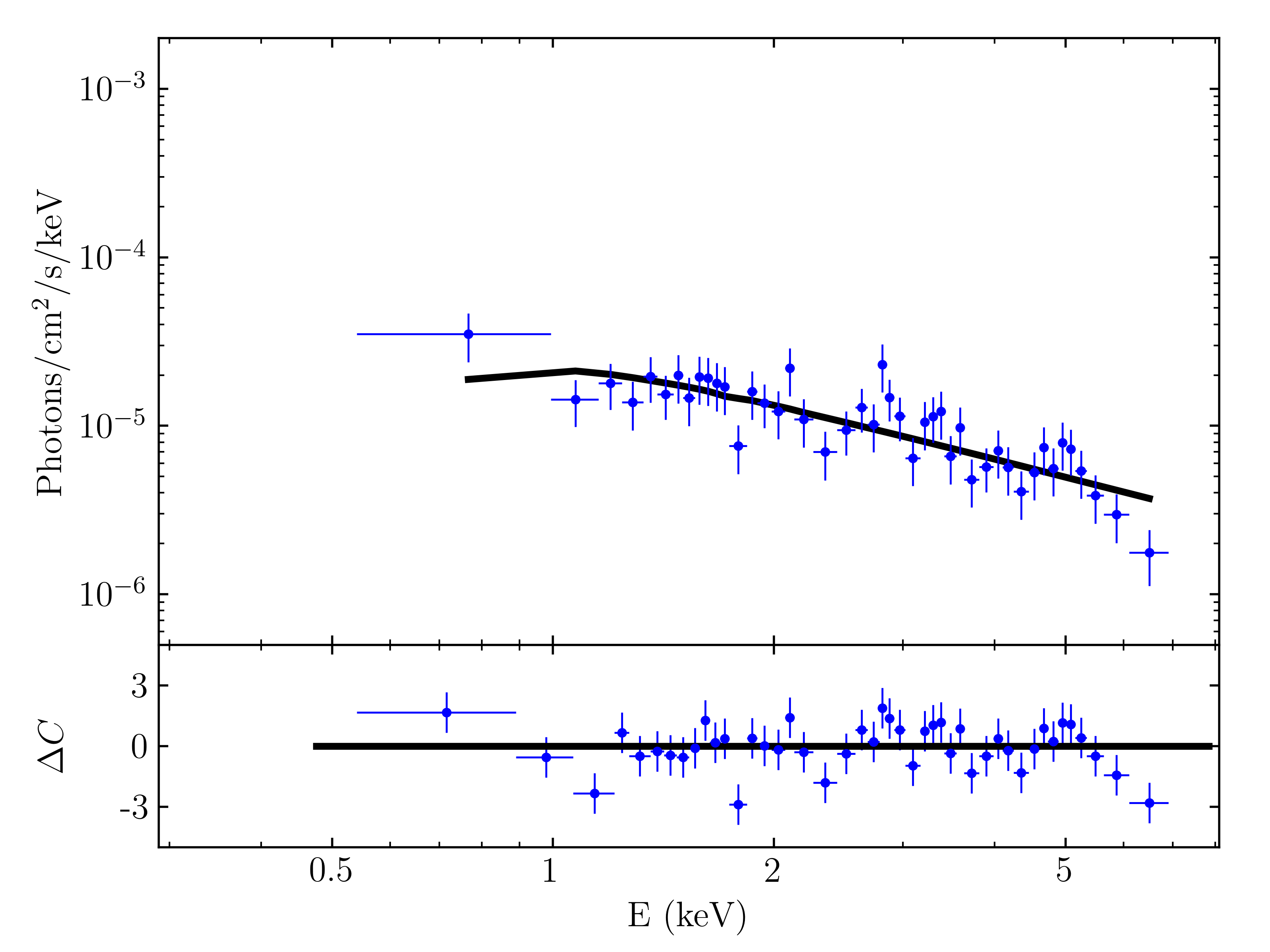}
     \caption{\chandra\ spectrum during ObsID 22649.}
     \label{fig:O12_CXO3}
 \end{figure}
 \begin{figure}[htbp!]
     \centering
     \includegraphics[width=0.95\columnwidth]{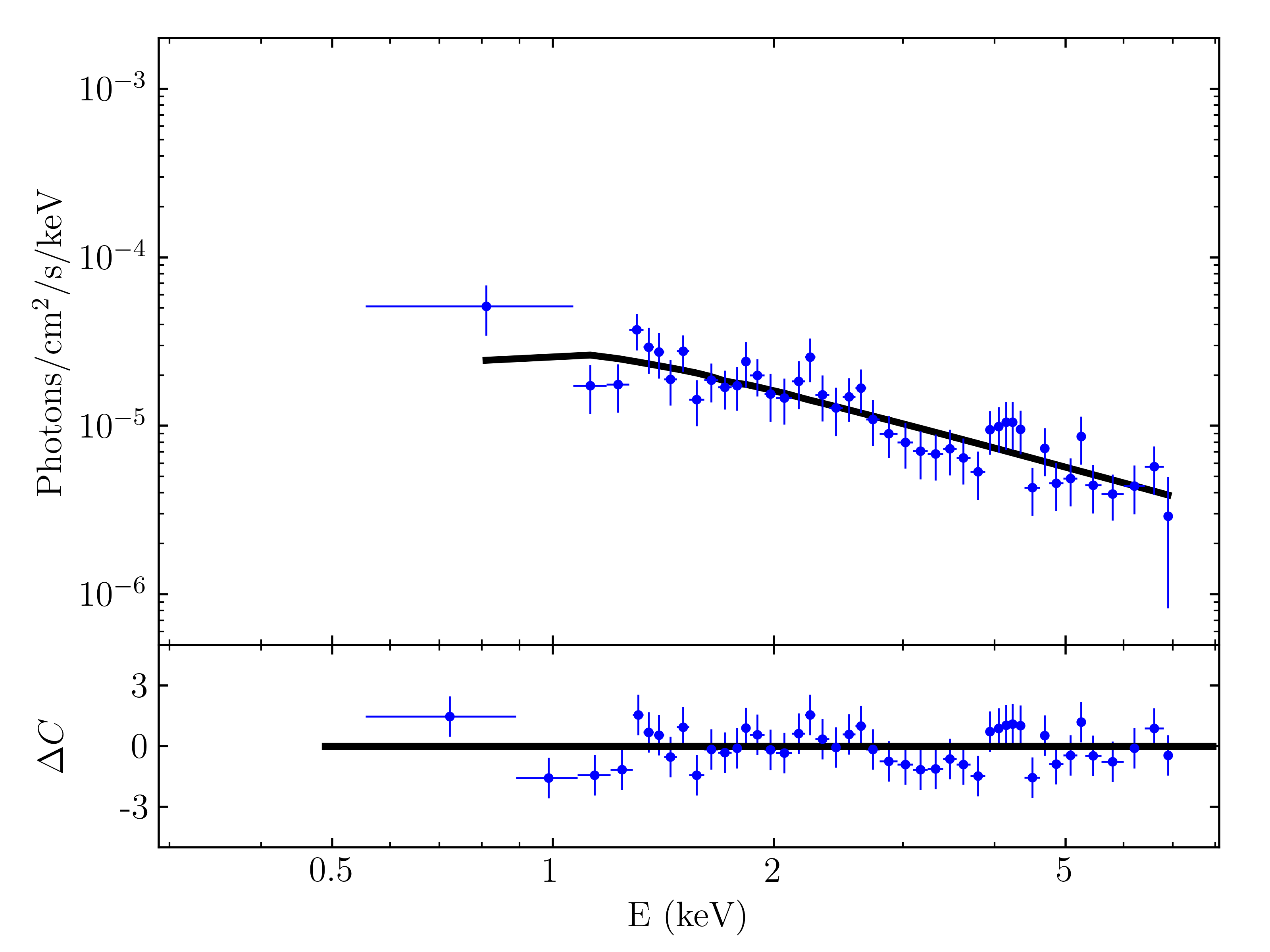}
     \caption{\chandra\ spectrum during ObsID 23182.}
     \label{fig:O13_CXO4}
 \end{figure}
 \begin{figure}[htbp!]
     \centering
     \includegraphics[width=0.95\columnwidth]{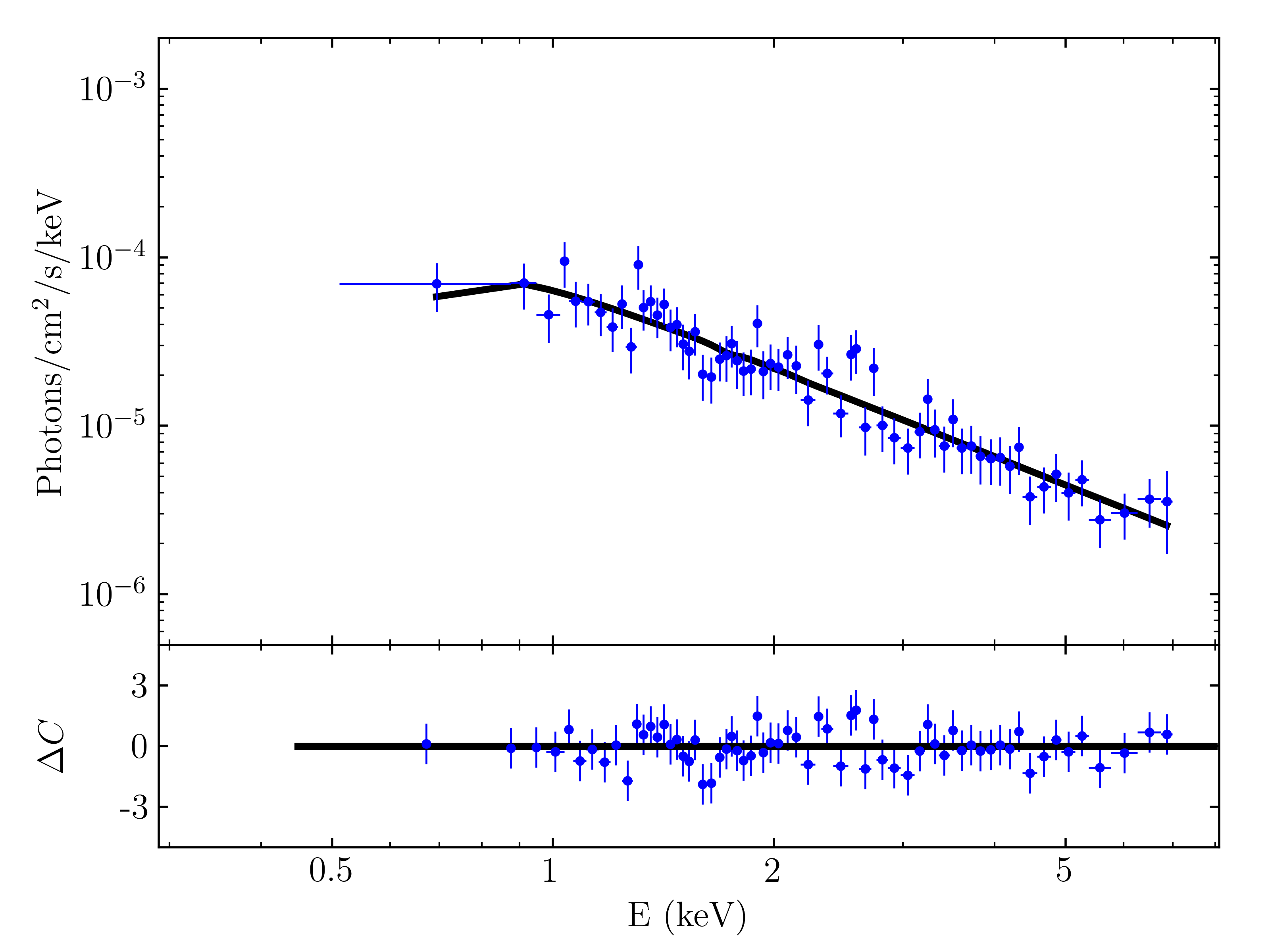}
     \caption{\chandra\ spectrum during ObsID 22930.}
     \label{fig:O14_CXO5}
 \end{figure}
 \begin{figure}[htbp!]
     \centering
     \includegraphics[width=0.95\columnwidth]{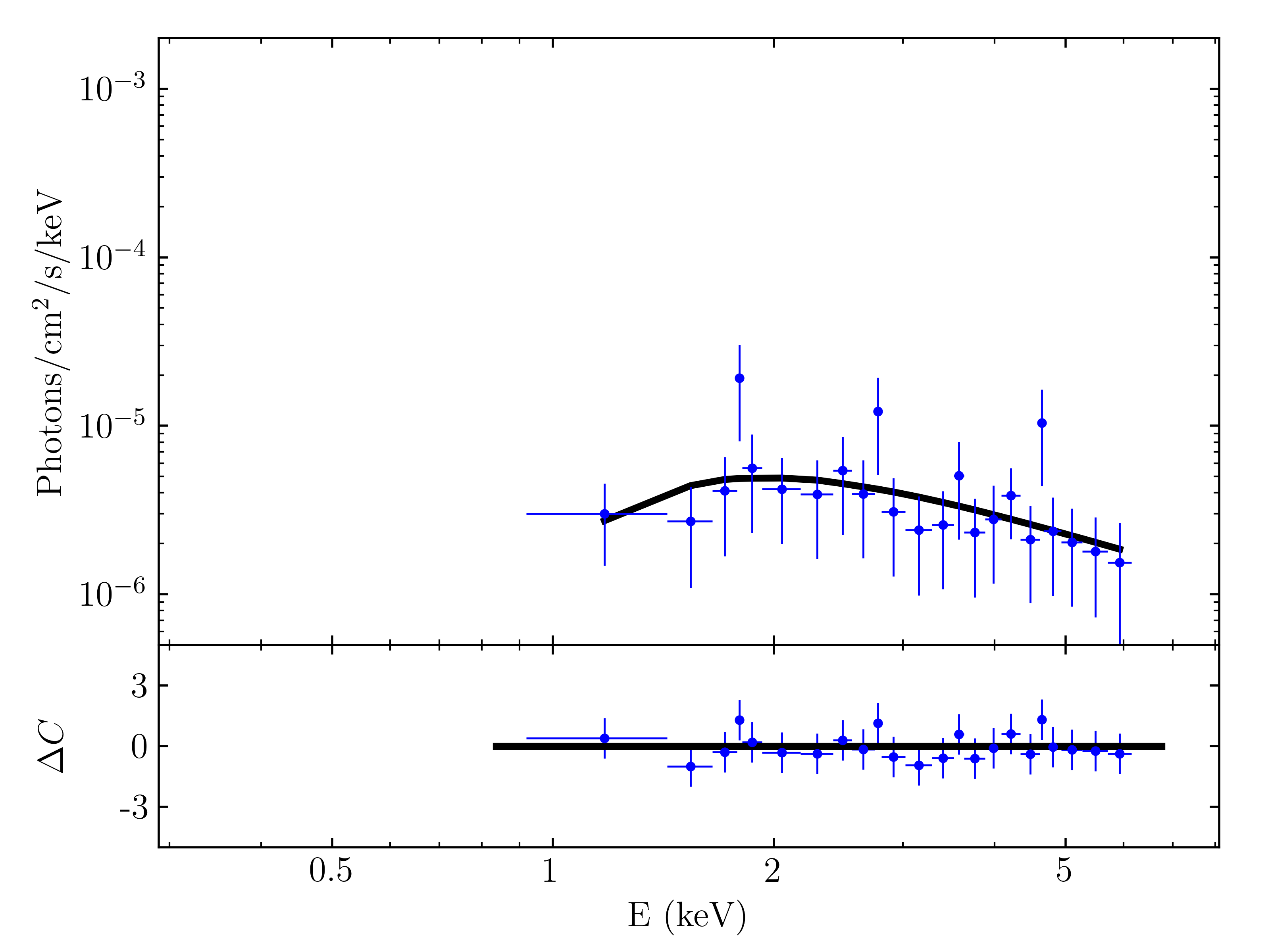}
     \caption{\chandra\ spectrum during ObsID 23361.}
     \label{fig:O15_CXO6}
 \end{figure}
 \begin{figure}[htbp!]
     \centering
     \includegraphics[width=0.95\columnwidth]{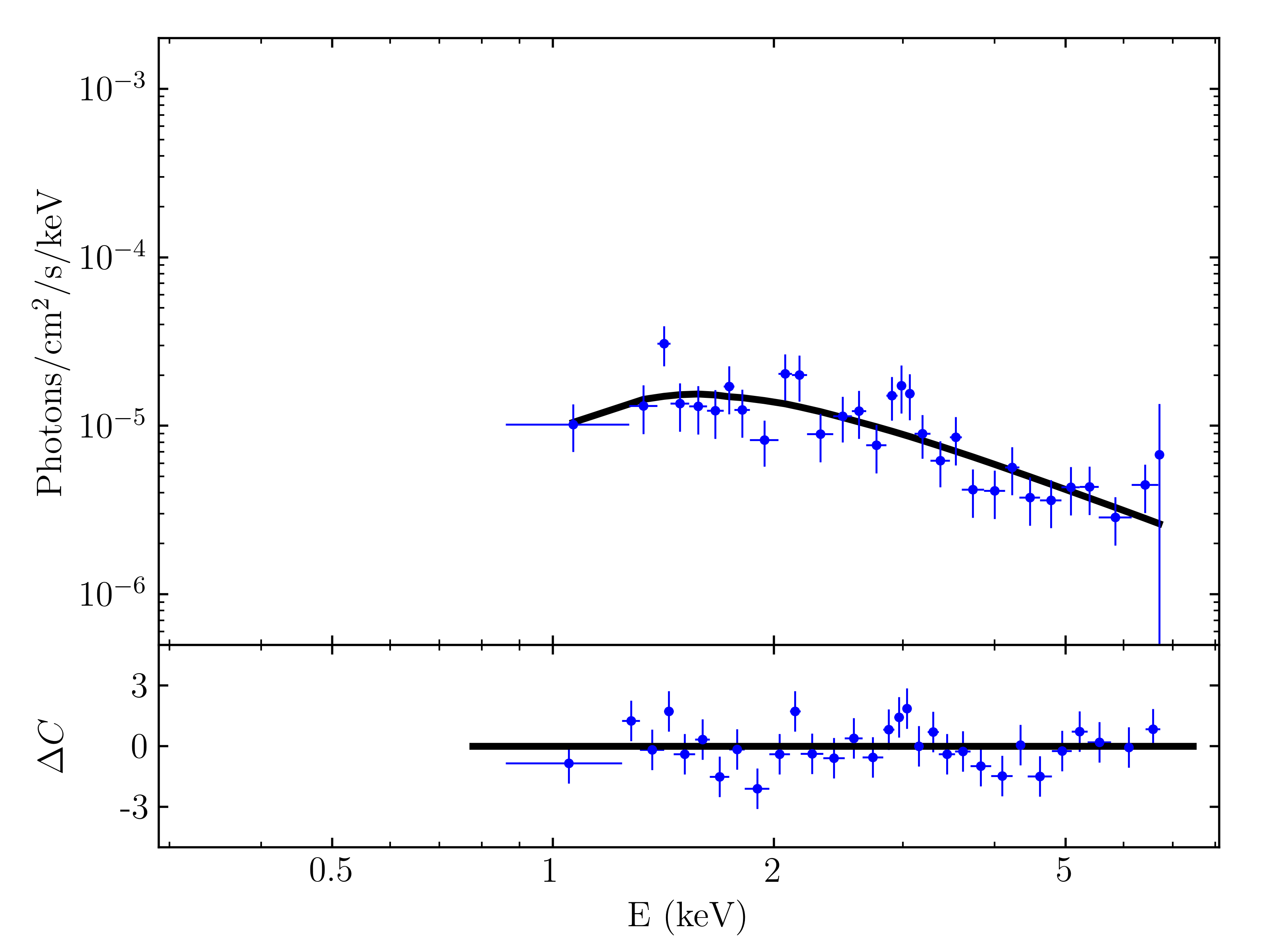}
     \caption{\chandra\ spectrum during ObsID 24853.}
     \label{fig:O16_CXO7}
 \end{figure}
 \begin{figure}[htbp!]
     \centering
     \includegraphics[width=0.95\columnwidth]{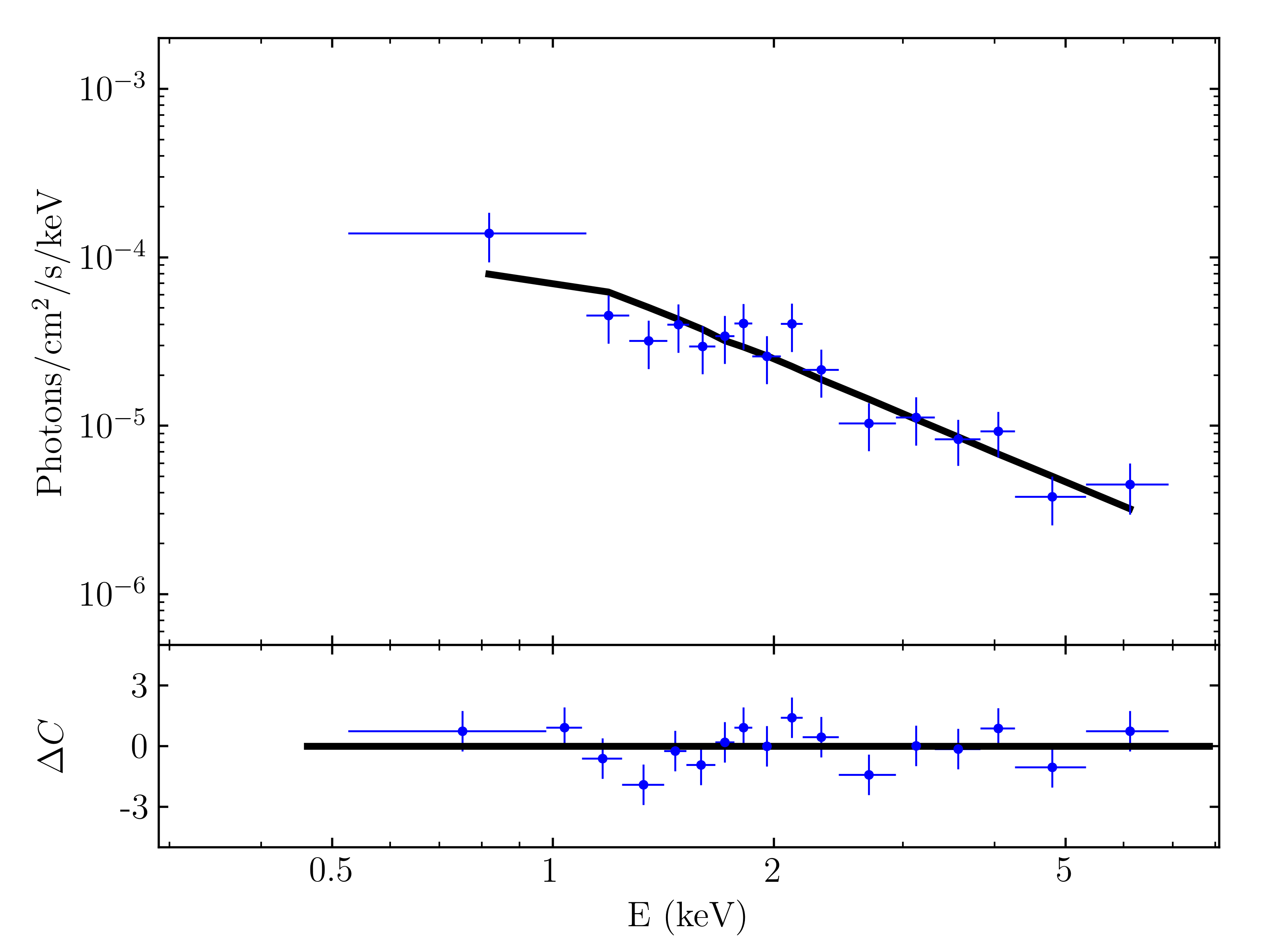}
     \caption{\chandra\ spectrum during ObsID 24854.}
     \label{fig:O17_CXO8}
 \end{figure}

\end{appendix}

\end{document}